\documentclass[
 reprint,
nofootinbib,
 amsmath,amssymb,
 aps,
]{revtex4-2}

\usepackage{graphicx}
\usepackage{dcolumn}
\usepackage{bm}
\usepackage{xcolor}
\usepackage{subfigure}
\usepackage{hyperref}
\usepackage[mathlines]{lineno}

\newcommand{\kfive}{\kappa_{5}^{2}}
\newcommand{\sech}{\operatorname{sech}}

\usepackage{graphicx}
\usepackage{hyperref}

\begin{document}

\title{Five-dimensional thick branes in the scalar representation of $f(Q,B)$ gravity}

\author{F. C. E. Lima}
\email{cleiton.estevao@ufabc.edu.br}
\affiliation{Centro de Matématica, Computação e Cognição (CMCC), Universidade Federal do ABC (UFABC), Av. dos Estados 5001, CEP 09210-580, Santo André, São Paulo, Brazil.}

\author{F. M. Belchior}
\email{fernandobelcks7@gmail.com}
\affiliation{Departamento de F\'isica, Universidade Federal da Para\'iba, Centro de Ci\^encias Exatas e da Natureza, CEP 58051-970, Jo\~ao Pessoa, Para\'iba, Brazil.}

\author{C. A. S. Almeida}
\email{carlos@fisica.ufc.br}
\affiliation{Departamento de F\'{\i}sica, Universidade Federal do Cear\'{a}, Centro de Ci\^encias, Campus do Pici, CEP 60.440-900, Fortaleza, Cear\'{a}, Brazil.}

\begin{abstract}
One examines a codimension-one thick braneworld in the scalar representation of $f(Q, B)$ gravity, where $Q$ is the nonmetricity scalar, and $B$ is the boundary term satisfying $\mathring R=Q-B$. We write the gravitational sector in terms of two auxiliary scalar fields, $\varphi=f_Q$ and $\psi=f_B$, and an interaction potential $U(\varphi,\psi)$ obtained by a Legendre transformation from the $f(Q,B)$ function. Working in the coincident gauge and assuming a five-dimensional warped geometry with four-dimensional Poincar\'{e} symmetry, we derive the complete set of brane equations sourced by a canonical bulk scalar field. To accomplish our purpose, we address two models of smooth asymptotically AdS thick brane with localized scalar profiles. Furthermore, we also analyze tensor perturbations and demonstrate the localization and stability of the massless graviton.
\end{abstract}

\keywords{Thick branes; symmetric teleparallel gravity; nonmetricity; boundary term; scalar-tensor representation.}

\maketitle

\section{Introduction}

The idea that our four-dimensional universe may be a hypersurface embedded in a higher-dimensional spacetime has led to a broad class of braneworld models \cite{Maartens,Kakushadze,Geng,Deffayet,Deffayet2,Langlois}. In the Randall-Sundrum scenario \cite{Randall1,Randall2}, the hierarchy between the gravitational and electroweak scales can be addressed by a warped extra dimension \cite{Randall1,Randall2}. The original construction assumes an infinitely thin brane, whose matter distribution is represented by a Dirac delta function. A more physically realistic framework replaces this singular source with a smooth domain-wall configuration generated by a bulk scalar field \cite{Bazeia1,Bazeia2,Bazeia3}. These regular configurations are commonly referred to as thick branes \cite{Gremm,deWolfe}. They smooth out the geometry near the brane core, provide a well-defined setting for investigating the localization of matter fields, and may exhibit a rich and nontrivial internal structure.

A second direction in which braneworld gravity can be generalized is to change the geometrical origin of the gravitational interaction \cite{Yang,Fu}. In the usual formulation of general relativity, gravity is described by curvature \cite{Wald}. Thus, one can construct equivalent formulations from torsion or nonmetricity \cite{Maluf,Aldrovandi2}. In the symmetric teleparallel description, the curvature and torsion of the independent affine connection vanish, while the nonmetricity tensor is nonzero \cite{Jarv}. The gravitational scalar $Q$ reproduces the Einstein dynamics up to a boundary contribution \cite{Nester}. Because nonlinear functions spoil this equivalence, $f(Q)$ gravity provides a genuine modification of general relativity \cite{Heisenberg,Capozziello}. A further extension is obtained by including the boundary term $B$ by considering modified $f(Q, B)$ gravity \cite{Capozziello,Paliathanasis,Capozziello2}. Within this framework, the combination $Q-B$ is particularly interesting, since it is a Ricci scalar built from the Levi-Civita connection \cite{Jarv}. Therefore, the special sector $f(Q-B)$ is dynamically related to curvature-based $f(R)$ gravity, with the $f(Q, B)$ function containing nonmetricity and boundary effects independently \cite{De,Capozziello,Paliathanasis,Capozziello2}.

Our purpose is to construct a five-dimensional thick braneworld in the scalar representation of the $f(Q,B)$ modified gravity. Instead of choosing a particular nonlinear function f at the beginning, we introduce two scalar fields, $\varphi\equiv f_Q$ and $\psi\equiv f_B$, and a scalar potential $U(\varphi,\psi)$. This representation is useful for brane physics because it separates the two geometrical sectors, where the scalar $\varphi$ controls the nonmetricity sector, while the scalar $\psi$ measures how the boundary term affects the bulk equations. Furthermore, the warp factor and the two geometric scalar profiles are specified, and the matter scalar, the matter potential, and the on-shell function $f(Q,B)$ are reconstructed from the field equations.

The paper is organized as follows. In Sec. \ref{geometry}, we review the geometrical ingredients of symmetric teleparallel gravity and present the scalar representation of $f(Q, B)$ gravity. In Sec. \ref{brane}, we obtain the field equations for a five-dimensional warped geometry and derive the brane system. In Sec. \ref{analytic_model}, we construct an analytic thick-brane model and discuss its physical constraints. Furthermore, we also examined a generalizing model with a deformable geometry in Sec. \ref{sec:deformed-brane}. In Sec. \ref{rrm}, we investigate the gravitational tensor perturbations, examining the localization and stability of the massless graviton. Our conclusions are given in Sec. \ref{conclusion}.

\section{Geometric framework and scalar representation of $f(Q,B)$ gravity}\label{geometry}

The study of the geometric framework and scalar representation of $f(Q, B)$ gravity is particularly relevant because the boundary term $B$ provides a dynamical bridge between nonmetricity- and curvature-based descriptions of gravity. While generic nonlinear functions of $Q$ depart from the dynamics of curvature-based modified gravity, the sector $f(Q-B)$ recovers the corresponding $f(\mathring{R})$ theory. Moreover, by introducing the auxiliary fields $\Phi=f_Q$ and $\Psi=f_B$, the Legendre transformation makes the additional geometric degrees of freedom explicit and converts the original higher-order formulation into a dynamically equivalent scalar representation. This reformulation facilitates the identification of the $f(Q)$ and $f(\mathring{R})$ limits, the analysis of physical consistency and stability conditions, and the reconstruction of exact gravitational backgrounds. The relevance of this framework is further supported by recent applications of boundary-corrected nonmetricity gravity to accelerating cosmological attractors \cite{Paliathanasis2}, observationally constrained late-time dynamics \cite{Lohakare}, black-hole and regular-black-hole solutions \cite{junior}, the extended geometric trinity of gravity \cite{Capozziello3}, and classical and quantum minisuperspace cosmology \cite{Battista}.

Let us start by adopting a five-dimensional manifold equipped with a metric $g_{MN}$ and an independent affine connection $\Gamma^{P}{}_{MN}$. Capital Latin indices run over $0,1,2,3,5$, and Greek indices run over the four-dimensional brane coordinates. Within the symmetric teleparallel framework, the connection is constrained by $R^{P}{}_{QMN}(\Gamma)=0$ and $T^{P}{}_{MN}(\Gamma)=0$. The nontrivial geometrical field strength is the nonmetricity tensor, i.e., $Q_{A MN}\equiv \nabla_A g_{MN}$. Therefore, its two traces are
\begin{align}
Q_A\equiv Q_A{}^{M}{}_{M} \quad \mathrm{and} \quad \widetilde Q_A\equiv Q^{M}{}_{A M}.
   \label{traces}
\end{align}
Within this framework, one defines the nonmetricity conjugate as 
\begin{align}\nonumber
P^{A}{}_{MN}=&\frac{1}{4}Q^{A}{}_{MN}+\frac{1}{2}Q_{(MN)}{}^{A}+\frac{1}{4}\left(Q^{A}-\widetilde Q^{A}\right)g_{MN}\\
-&\frac14\delta^{A}{}_{(M}Q_{N)} .
\label{conjugate}
\end{align}
Therefore, the nonmetricity scalar will be $Q\equiv Q_{A MN}P^{A MN}$. Naturally, when adopting these conventions, the Levi-Civita Ricci scalar and the nonmetricity scalar are related by $\mathring{R}= Q-B$ \footnote{In this paper, $\mathring{R}$ is the Ricci scalar constructed exclusively from the Levi-Civita connection $\mathring{\Gamma}^{P}{}{MN}$, and the metric signature is $\eta_{\mu\nu}=\mathrm{diag}(-1,1,1,1)$.}, where $B$ is the boundary term given by
\begin{align}\label{e3}
B\equiv \mathring\nabla_A\left(Q^A-\widetilde Q^A\right).    
\end{align}

Therefore, once both curvature and torsion vanish, the affine connection is purely gauge and carries only inertial information. Thus, in the coincident gauge, we choose coordinates for which $\Gamma^{P}{}_{MN}=0$. This gauge is particularly efficient in thick-brane calculations, given that it does not imply a trivial geometry, since gravity remains encoded in the nonmetricity tensor, which is nonzero because the metric depends on the coordinates.

In this scenario, the five-dimensional $f(Q,B)$ action coupled to a matter Lagrangian density $(\mathcal{L}_m)$ is
\begin{align}
   S=\int d^5x\sqrt{-g}\left[\frac{1}{2\kfive}f(Q,B)+\mathcal L_m\right].
   \label{original_action}
\end{align}
where $f(Q,B)$ is a well-behaved function, and the subscripts $Q$ and $B$ are partial derivatives with respect to these fields $Q$ and $B$, e.g., $f_{QB}=\frac{\partial^2 f}{\partial Q\partial B}$ and $f_{BQ}=\frac{\partial^2 f}{\partial B\partial Q}$. Furthermore, we assume that the Hessian of $f(Q,B)$ concerning the $Q$ and $B$ is nondegenerate \cite{Gakis,Matheus}, i.e., 
\begin{align}
   \det\begin{pmatrix}
        f_{QQ} & f_{QB}\\ f_{BQ} & f_{BB} \end{pmatrix}\ne 0,
   \label{hessian_condition}
\end{align}
Besides, let us implement the Legendre variables $\Phi=f_Q$ and $\Psi=f_B$ \cite{Matheus}, where the geometric scalar potential is
\begin{align}
   U(\Phi,\Psi)=\Phi Q+\Psi B-f(Q,B).
   \label{legendre_potential}
\end{align}
Therefore, the dynamically equivalent scalar representation is
\begin{align}
    S=\int d^5x\sqrt{-g}\left[\frac{1}{2\kfive}\left(\Phi Q+\Psi B-U(\Phi,\Psi)\right)+\mathcal L_m\right].
   \label{scalar_action}
\end{align}
Naturally, by varying $U$ concerning the $\Phi$ and $\Psi$, one obtains the nonmetricity and the boundary term, viz., $U_\Phi=Q$ and $U_\Psi=B$. Thus, when Eq. \eqref{hessian_condition} fails, the scalar representation may still exist, but it describes a constrained or partially degenerate sector \footnote{Throughout this work, we will assume the regular case}. Naturally, a useful relation follows directly from Eq. \eqref{e3}, i.e.,  
\begin{align}
   \Phi Q+\Psi B-U=(\Phi+\Psi)Q-\Psi\mathring R-U.
   \label{rewritten_action}
\end{align}
Thereby, one can conclude that the first term behaves as a scalar-coupled nonmetricity action, and the second term is an ordinary nonminimally coupled Ricci scalar with coefficient $-\Psi$.

The metric equations may be obtained either by varying Eq. \eqref{scalar_action} directly or by using Eq. \eqref{rewritten_action}. Thus, its compact form is
\begin{align}\nonumber
\kfive T_{MN} = &\,\mathcal E^{(Q)}_{MN}[\Phi+\Psi]-\Psi\,\mathring G_{MN}-\left(g_{MN}\mathring\Box-\mathring\nabla_M\mathring\nabla_N\right)\Psi\\
+&\frac12 U g_{MN},
   \label{field_equations_compact}
\end{align}
where $\mathcal E^{(Q)}_{MN}[F]$ is the metric variation of the scalar-coupled nonmetricity term $FQ$. Therefore, in the coincident gauge, this variation boils down to
\begin{align}\nonumber
\mathcal E^{(Q)}_{MN}[F]=&\frac{2}{\sqrt{-g}}\partial_A\left(\sqrt{-g}F P^{A}{}_{MN}\right)-\frac12 FQ g_{MN}+F\times\\ &\left(P_{MAB}Q_N{}^{AB}-2Q_{ABM}P_N{}^{AB}\right).
\label{EQF_general}
\end{align}

Equations~\eqref{field_equations_compact} and \eqref{EQF_general} make all limiting cases transparent. First, if $\Psi=0$ the system reduces to the scalar representation of $f(Q)$ gravity. Second, if $f=f(Q-B)$ then $\Phi=f_{Q}=F$ and $\Psi=f_B=-F$, so that Eq.~\eqref{scalar_action} becomes
\begin{align}
   S=\int d^5x\sqrt{-g}\left[\frac{1}{2\kfive}\left(F\mathring R-U(F)\right)+\mathcal L_m\right],
   \label{fR_scalar_action}
\end{align}
which is the standard scalar representation of $f(\mathring R)$ gravity. This check is important and will be used below to verify the signs of the brane equations.

\section{On the five-dimensional braneworld}\label{brane}

To investigate the braneworld, we consider that the spacetime
manifold is described by the five-dimensional line element, viz.,
\begin{align}
   ds^2=e^{2A(y)}\eta_{\mu\nu}dx^\mu dx^\nu+dy^2,
   \label{brane_metric}
\end{align}
with $\eta_{\mu\nu}=\mathrm{diag}(-1,1,1,1)$. In the coincident gauge, Eq.~\eqref{brane_metric} boils down to
\begin{align}
Q=12A'^2 \quad \mathrm{and} \quad B=8A''+32A'^2,
   \label{QB_brane}
\end{align}
which leads us to
\begin{align}
   \mathring{R}=Q-B=-8A''-20A'^2.
   \label{R_brane}
\end{align}

The geometric scalars depend only on the extra-dimensional coordinate, i.e., $\Phi=\Phi(y)$ and $\Psi=\Psi(y)$. Within this framework, the matter sector is described by the Lagrangian matter density ($\mathcal{L}_\mathrm{m}$), viz., \begin{align}
\mathcal{L}_\mathrm{m}=-\frac12 g^{MN}\partial_{M}\chi\partial_{N}\chi-V(\chi),
\label{matter_lagrangian}
\end{align}
with $\chi=\chi(y)$ indicating that the matter field depends only on the extra-dimensional coordinate. Meanwhile, the nonzero components of the energy-momentum tensor are
\begin{align}
    T_{\mu\nu}=-g_{\mu\nu}\left(\frac12\chi'^2+V\right) \quad \mathrm{and} \quad T_{55}=\frac12\chi'^2-V.
\label{energy_momentum}
\end{align}

Substituting Eq. \eqref{field_equations_compact}, one obtains the two independent gravitational equations, viz., 
\begin{align}
\kfive\left(-\frac{1}{2}\chi'^2-V\right)&=3\Phi A''+6\Phi A'^2+3A'\Phi'-\Psi''+\frac{1}{2}U,
\label{munu_equation}
\end{align}
and
\begin{align}
\kfive\left(\frac{1}{2}\chi'^2-V\right)&=6\Phi A'^2-4A'\Psi'+\frac{1}{2}U.
\label{yy_equation}
\end{align}

The matter equation is
\begin{align}
   \chi''+4A'\chi'=V_\chi.
   \label{matter_equation}
\end{align}
Once the auxiliary equations and Eqs.~\eqref{munu_equation}-\eqref{yy_equation} are applied, Eq.~\eqref{matter_equation} results from the conservation law. We also need to check how the variation applies to the flat, torsionless connection. Using $\Phi Q+\Psi B=(\Phi+\Psi)Q-\Psi\mathring R$, we see that the Levi-Civita term $\Psi\mathring R$ does not affect the independent symmetric-teleparallel connection. As a result, the connection equation matches that in scalar-coupled $f(Q)$ gravity, with $f_Q$ replaced by $F=\Phi+\Psi$. For the flat static brane ansatz in the coincident gauge, and with $F=F(y)$, the only possibly nontrivial part is linked to a four-dimensional derivative of a function of $y$, but this always vanishes. This means the connection equation is automatically satisfied in the usual $f(Q)$ thick-brane ansatz.

By algebraic manipulations of the equations \eqref{munu_equation} and \eqref{yy_equation}, one obtains
\begin{align}
    \kfive\chi'^2=-3\Phi A''-3A'\Phi'-4A'\Psi'+\Psi'',
    \label{chip2_formula}
\end{align}
and
\begin{align}\nonumber
\kfive V(y)=&-\frac12[(3A''+12A'^2)\Phi+3A'\Phi'-4A'\Psi'\\
-&\Psi''+U].
   \label{V_formula}
\end{align}
In this case, the on-shell geometric potential is determined by
\begin{align}
   \frac{dU}{dy}=U_\Phi\Phi'+U_\Psi\Psi'=Q\Phi'+B\Psi'.
   \label{U_quadrature}
\end{align}
Once $A(y)$, $\Phi(y)$, and $\Psi(y)$ are specified, Eqs. [\eqref{chip2_formula}-\eqref{U_quadrature}] reconstruct the brane source. Thus, the scalar field is
\begin{align}
    \chi(y)=\chi(0)+\int_0^y d\bar{y}\, \sqrt{\frac{-3\Phi A''-3A'\Phi'-4A'\Psi'+\Psi''}{\kfive}}.
\label{chi_quadrature}
\end{align}
For a braneworld that supports domain walls, we adopt an antisymmetric matter sector $\chi(-y)=-\chi(y)$. Within this conjecture, the physical energy density contains the vacuum contribution. Thus, one defines the physical energy density as
\begin{align}
   \rho(y)=\mathrm{e}^{2A}\left(\frac12\chi'^2+V\right),
   \label{rho_full}
\end{align}
 and the localized density, with the asymptotic vacuum subtracted, as
\begin{align}
    \rho_{\rm loc}(y)=\mathrm{e}^{2A}\left[\frac12\chi'^2+V-V_\infty\right],
   \label{rho_local}
\end{align}
where
\begin{align}
    V_\infty\equiv \lim_{|y|\to\infty}V(y).
\end{align}
For asymptotically AdS branes, $\rho(y)$ may have a negative tail, while $\rho_{\mathrm{loc}}(y)$ isolates the finite domain-wall contribution. Now, let us announce some consistency checks, viz., the $f(Q)$ limit is obtained as $\Psi\to 0$. Meanwhile, the Eqs. \eqref{munu_equation} and \eqref{yy_equation} boil down to
\begin{align}
    \kfive\left(-\frac12\chi'^2-V\right)&=3\Phi A''+6\Phi A'^2+3A'\Phi'+\frac12U
\end{align}
and
\begin{align}
\kfive\left(\frac12\chi'^2-V\right)&=6\Phi A'^2+\frac12U,
\end{align}
which agree with the scalar form of the $f(Q)$ brane equations and lead us to
\begin{align}
   \kfive\chi'^2=-3\Phi A''-3A'\Phi'.
   \label{fQ_check}
\end{align}

The curvature sector is obtained by setting $\Psi=-\Phi= -F$. Then the action reduces to $F\mathring R-U(F)$, and Eqs.~\eqref{munu_equation}--\eqref{yy_equation} give
\begin{align}
   \kfive\left(-\frac12\chi'^2-V\right)&=F(3A''+6A'^2)+F''+3A'F'+\frac12U
\end{align}
and
\begin{align}
   \kfive\left(\frac12\chi'^2-V\right)&=6FA'^2+4A'F'+\frac12U.
\end{align}
Thus, one concludes that
\begin{align}
   \kfive\chi'^2=-3F A''+A'F'-F''.
   \label{fR_check}
\end{align}
This is the known scalar-tensor structure of thick branes in $f(\mathring R)$ gravity. These two limits provide a useful sign check for the $\Psi''$ and $A'\Psi'$ terms in Eqs.~\eqref{munu_equation} and \eqref{yy_equation}.

The calculation has been audited in four independent ways. First, the geometrical identity $Q-B=\mathring R$ reproduces the direct Levi-Civita result $\mathring R=-8A''-20A'^2$. Second, the limit $\Psi=0$ gives exactly the scalar form of the $f(Q)$ brane equations used in Ref. \cite{Fu}. Third, the constrained sector $\Psi=-\Phi$ gives the scalar-tensor $f(\mathring R)$ brane equations. Fourth, the matter equation follows after differentiating Eqs.~\eqref{munu_equation}-\eqref{yy_equation} and using $U'=Q\Phi'+B\Psi'$. Thus, the reconstructed solution is not overdetermined.

A direct approach would start from a specified function $f(Q,B)$ and then solve the resulting higher-derivative background equations. The scalar representation suggests a different procedure. In this case, we should choose a regular warp factor $A(y)$ with the desired asymptotics and choose smooth profiles $\Phi(y)$ and $\Psi(y)$ such that $\Phi(y)>0$. After that, we compute $Q(y)$ and $B(y)$ from Eq.~\eqref{QB_brane} and obtain $U(y)$ from Eq. \eqref{U_quadrature}. The next step is to obtain $\chi(y)$ and $V(y)$ from Eqs. \eqref{chi_quadrature} and \eqref{V_formula}, and compute the on-shell gravitational function
\begin{align}
   f_{\rm on}(y)=\Phi(y)Q(y)+\Psi(y)B(y)-U(y).
   \label{f_on_shell}
\end{align}

This method reconstructs $f$ only along the curve traced by the brane in the $(Q,B)$ plane. A unique two-variable function $f(Q,B)$ requires an off-shell extension. Naturally, the extension is not unique, because a one-dimensional background curve cannot determine a two-dimensional function globally. This is not a defect of the method; it is the same kind of nonuniqueness that appears in many reconstruction problems in modified gravity. What is physically determined by the solution is the on-shell Lagrangian, the first derivatives $f_Q$ and $f_B$ along the curve, and the compatibility condition
\begin{align}
   d f_{\rm on}=\Phi\,dQ+\Psi\,dB.
   \label{df_on_shell}
\end{align}
Indeed, differentiating Eq.~\eqref{f_on_shell} and using Eq.~\eqref{U_quadrature} gives Eq.~\eqref{df_on_shell}. Any off-shell extension satisfying these data in a neighborhood of the curve produces the same background brane.

There are three elementary physical requirements:
\begin{align}
   \Phi(y)>0,
   \qquad
   \chi'^2(y)\ge 0,
   \qquad
   \int_{-\infty}^{\infty}dy\,e^{2A(y)}\Phi(y)<\infty.
   \label{physical_conditions}
\end{align}
Although the independent-connection equation involves the combination $F=\Phi+\Psi$, this quantity should not be identified with the effective kinetic coefficient of the tensor perturbations. Indeed, using $B=Q-\mathring{R}$, the gravitational sector can be written as
\begin{align}
\Phi Q+\Psi B=(\Phi+\Psi)Q-\Psi\mathring{R}.
\end{align}
In the transverse-traceless tensor sector, the $Q$ and $\mathring{R}$ terms generate the same principal graviton kinetic operator, since they differ only by a boundary contribution. Therefore, the contribution proportional to $\Psi$ in the nonmetricity term is canceled by that arising from $-\Psi\mathring{R}$, and the quadratic action contains
\begin{align}
S_{\rm T}^{(2)}\supset \frac{1}{8\kappa_{5}^{2}}\int d^{4}x\,dy\,
e^{2A(y)}\Phi(y)\partial_{\rho}h_{\mu\nu}\partial^{\rho}h^{\mu\nu}.
\end{align}
Hence, $\Phi(y)>0$ is the no-ghost condition in the tensor sector, whereas the effective four-dimensional Planck mass is proportional to
\begin{align}
M_{\rm Pl}^{2}\propto \int_{-\infty}^{+\infty}dy\,e^{2A(y)}\Phi(y).
\end{align}
Naturally, this conclusion is consistent with the $f(\mathring{R})$ limit, for which $\Psi=-\Phi$ and $\Phi+\Psi=0$, while the tensor kinetic coefficient remains $\Phi$.

\subsection{A smooth fundamental brane}\label{analytic_model}

\begin{figure}[ht!]
    \centering
    \subfigure[The warp factor varying the parameter $k$.]{\includegraphics[width=4.3cm,height=3.7cm]{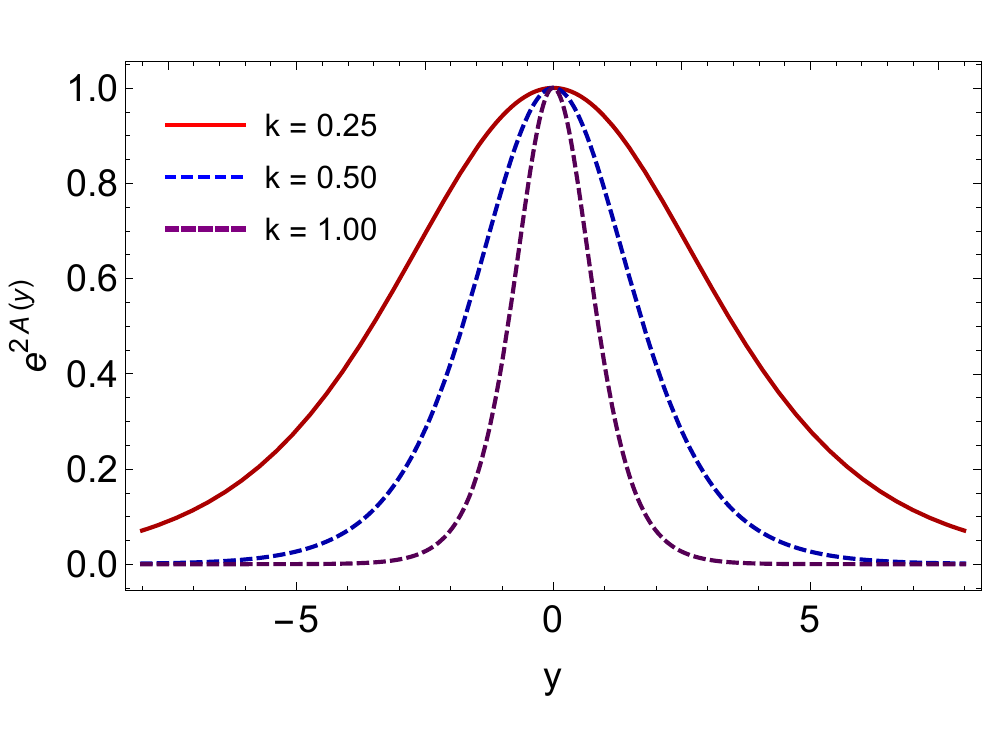}}\hfill
    \subfigure[The warp factor varying the parameter $p$.]{\includegraphics[width=4.3cm,height=3.7cm]{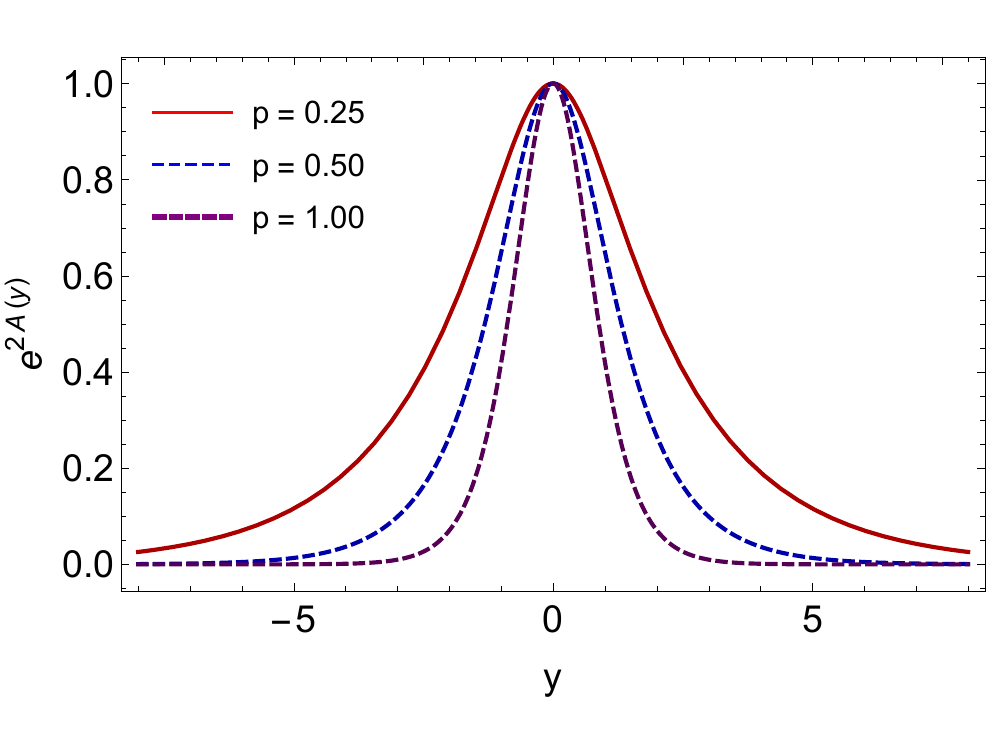}}
    \caption{Profile of the warp factor vs. the extra-dimensional coordinate $y$.}
    \label{Fig1}
\end{figure}

To construct a regular thick-brane background, we adopt a smooth and $\mathbb{Z}_2$-symmetric warp function that reproduces the essential geometric properties of a domain wall. Particularly, the geometry must be nonsingular at the brane core, localized around $y=0$, and asymptotically approach an AdS$_5$ spacetime \cite{Gremm}. A convenient choice satisfying these requirements is the well-known hyperbolic warp factor
\begin{align}
    A(y)=-p\ln\cosh(ky),
    \label{fundamental_warp}
\end{align}
where $k>0$ sets the inverse thickness of the brane, while $p>0$ controls the strength of the gravitational warping \footnote{One can find profiles of the warp factor in Fig. \ref{Fig1}(a) and \ref{Fig1}(b).}. Indeed, the corresponding exponential factor, $\mathrm{e}^{2A(y)}=\sech^{2p}(ky)$, is finite and maximal at the brane core and decays exponentially in the bulk. Moreover, $A'(0)=0$, ensuring a smooth reflection-symmetric geometry, whereas $A(y)\sim -pk\vert y\vert$ for $\vert y\vert\to \infty$, which guarantees the desired asymptotically AdS behavior and favors the localization of the effective four-dimensional gravitational interaction \cite{Gremm,Csaki}.

\begin{figure}[ht!]
    \centering
    \subfigure[Profile of $Q(y)$ for different values of the parameter $k$.]{\includegraphics[width=4.3cm,height=3.7cm]{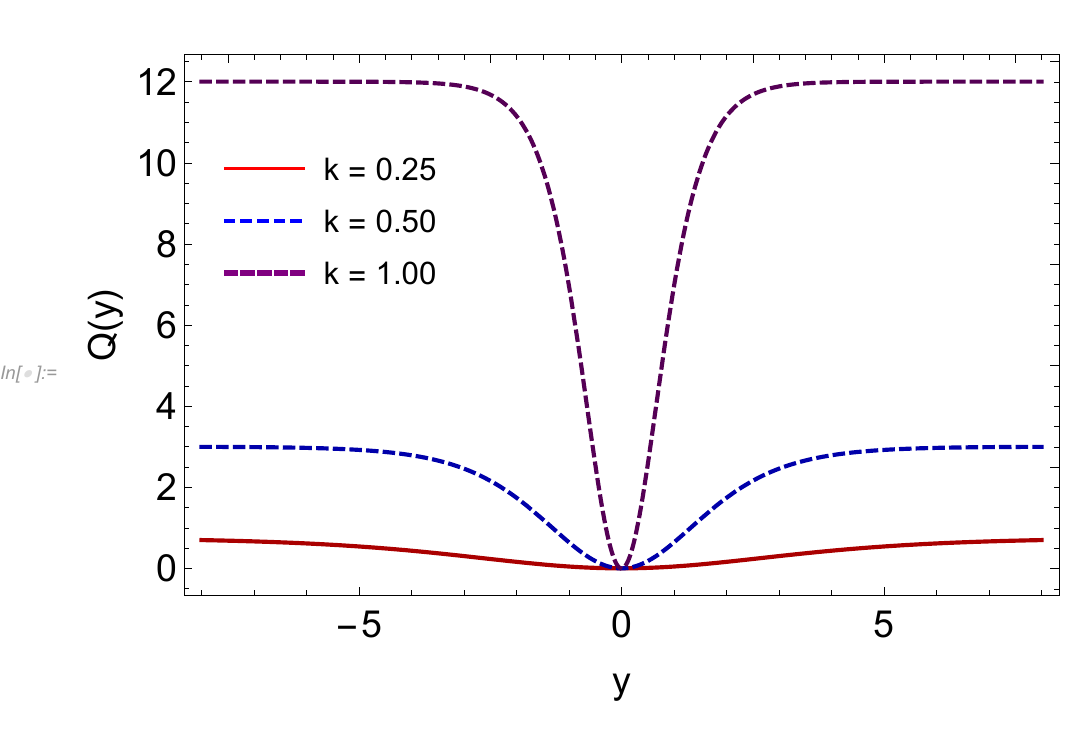}}\hfill
    \subfigure[Profile of $Q(y)$ for different values of the parameter $p$.]{\includegraphics[width=4.3cm,height=3.7cm]{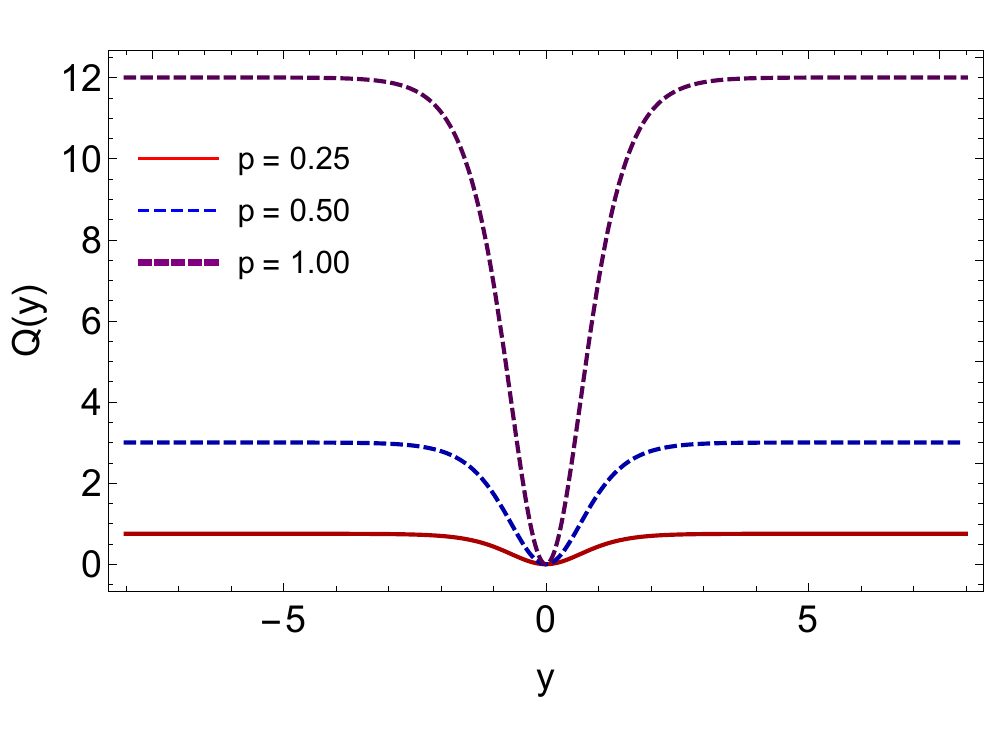}}
    \subfigure[The boundary term for different values of the parameter $k$.]{\includegraphics[width=4.3cm,height=3.7cm]{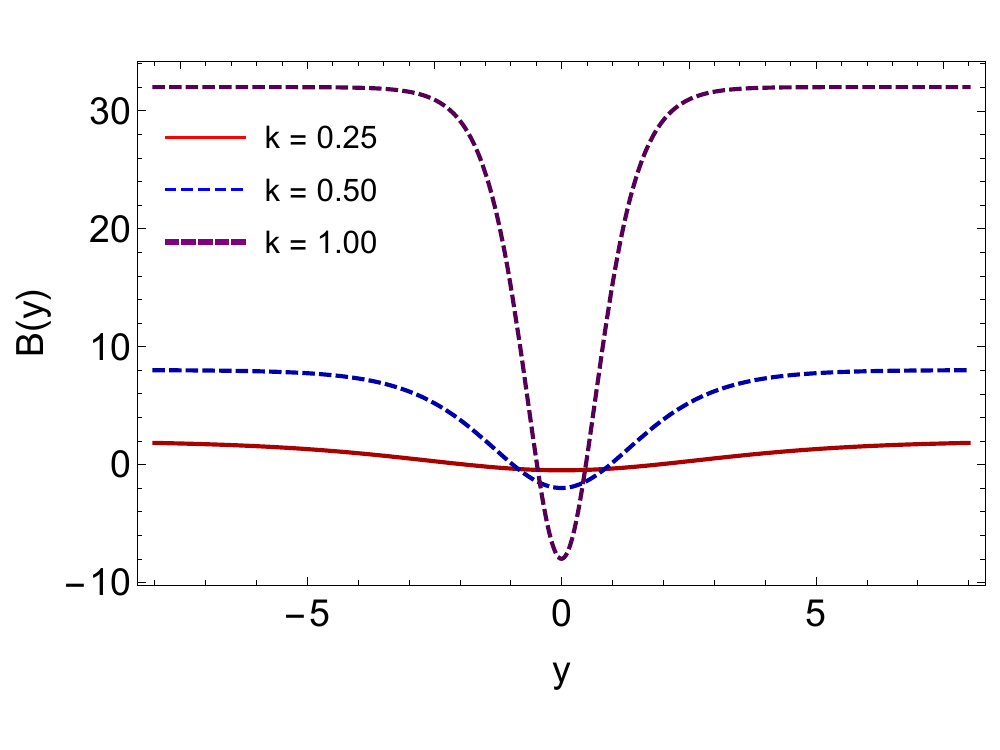}}\hfill
    \subfigure[The boundary term different values of the parameter $p$.]{\includegraphics[width=4.3cm,height=3.7cm]{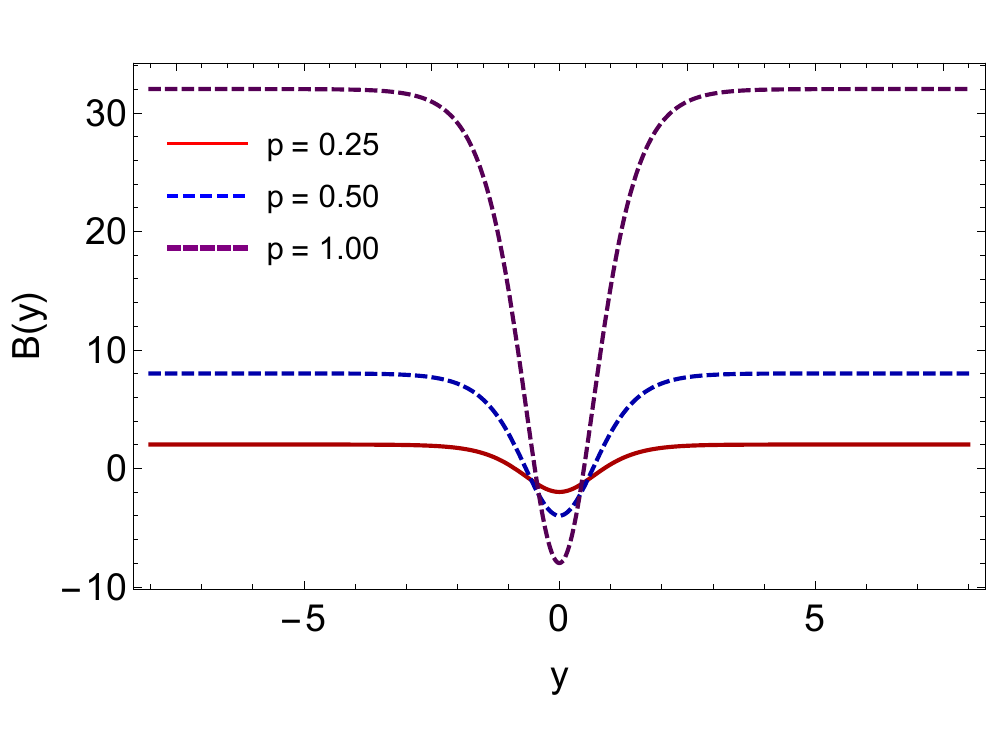}}
    \caption{Profiles of the nonmetricity and boundary scalars vs. the extra-dimensional coordinate $y$.}
    \label{Fig2}
\end{figure}

Within this framework, the corresponding nonmetricity and boundary scalars are
\begin{align}
   Q(y)=12p^2k^2\tanh^2(ky),
   \label{Q_fundamental}
\end{align}
and
\begin{align}
   B(y)=-8pk^2\sech^2(ky)+32p^2k^2\tanh^2(ky).
   \label{B_fundamental}
\end{align}
Figures \ref{Fig2}[(a)–(b)] and \ref{Fig2}[(c)–(d)] display, respectively, the profiles of the nonmetricity scalar $Q(y)$ and the boundary term $B(y)$ varying the geometric parameters $k$ and $p$. Both quantities are regular and $\mathbb{Z}_{2}$-symmetric with respect to the brane core. At $y=0$, the nonmetricity scalar vanishes, $Q(0)=0$. Consequently, $A'(0)=0$, whereas the boundary term reaches the negative value $B(0)=-8pk^{2}$, signaling the localized contribution associated with the nonvanishing second derivative of the warp function at the core. Far from the brane, the scalars approach the constant positive values $Q\to 12p^2k^2$ and $B\to 32p^2k^2$, consistently with the asymptotically AdS$_5$ geometry. Figures \ref{Fig2}[(c)-(d)] further show that $B(y)$ evolves from a negative minimum at the brane center, crosses zero on both sides of the core, and tends monotonically to a positive asymptotic plateau. Increasing $k$ enhances the magnitudes of $Q$ and $B$ and narrows their transition regions. Naturally, this behavior indicates a thinner and more strongly localized geometric structure. By contrast, increasing $p$ predominantly amplifies their magnitudes without significantly changing the characteristic width of the profiles. Particularly, the depth of the central minimum of $B$ grows linearly with $p$, whereas its asymptotic value grows quadratically. These results show that $k$ controls both the localization scale and the intensity of the geometric scalars, while $p$ mainly determines the strength of the bulk. 

\begin{figure}[ht!]
    \centering
    \subfigure[$\Phi(y)$ varying the parameter $k$.]{\includegraphics[width=4.3cm,height=3.5cm]{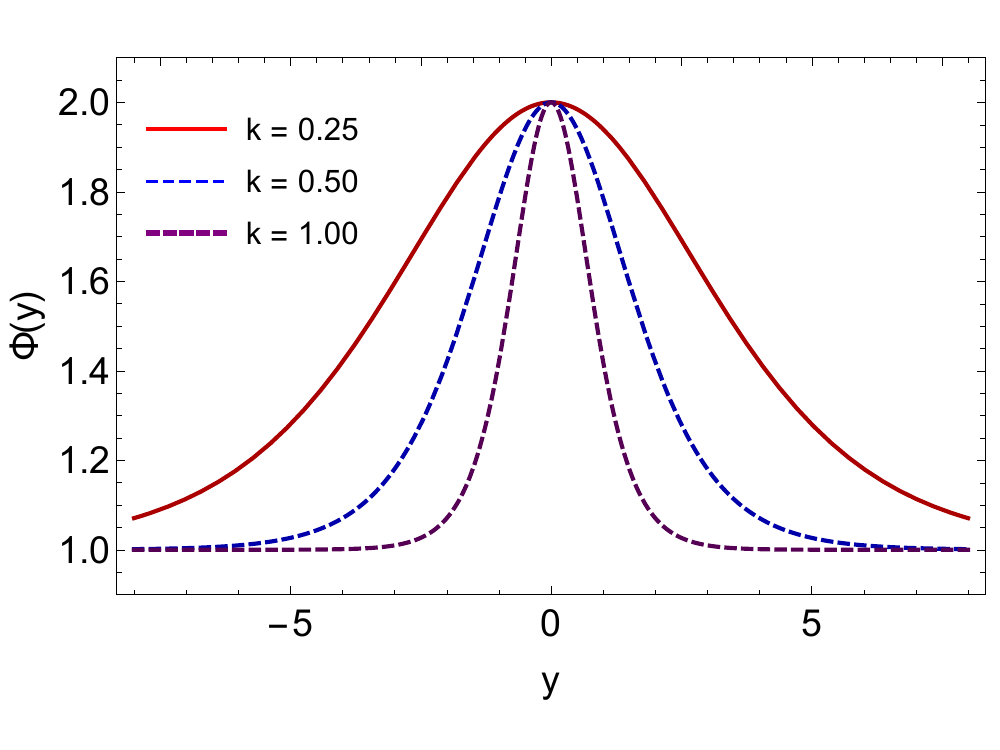}}\hfill
    \subfigure[$\Phi(y)$ varying the parameter $\alpha$.]{\includegraphics[width=4.3cm,height=3.5cm]{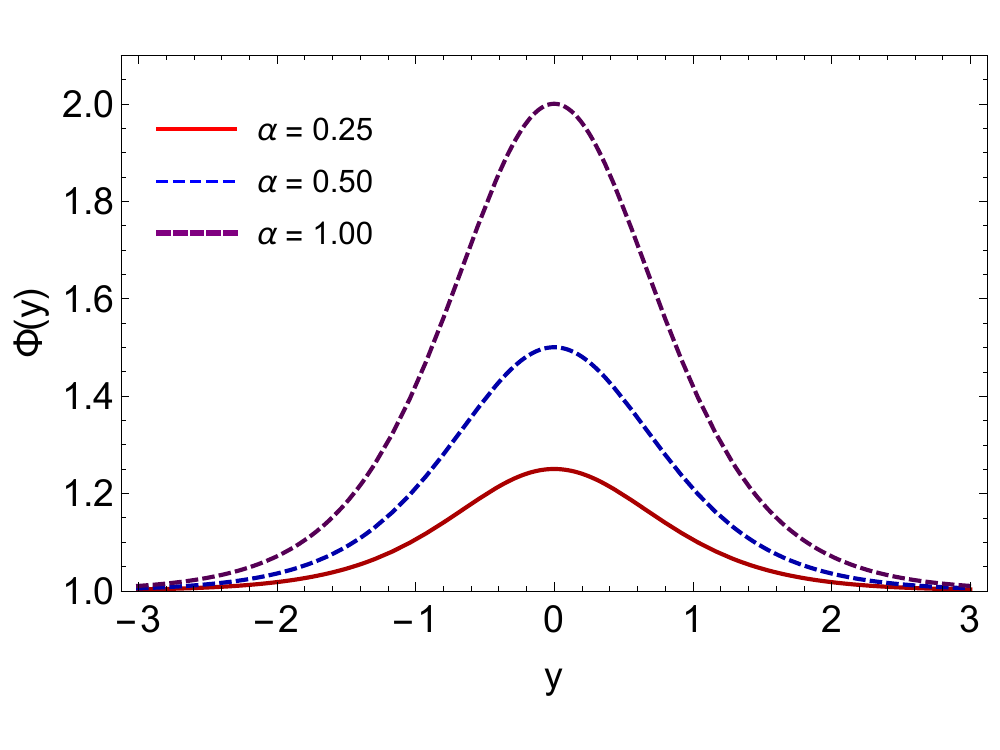}}\\
    \subfigure[$\Psi(y)$ varying the parameter $k$.]{\includegraphics[width=4.3cm,height=3.5cm]{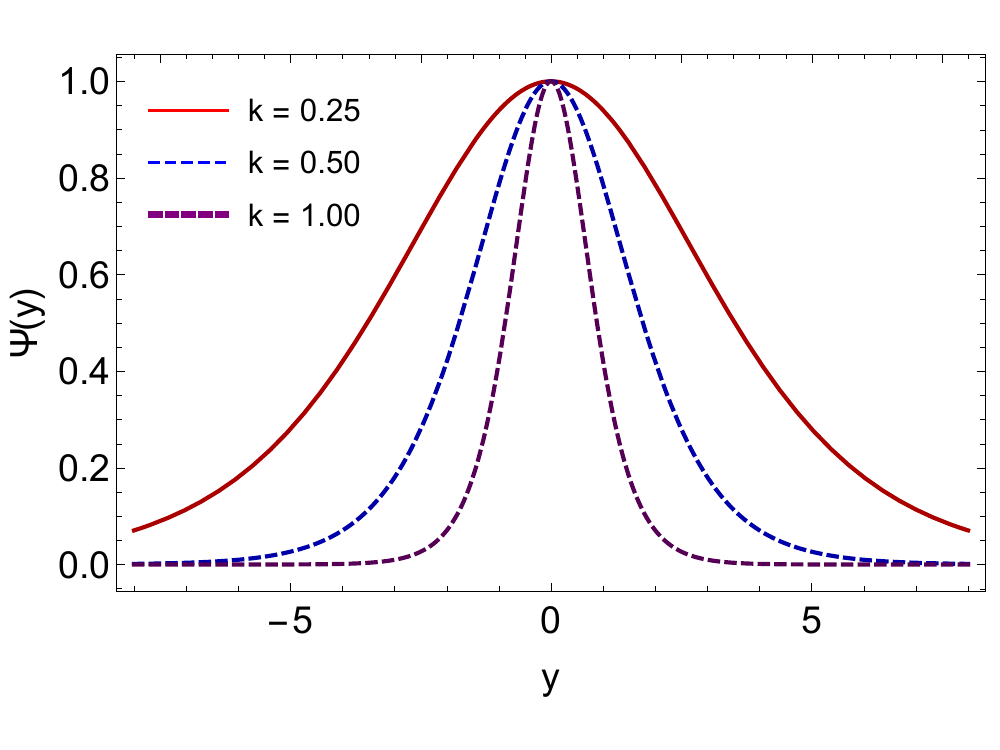}}\hfill
    \subfigure[$\Psi(y)$ varying the parameter $\beta$.]{\includegraphics[width=4.3cm,height=3.5cm]{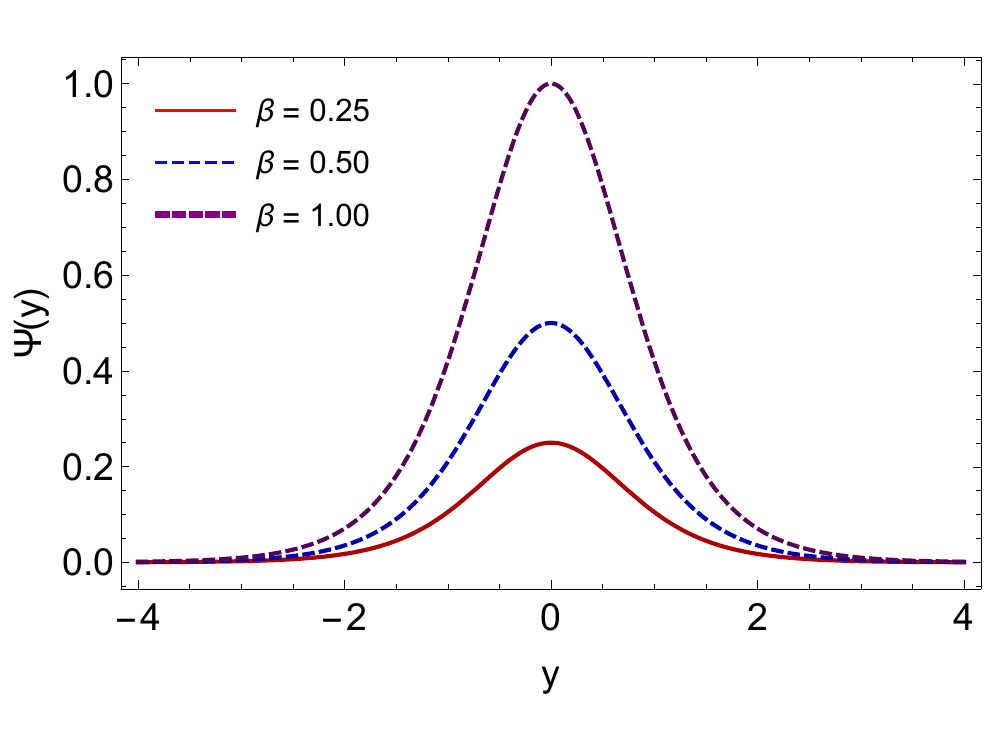}}
    \caption{Profile of geometric scalars $\Phi(y)$ and $\Psi(y)$ vs. the extra-dimensional coordinate $y$.}
    \label{Fig3}
\end{figure}

Meanwhile, the geometric scalars are
\begin{align}
   \Phi(y)=1+\alpha\sech^2(ky)   \quad \mathrm{and} \quad  \Psi(y)=\beta\sech^2(ky),
   \label{PhiPsi_model}
\end{align}
where $\alpha>-1$ guarantees $\Phi>0$. In Figs. \ref{Fig3}(a)–\ref{Fig3}(b), we plot the auxiliary geometric scalar $\Phi(y)$ varying $k$ and $\alpha$. Meanwhile, Figs. \ref{Fig3}(c)–\ref{Fig3}(d) display the corresponding profiles of $\Psi(y)$ under variations of $k$ and $\beta$. Both fields are smooth, $\mathbb{Z}_{2}$-symmetric, and localized around the brane core. Increasing $k$ narrows the profiles of both scalars, indicating that this parameter controls their localization scale along the extra dimension, while leaving their central amplitudes unchanged. By contrast, $\alpha$ directly controls the height of the localized correction to $\Phi$, since $\Phi(0)=1+\alpha$, whereas $\beta$ determines the amplitude of the boundary-sector field through $\Psi(0)=\beta$, without significantly modifying the characteristic width. Far from the brane, the fields $\Phi\to1$ and $\Psi\to 0$, showing that the modified geometric contributions are concentrated near the domain wall and that the standard gravitational coupling is asymptotically recovered. Moreover, the condition $\alpha>-1$ guarantees $\Phi(y)>0$ throughout the bulk, thereby avoiding a ghost-like graviton kinetic term.

\begin{figure}[ht!]
    \centering
    \subfigure[$U(y)$ varying the parameters $k$.]{\includegraphics[width=4.3cm,height=3.5cm]{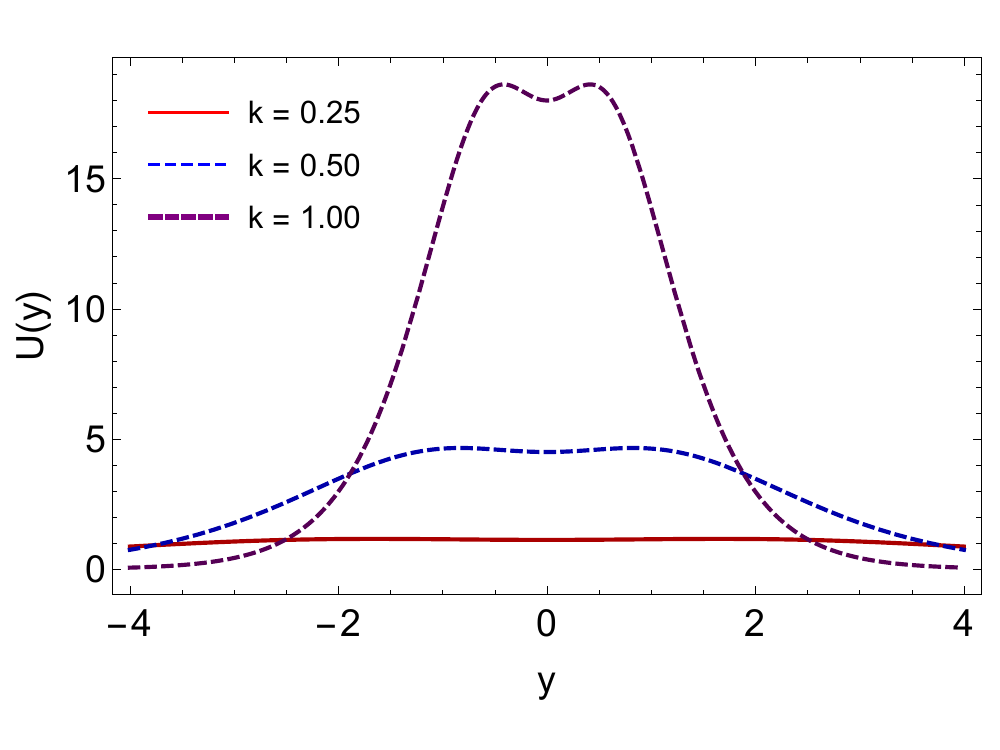}}\hfill
    \subfigure[$U(y)$ varying the parameters $p$.]{\includegraphics[width=4.3cm,height=3.5cm]{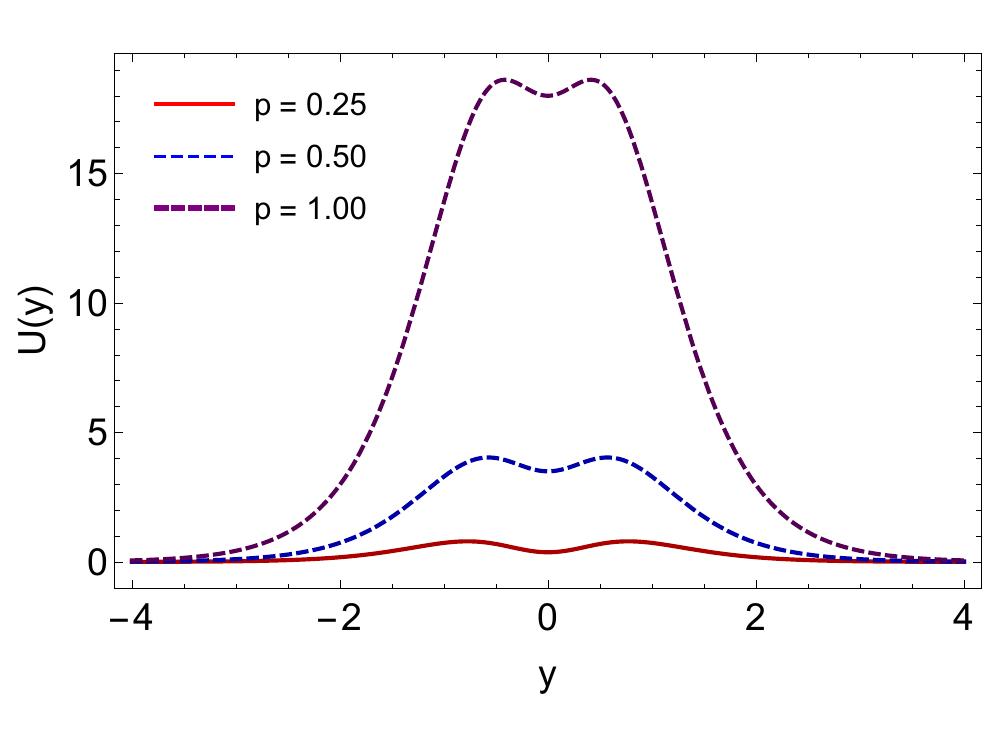}}\\
    \subfigure[$U(y)$ varying the parameters $\alpha$.]{\includegraphics[width=4.3cm,height=3.5cm]{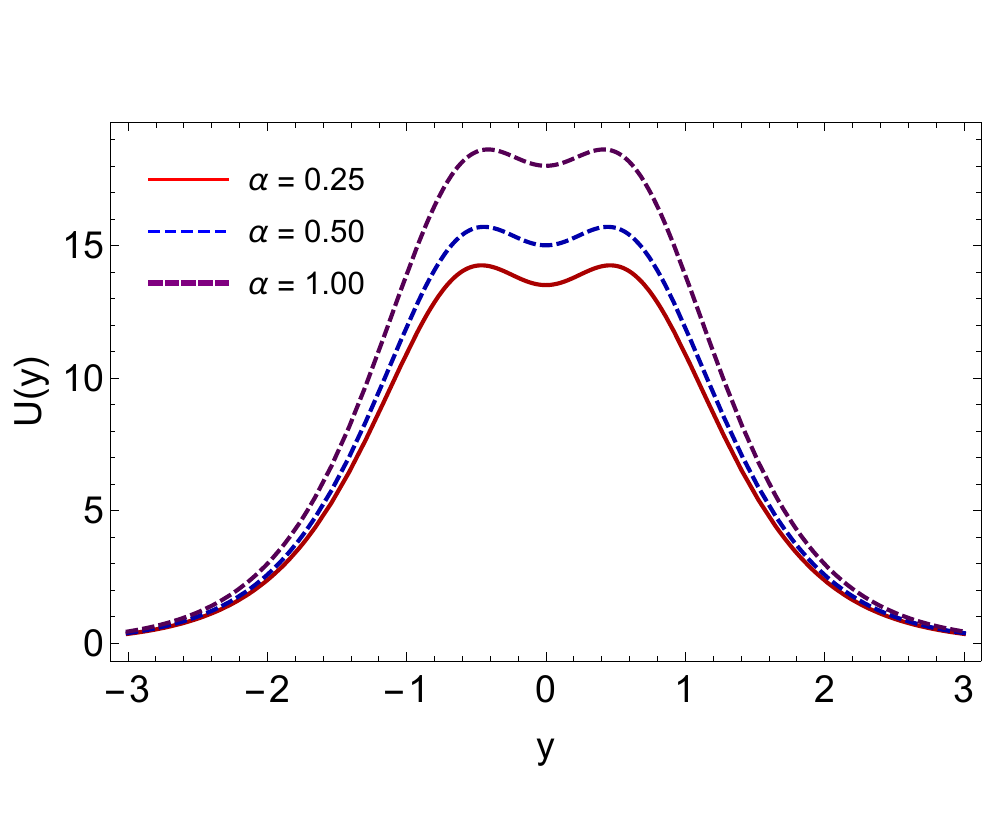}}\hfill
    \subfigure[$U(y)$ varying the parameters $\beta$.]
    {\includegraphics[width=4.3cm,height=3.5cm]{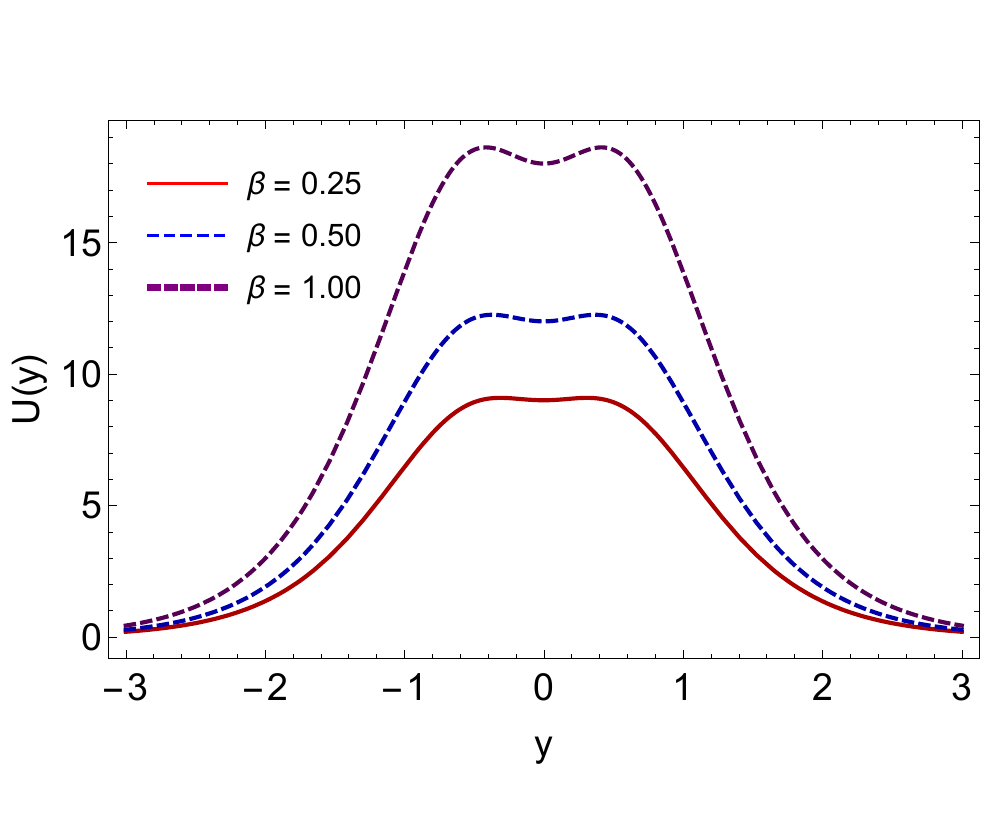}}
    \caption{Profile of the geometric scalar potential $U(y)$ vs. the extra-dimensional coordinate $y$ with $U_{\infty}=0$.}
    \label{Fig4}
\end{figure}

Naturally, the potential $U(y)$ is obtained from
\begin{align}
   U'(y)=12A'^2\Phi'(y)+\left(8A''+32A'^2\right)\Psi'(y).
   \label{U_model_derivative}
\end{align}
For the profiles in Eqs. \eqref{fundamental_warp} and \eqref{PhiPsi_model}, by writing $s(y)=\sech^2(ky)$, one obtains
\begin{align}
   U(y)=k^2\left[C\,s-\frac{C+D}{2}s^2\right]+U_\infty.
   \label{U_analytic_model}
\end{align}
where $C=4p^2(3\alpha+8\beta)$, $D=8p\beta$, $s\equiv s(y)=\mathrm{sech}^2(ky)$, and $U_\infty$ is an integration constant corresponding to the asymptotic value of the geometric potential. Furthermore, one highlights that this integration constant is physically a bulk cosmological contribution. In the Figs. \ref{Fig4}[(a)-(d)] displays the potential $U(y)$\footnote{Here, we choose $U_\infty=0$, which fixes $U(|y|\to\infty)=0$.}.
In Figs. \ref{Fig4}[(a)–(d)], we plot the geometric scalar potential $U(y)$ vs. the extra-dimensional coordinate for different values of $k$, $p$, $\alpha$, and $\beta$, respectively, setting $U_{\infty}=0$. The behavior of this potential is consistent with the properties of the auxiliary field $\Phi(y)$ displayed in Figs. \ref{Fig3}(a) and \ref{Fig3}(b). Specifically, increasing $k$ makes the profile of $\Phi(y)$ narrower without changing its central value, showing that $k$ determines the localization scale of the geometric modification around the brane. Conversely, varying $\alpha$ changes the amplitude at the brane core, $\Phi(0)=1+\alpha$, while leaving the characteristic width essentially unchanged. In all cases, $\Phi(y)$ remains smooth, positive, and $\mathbb{Z}_{2}$-symmetric, approaching its general-relativistic value $\Phi\to 1$ far from the brane. Therefore, $k$ controls how strongly the modified gravitational sector is confined along the extra dimension, whereas $\alpha$ controls the intensity of its localized contribution. 

Meanwhile, by adopting Eq. \eqref{chip2_formula}, the derivative of the brane-generating scalar field is
\begin{align}
   \chi'^2=k^2\{[3p-6\alpha p+\beta(4-8p)]s+[9\alpha p+\beta(8p-6)]s^2\},
   \label{chip2_simplified}
\end{align}
with $s=\sech^2(ky)$ and $\kfive=1$.

\begin{figure}[ht!]
    \centering
    \subfigure[$\chi(y)$ varying the parameters $k$.]{\includegraphics[width=4.3cm,height=3.7cm]{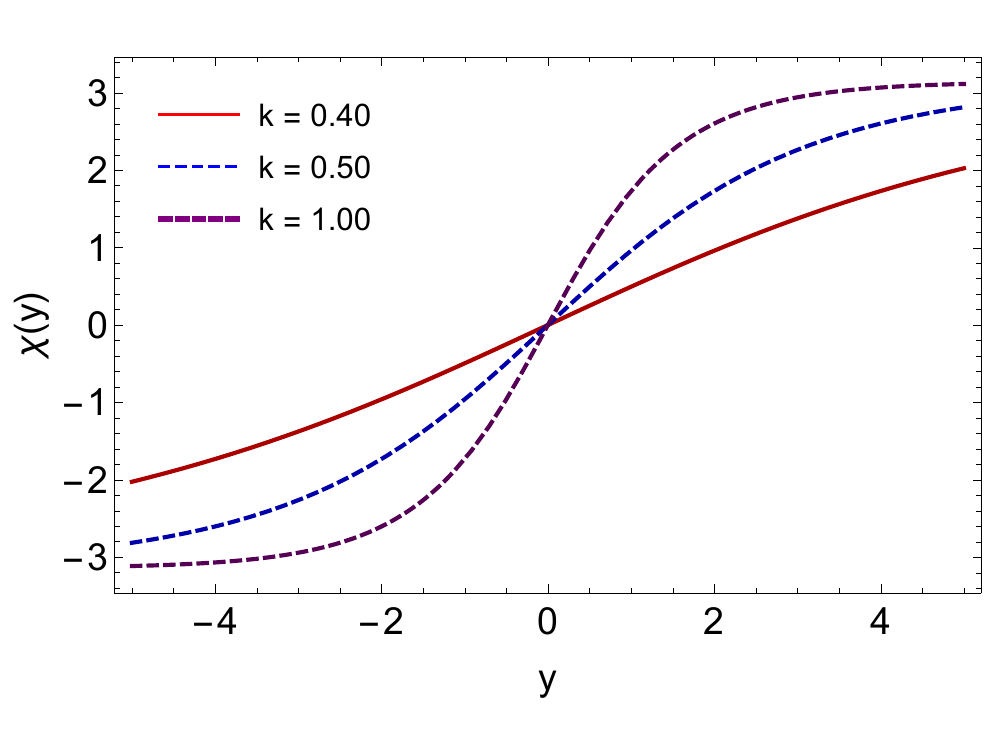}}\hfill
    \subfigure[$\chi(y)$ varying the parameters $p$.]{\includegraphics[width=4.3cm,height=3.7cm]{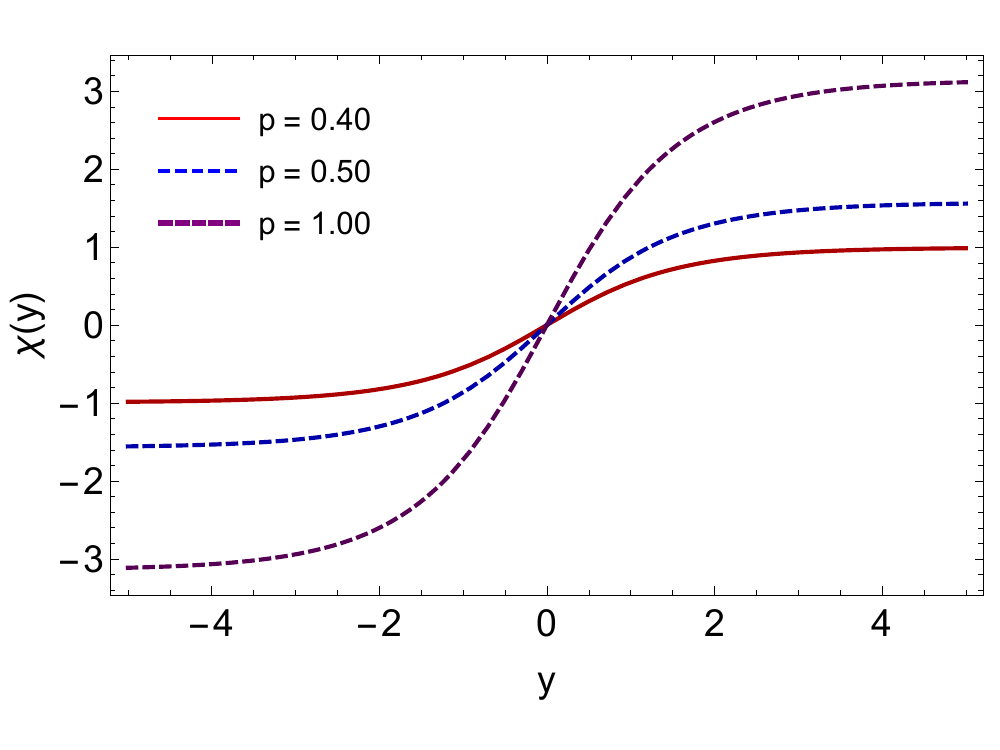}}\\
    \subfigure[$\chi(y)$ varying the parameters $\alpha$.]{\includegraphics[width=4.3cm,height=3.7cm]{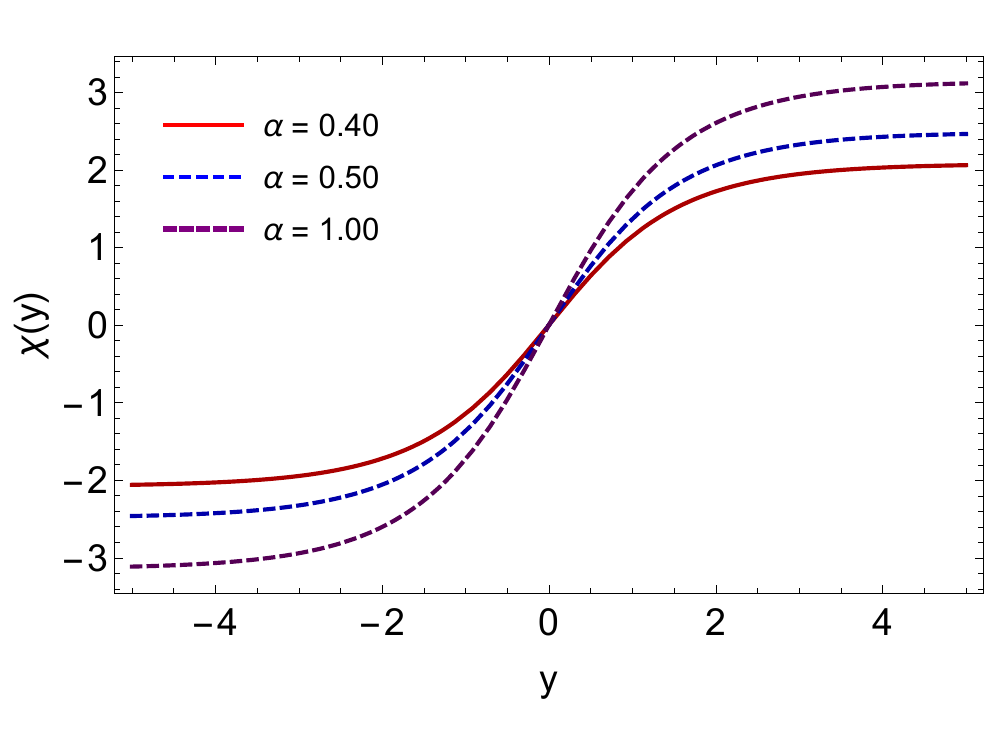}}\hfill
    \subfigure[$\chi(y)$ varying the parameters $\beta$.]
    {\includegraphics[width=4.3cm,height=3.7cm]{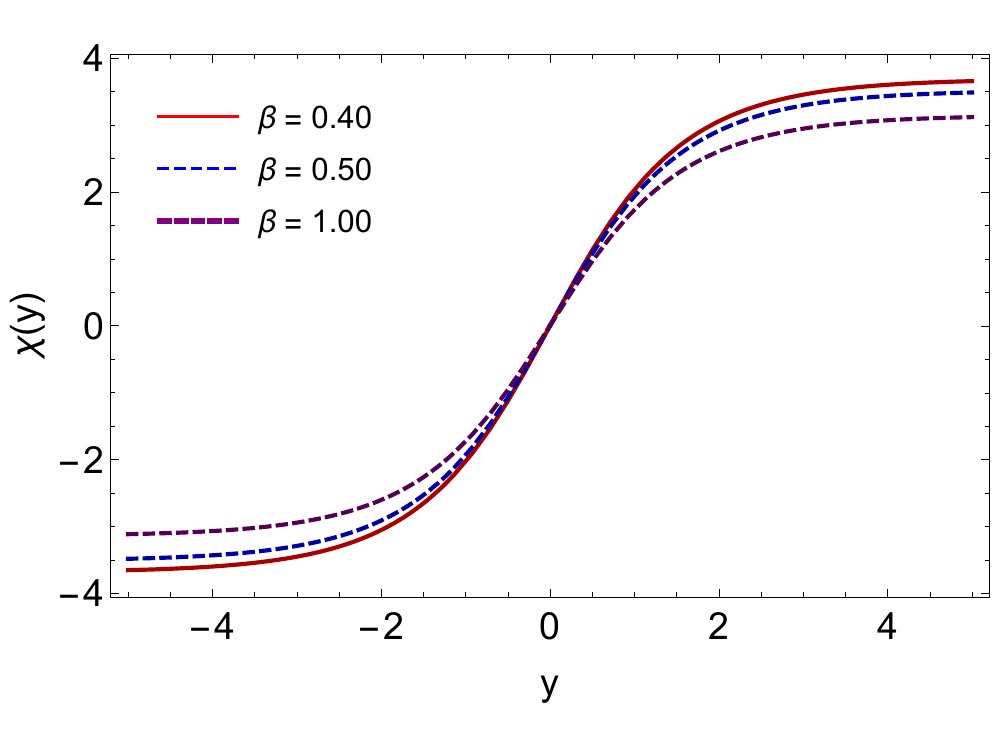}}
    \caption{Profile of the brane-generating scalar field $\chi(y)$ vs. the extra-dimensional coordinate $y$.}
    \label{Fig5}
\end{figure}

A physically admissible brane-generating scalar-field configuration requires  $\chi'(y)\geq 0$  throughout the bulk. Since $0<s(y)\leq 1$ , this requirement is fulfilled provided that
\begin{align}
    3p-6\alpha p+4\beta(1-2p)\geq 0
\end{align}
and
\begin{align}
\qquad 3p(1+\alpha)-2\beta\geq 0.
\end{align}
Thus, at the brane core, where $s(0)=1$, Eq. \eqref{chip2_simplified} reduces to
\begin{align}
   \chi'^2(0)=3p(1+\alpha)k^2-2\beta k^2,
   \label{chip2_core}
\end{align}
while in the asymptotic region it decays exponentially.One highlights that Eq. \eqref{chip2_simplified} must be solved numerically to determine the scalar field $\chi(y)$. Therefore, we solve Eq. \eqref{chip2_simplified} numerically and display its solutions in Figs. \ref{Fig5}[(a)–(d)].

Figs. \ref{Fig5}[(a)–(d)] display the numerical profiles of the brane-generating scalar field $\chi(y)$ varying the parameters $k$, $p$, $\alpha$, and $\beta$, respectively. In all cases, the field exhibits a smooth and monotonic kink-like configuration, satisfying $\chi(-y)=-\chi(y)$, crossing $\chi(0)=0$ at the brane core, and approaching finite asymptotic values on both sides of the extra dimension. As shown in Fig. \ref{Fig5}(a), increasing $k$ makes the transition around $y=0$ sharper and increases the separation between the asymptotic field values, consistently with a thinner and more strongly localized domain wall. Meanwhile, by increasing $p$, one notes in Fig.  \ref{Fig5}(b) that stronger gravitational warping requires a more pronounced scalar-field configuration to support the brane. Fig. \ref{Fig5}(c) shows that increasing $\alpha$, which controls the localized nonmetricity-sector correction through $\Phi(y)$, also enhances the amplitude and steepness of the kink. By contrast, Fig.  \ref{Fig5}(d) demonstrates that increasing $\beta$ reduces the asymptotic amplitude and slightly smooths the transition, revealing that the boundary-sector scalar $\Psi(y)$ counteracts the formation of a stronger domain wall. Therefore, $k$, $p$, and $\alpha$ reinforce the localization and topological field excursion, whereas $\beta$ produces a suppressive contribution, while preserving the regular single-kink structure of the fundamental brane. 

\begin{figure}[ht!]
    \centering
    \subfigure[$\rho(y)$ varying the parameters $k$.]{\includegraphics[width=4.3cm,height=3.7cm]{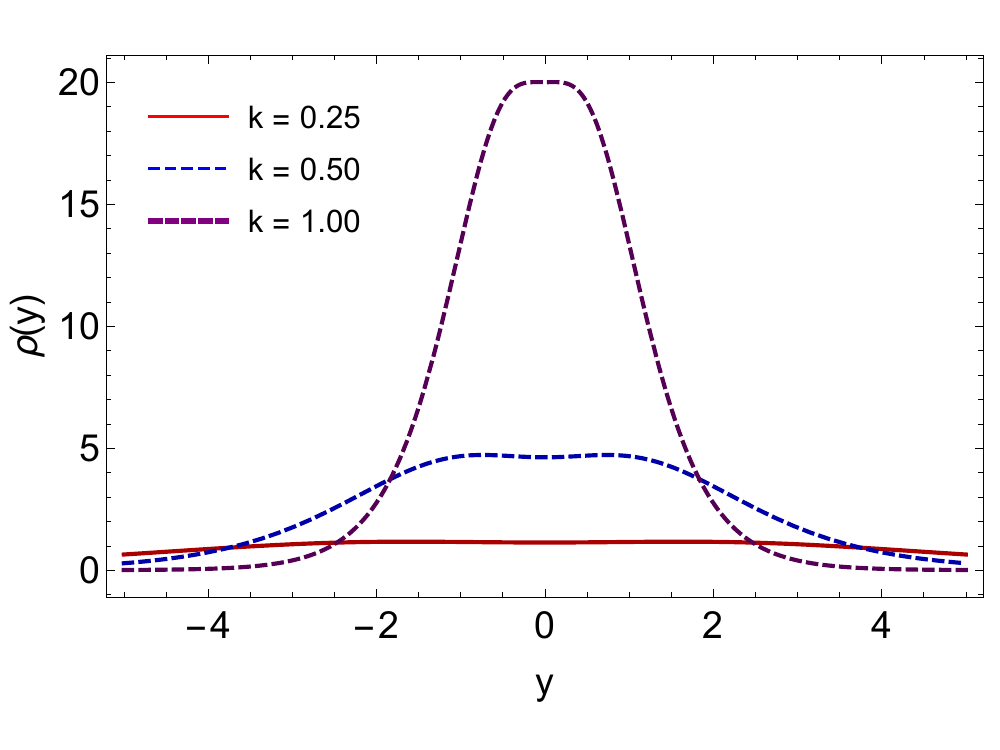}}\hfill
    \subfigure[$\rho(y)$ varying the parameters $p$.]{\includegraphics[width=4.3cm,height=3.7cm]{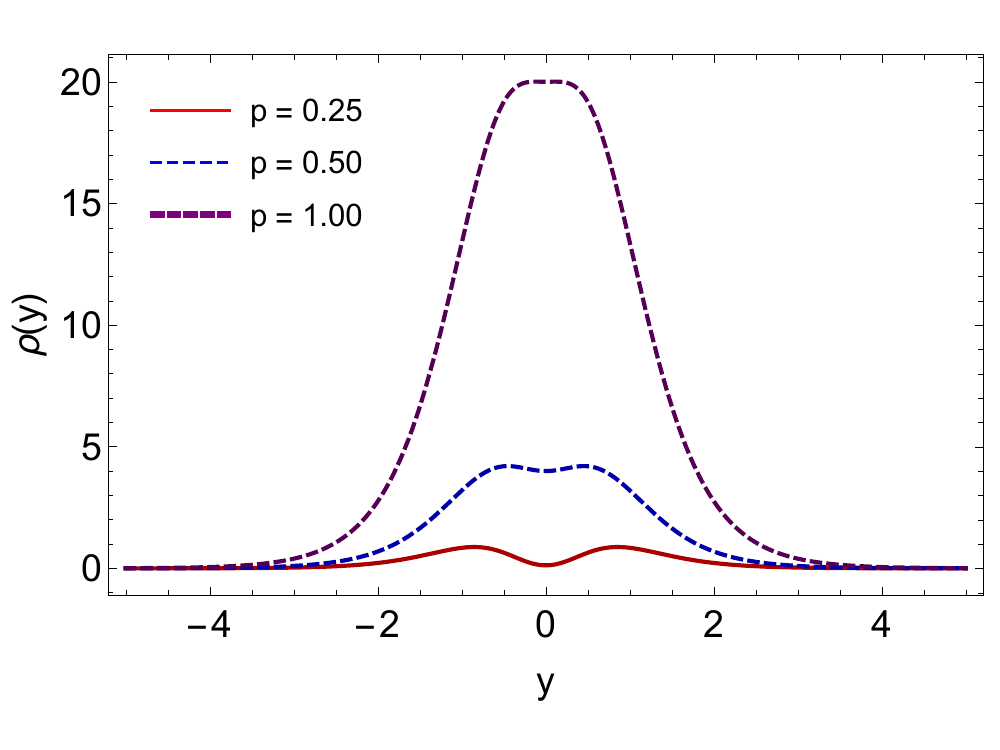}}\\
    \subfigure[$\rho(y)$ varying the parameters $\alpha$.]{\includegraphics[width=4.3cm,height=3.7cm]{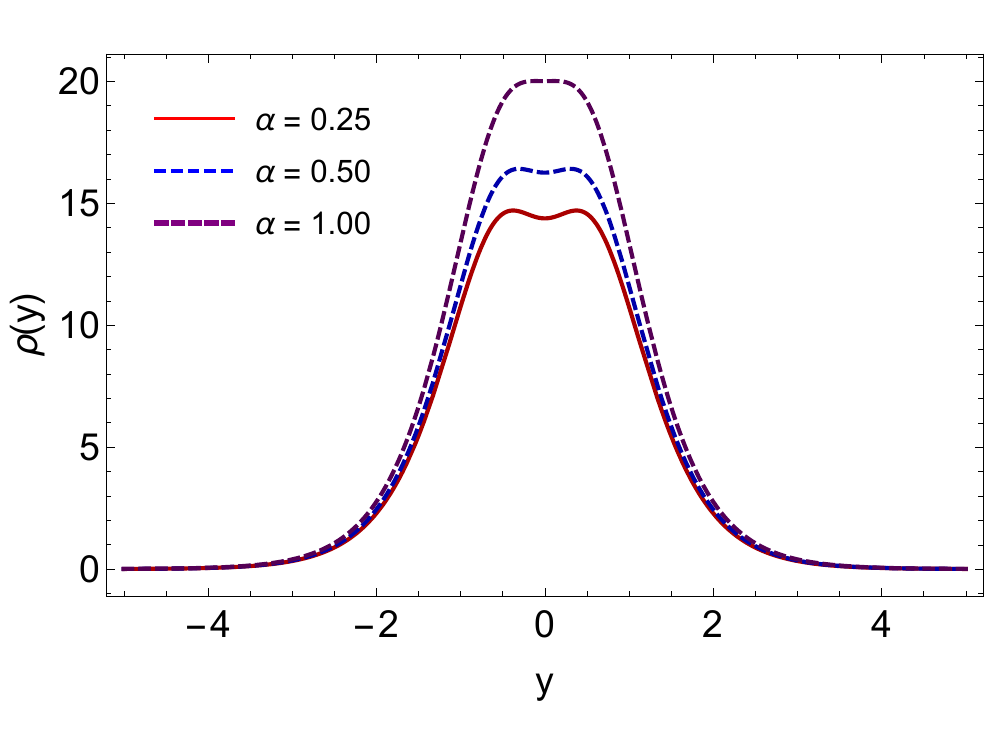}}\hfill
    \subfigure[$\rho(y)$ varying the parameters $\beta$.]
    {\includegraphics[width=4.3cm,height=3.7cm]{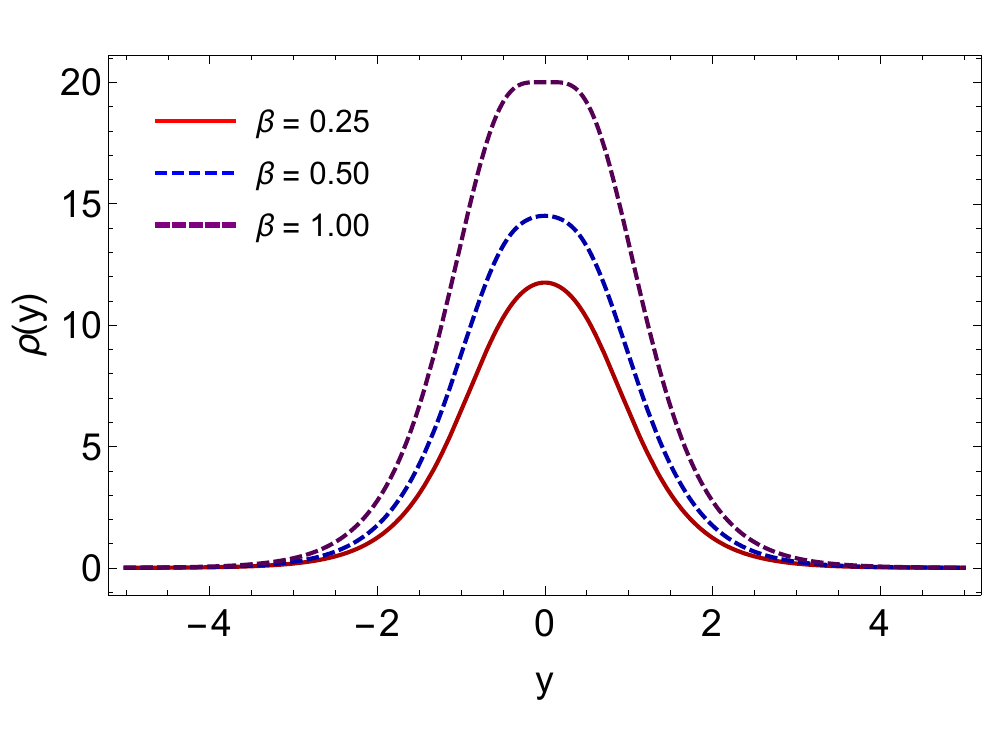}}
    \caption{Profile of the energy density $\rho(y)$ vs. the extra-dimensional coordinate $y$.}
    \label{Fig6}
\end{figure}

Finally, the energy density is numerically evaluated from Eq. \eqref{rho_full}, using the scalar-field derivative given in Eq. \eqref{chip2_simplified} and the matter potential reconstructed from Eq. \eqref{V_formula}. In this computation, we substitute the warp function in Eq. \eqref{fundamental_warp}, the auxiliary scalar profiles in Eq. \eqref{PhiPsi_model}, and the geometric potential $U(y)$ given in Eq. \eqref{U_analytic_model}, setting $\kfive=1$ and $U_\infty=0$. The resulting numerical profiles of the total energy density are displayed in Figs. \ref{Fig6}[(a)–(d)], where the parameters $k$, $p$, $\alpha$, and $\beta$ are varied, respectively.

Figs. \ref{Fig6}[(a)-(d)] shows that the total energy density is smooth, positive, and $\mathbb{Z}_2$-symmetric, with a single maximum at the brane core $y=0$, confirming the formation of a regular fundamental thick brane. As displayed in Fig. \ref{Fig6}(a), increasing $k$ raises the central peak and narrows the energy-density profile, consistently with the interpretation of $k$ as the inverse thickness of the brane: larger values of $k$ produce a thinner domain wall whose energy is more strongly concentrated around the core. Figure \ref{Fig6}(b) shows that increasing $p$ also enhances the central energy density and suppresses its distribution away from $y=0$. This behavior follows from the stronger gravitational warping generated by larger $p$, which increases the confinement of the scalar source near the brane. Thus, while $k$ predominantly controls the characteristic localization length, $p$ determines the strength of the warped geometry and reinforces the concentration of energy around the domain wall.

\subsection{Deformed brane with internal structure}
\label{sec:deformed-brane}

Let us investigate whether a localized deformation of the warp function can generate an internal structure in the brane without changing the asymptotic bulk geometry. To accomplish our purpose, we consider the warp function
\begin{align}
    A_{\delta}(y)=-p\ln\!\left[\cosh(ky)\right]+\delta\tanh^{2}(ky),
 \label{eq:deformed-warp}
\end{align}
where $\delta\geq 0$.

Since the deformation is bounded and even under $y\rightarrow -y$, the geometry remains smooth and $\mathbb{Z}_{2}$-symmetric. Moreover, $A_{\delta}(0)=0$, whereas
\begin{align}
        A_{\delta}(y)\simeq -pk|y|+\delta+p\ln 2 \qquad \mathrm{when} \qquad \vert y\vert\rightarrow\infty .
 \label{eq:deformed-asymptotic}
\end{align}
Therefore, the deformation changes only the overall asymptotic
normalization and does not modify the exponential decay responsible for the localization of the effective four-dimensional gravitational interaction.

\begin{figure}[t]
    \centering
    \subfigure[The warp factor varying the parameter $k$.]{\includegraphics[width=2.8cm,height=3.5cm]{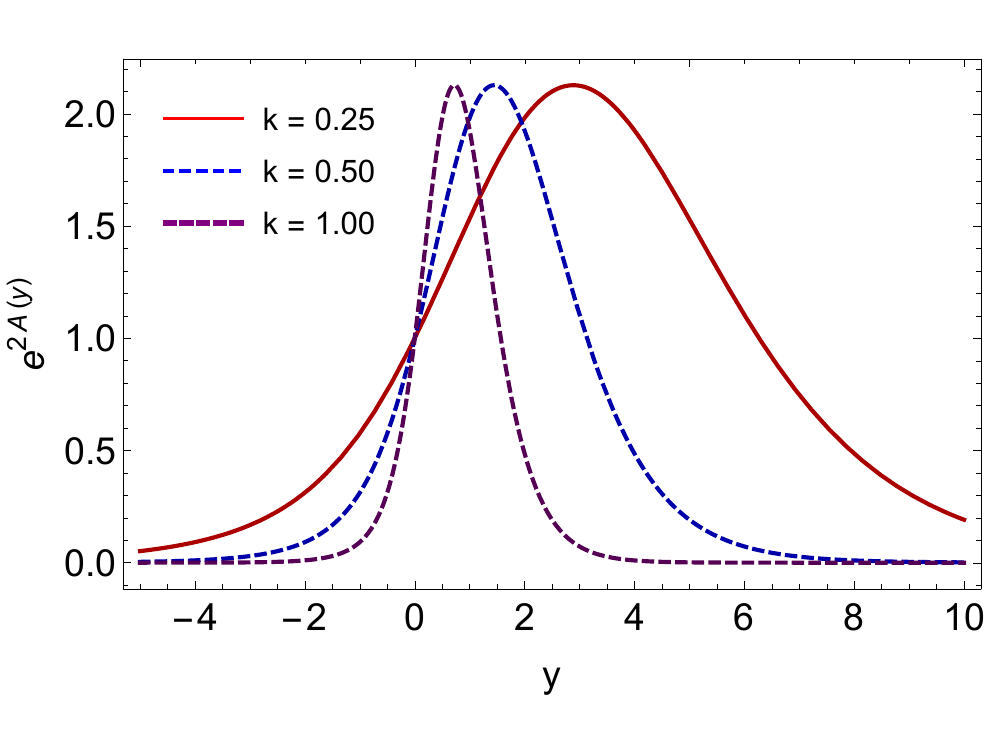}}\hfill
    \subfigure[The warp factor varying the parameter $p$.]{\includegraphics[width=2.8cm,height=3.5cm]{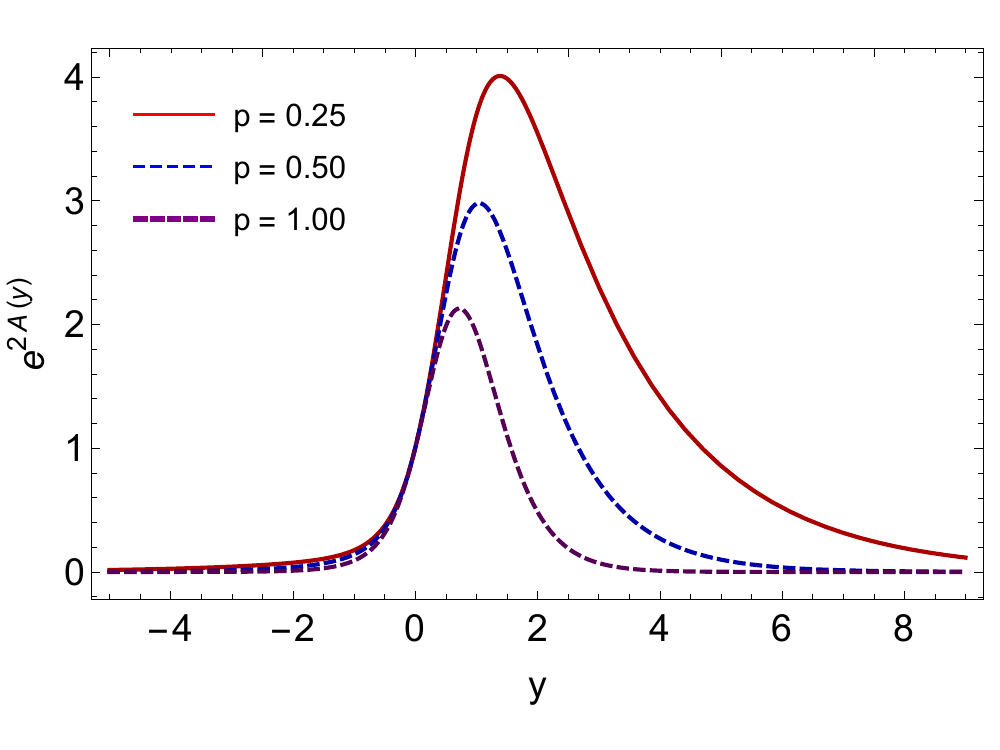}}\hfill
    \subfigure[The warp factor varying the parameter $\delta$.]{\includegraphics[width=2.8cm,height=3.5cm]{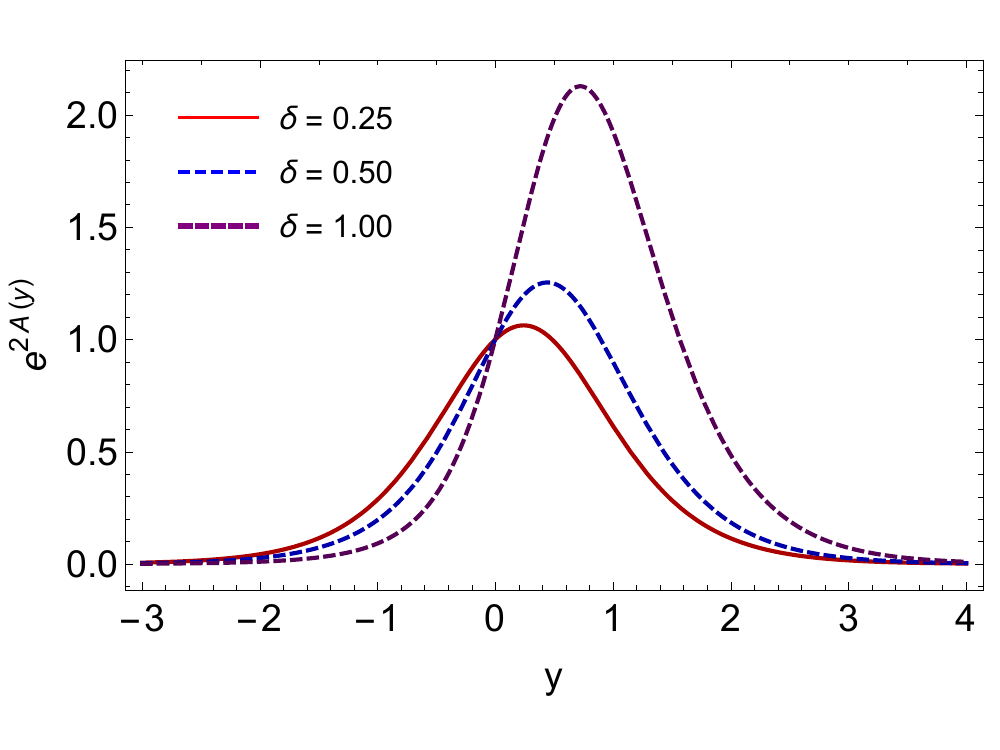}}
    \caption{Profile of the warp factor $e^{2A(y)}$ vs. the extra-dimensional coordinate $y$.}
    \label{Fig9}
\end{figure}

Note that the first two derivatives of the warp function are
\begin{align}
    A_{\delta}'(y)=&k\tanh(ky) \left[-p+2\delta\sech^{2}(ky)\right], \label{eq:deformed-first-derivative}\\
    A_{\delta}''(y)=&k^{2}\sech^{2}(ky)\left[-p-4\delta+6\delta\sech^{2}(ky)\right],
    \label{eq:deformed-second-derivative}
\end{align}
which lead us to
\begin{align}
    A_{\delta}''(0)=k^{2}(2\delta-p).
    \label{eq:deformed-core-second-derivative}
\end{align}
Consequently, the qualitative structure of the warp factor is
determined by the ratio $\delta/p$. For $\delta<p/2$, the brane core remains a local maximum of $\mathrm{e}^{2A_{\delta}}$, and the geometry describes a single thick brane. At the critical value $\delta=p/2$, the quadratic contribution around the origin vanishes and the warp factor develops a flattened core. For $\delta>p/2$, the point $y=0$ becomes a local minimum and two symmetric maxima appear at
\begin{align}
    y_{\pm}=\pm\frac{1}{k}\mathrm{arcosh}\sqrt{\frac{2\delta}{p}} \qquad \mathrm{with} \qquad \delta>\frac{p}{2}.
    \label{eq:split-maxima}
\end{align}
This transition provides a direct geometrical criterion for brane splitting. Fig. \ref{Fig9}[(a)-(c)] illustrates the dependence of the warp factor on $k$, $p$, and $\delta$. Increasing $k$ reduces the characteristic width of the configuration, while $p$ controls its asymptotic suppression. The deformation parameter $\delta$ modifies the structure near the brane core and drives the transition from a single-peak profile to a two-peak configuration.

\begin{figure}[ht!]
    \centering
    \subfigure[$Q(y)$ for different values of the parameter $k$.]{\includegraphics[width=2.8cm,height=3.5cm]{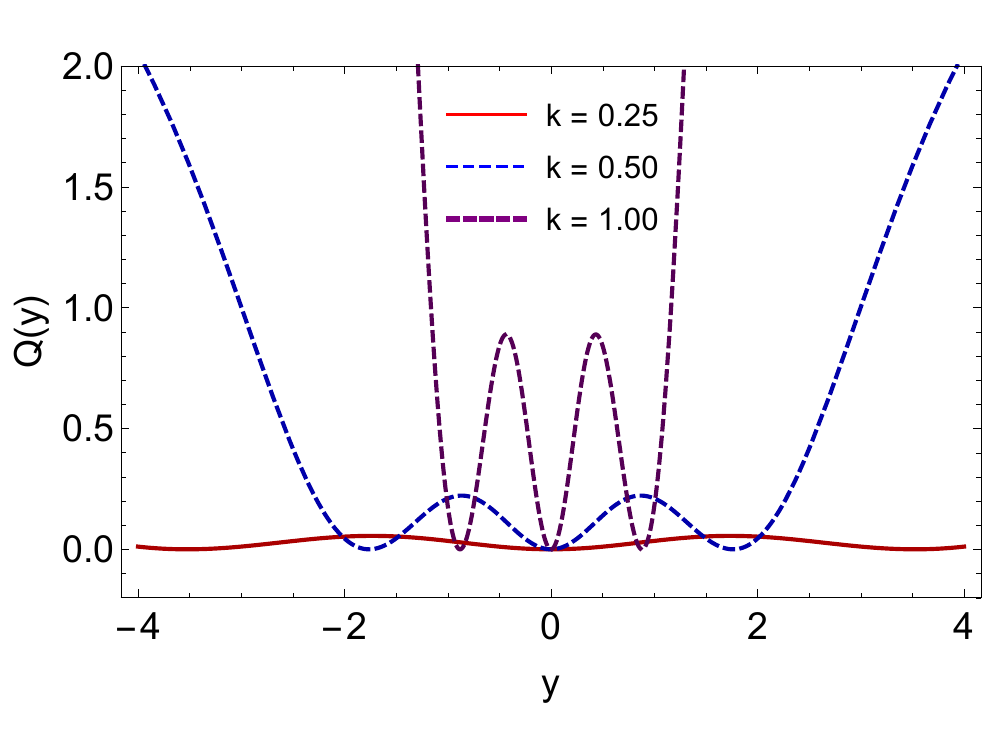}}\hfill
    \subfigure[$Q(y)$ for different values of the parameter $p$.]{\includegraphics[width=2.8cm,height=3.5cm]{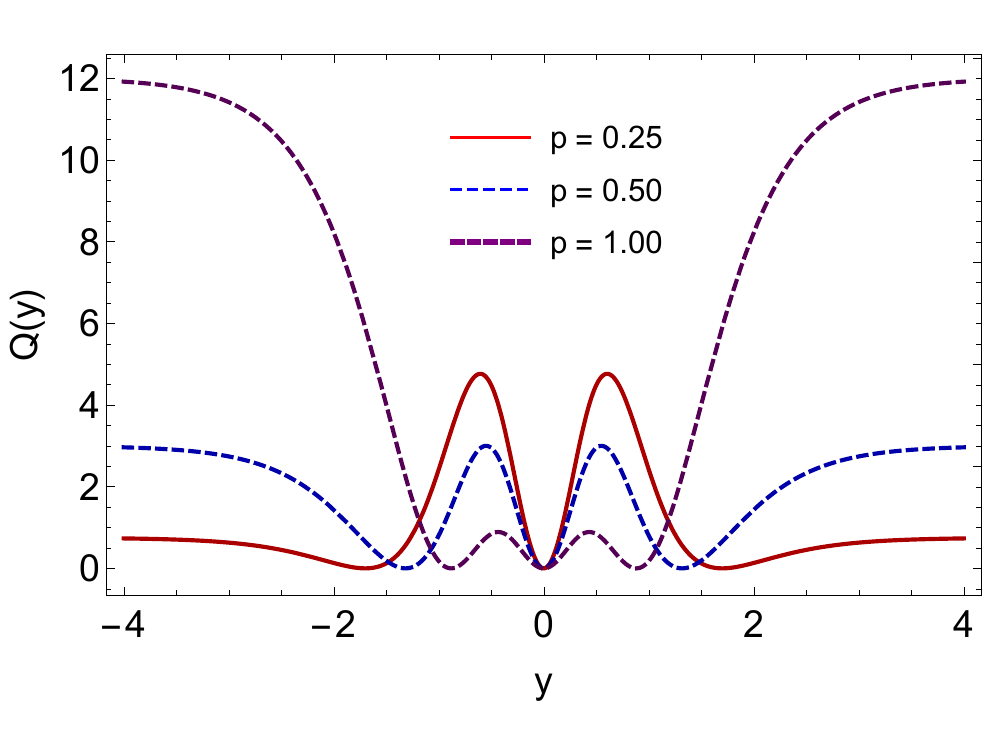}}\hfill
    \subfigure[$Q(y)$ for different values of the parameter $\delta$.]{\includegraphics[width=2.8cm,height=3.5cm]{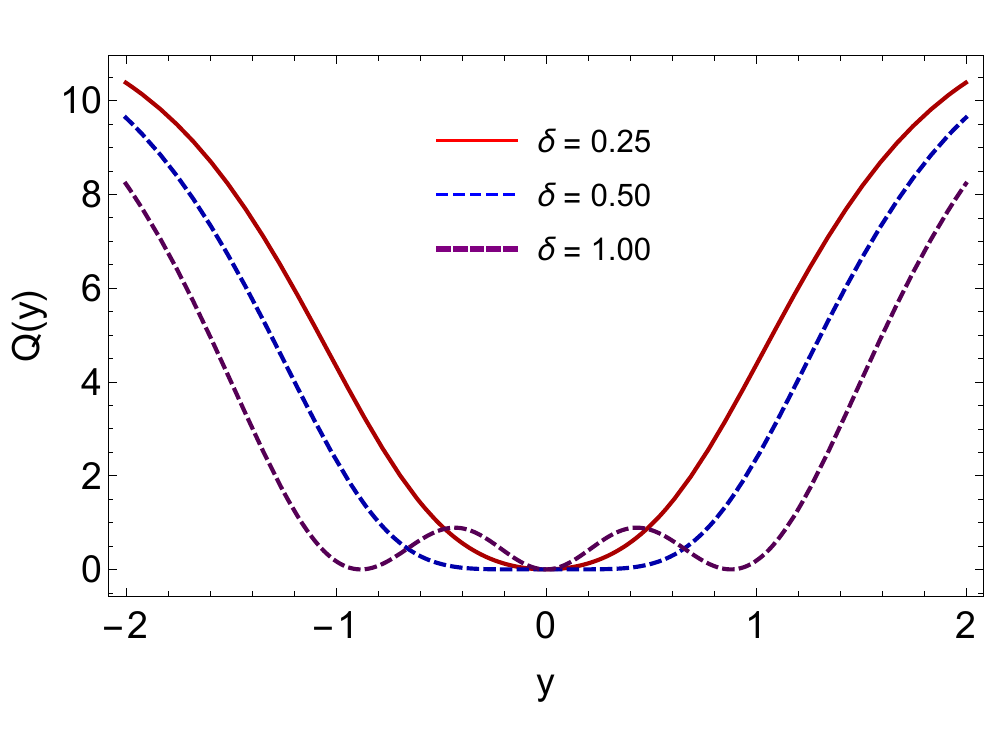}}\\
    \subfigure[$B(y)$ for different values of the parameter $k$.]{\includegraphics[width=2.8cm,height=3.5cm]{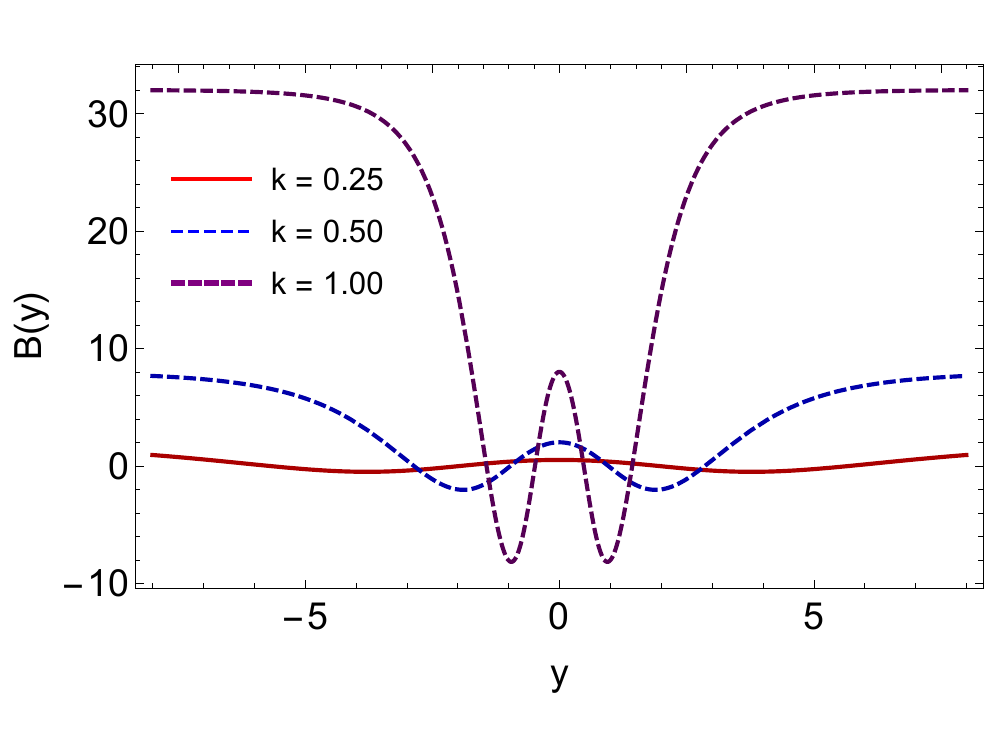}}\hfill
    \subfigure[$B(y)$ for different values of the parameter $p$.]{\includegraphics[width=2.8cm,height=3.5cm]{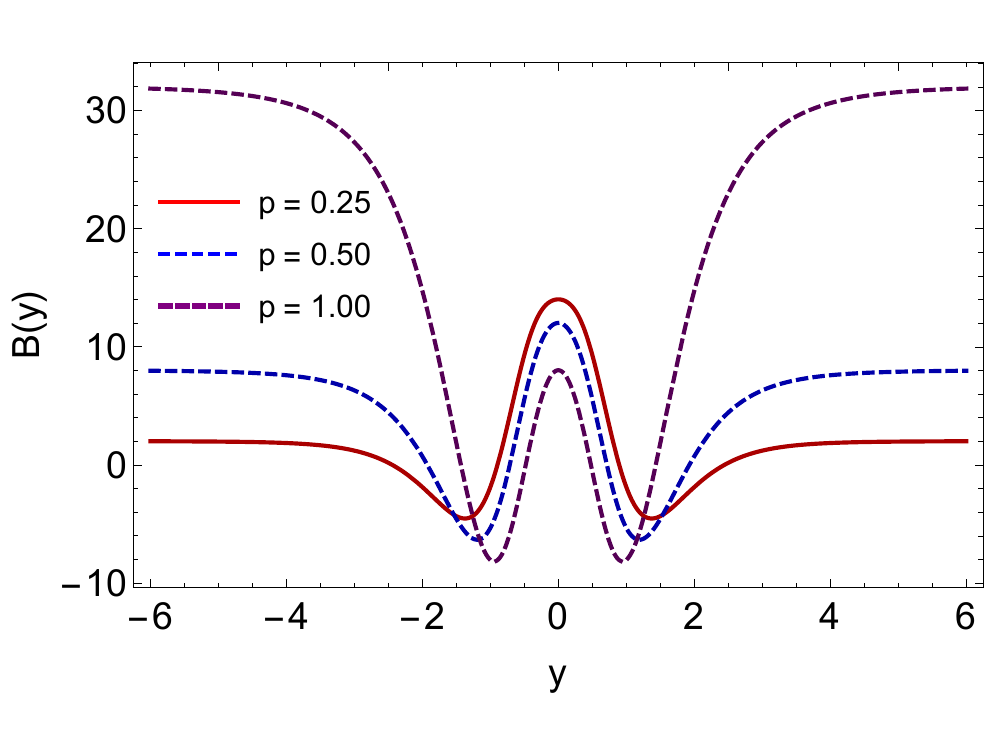}}\hfill
    \subfigure[$B(y)$ for different values of the parameter $\delta$.]{\includegraphics[width=2.8cm,height=3.5cm]{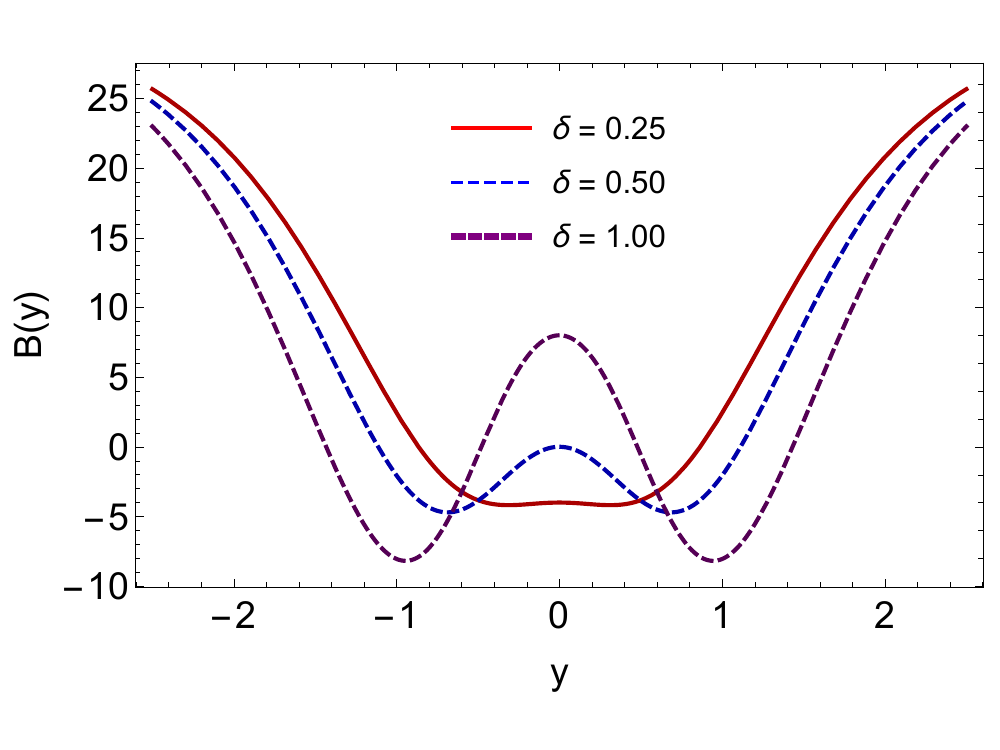}}
    \caption{Profiles of the nonmetricity and boundary scalars vs. the extra-dimensional coordinate $y$.}
    \label{Fig10}
\end{figure}

By adopting $s(y)=\sech^2(ky)$ together with the warp factor announced in Eq.~\eqref{eq:deformed-warp}, the nonmetricity scalar and the boundary term become
\begin{align}
    Q(y)=12k^{2}\bigl[1-s(y)\bigr]\bigl[p-2\delta s(y)\bigr]^{2},
 \label{eq:deformed-Q}
\end{align}
and
\begin{align}\nonumber
    B(y)=&8k^{2}s(y)\left[-p-4\delta+6\delta s(y)\right]+32k^{2}\bigl[1-s(y)\bigr]\times\\
    &\bigl[p-2\delta s(y)\bigr]^{2}.
 \label{eq:deformed-B}
\end{align}
We display the behavior of these quantities in Figs. \ref{Fig10}[(a)-(c)] and \ref{Fig10}[(d)-(f)], respectively.

As shown in Fig. \ref{Fig10}[(a)-(c)], the nonmetricity scalar preserves its regular and even character for all values of the deformation parameter. At the same time, its internal profile undergoes a qualitative modification as $\delta$ increases. For $\delta\leq p/2$, $Q(y)$ exhibits the usual single-peak distribution centered around the brane, indicating that the nonmetricity is predominantly localized within a single gravitational core. However, once $\delta>p/2$, two additional zeros emerge at $y=y_{\pm}$, precisely where the warp function develops new extrema. Consequently, the nonmetricity density redistributes into a multipeak structure, reflecting the appearance of internal layers inside the thick brane. Since the asymptotic value $Q_{\infty}=12p^{2}k^{2}$ remains unchanged, this deformation does not modify the bulk geometry. Instead, it reorganizes the localization of the gravitational degrees of freedom in the neighborhood of the brane.

Figs. \ref{Fig10}[(d)-(f)] further reinforces this picture through the behavior of the boundary term. Unlike the nonmetricity scalar, the boundary scalar is highly sensitive to the deformation at the brane core, where $B(0)=8k^{2}(2\delta-p)$ changes sign exactly at the critical value $\delta=p/2$. This behavior provides a clear geometrical signature of the transition from a fundamental to a split-brane configuration. For $\delta<p/2$, the negative central value of $B$ indicates that the boundary contribution is concentrated around a single core, whereas for $\delta>p/2$ the central region becomes positive with an extremum that moves away from the origin, consistently following the emergence of the internal structure observed in the warp factor and in $Q(y)$. Since the asymptotic limit $B_{\infty}=32p^{2}k^{2}$ is independent of $\delta$, the deformation affects only the local geometry while preserving the asymptotically AdS$_5$ character of the five-dimensional spacetime. Collectively, Figs. \ref{Fig10}[(a)-(c)] and \ref{Fig10}[(d)-(f)] demonstrate that the parameter $\delta$ controls the internal organization of the brane without altering the global properties of the bulk, making it a genuine geometrical mechanism for inducing brane splitting within the scalar representation of $f(Q, B)$ gravity.

Furthermore, one notes that the parameter $\delta$ changes the geometry only in the vicinity of the domain wall, leaving the asymptotic values of the geometrical scalars unchanged. In the split regime, $\delta>p/2$, the nonmetricity scalar also vanishes at $y=y_{\pm}$, since these points correspond to additional extrema of the warp function. Consequently, $Q(y)$ develops a multi-peak profile. Meanwhile, the sign change of $B(0)$ at $\delta=p/2$ shows that the boundary sector is directly sensitive to the transition between the fundamental and split configurations. These properties are displayed in Figs. \ref{Fig10}[(a)-(c)] and \ref{Fig10}[(d)-(f)].

Within this framework, the geometric scalars retain their original form, viz.,
\begin{align}
 \Phi(y)=1+\alpha s(y) \qquad \mathrm{and} \qquad \Psi(y)=\beta s(y),
 \label{eq:deformed-auxiliary-fields}
\end{align}
with $\alpha>-1$, so that the effective tensor kinetic coefficient remains positive throughout the bulk. Substituting
Eqs. \eqref{eq:deformed-Q}, \eqref{eq:deformed-B}, and
\eqref{eq:deformed-auxiliary-fields} into Eq. \eqref{U_quadrature}, the geometric potential can be obtained analytically. In this case, the geometric potential is
\begin{align}\nonumber
    U(y)=&k^{2}\Bigg\{4(3\alpha+8\beta) p^{2}s-[2(3\alpha+8\beta)\left(p^{2}
    +4p\delta\right)\\ \nonumber
    +&4\beta(p+4\delta)]s^{2}+\left[\frac{16}{3}(3\alpha+8\beta)\delta(p+\delta)+16\beta\delta\right]s^{3}\\
    -&4(3\alpha+8\beta)\delta^{2}s^{4}\Bigg\}+U_\infty.
 \label{eq:deformed-U}
\end{align}

\begin{figure}[ht!]
    \centering
    \subfigure[$U(y)$ for different values of the parameter $k$.]{\includegraphics[width=2.8cm,height=3.5cm]{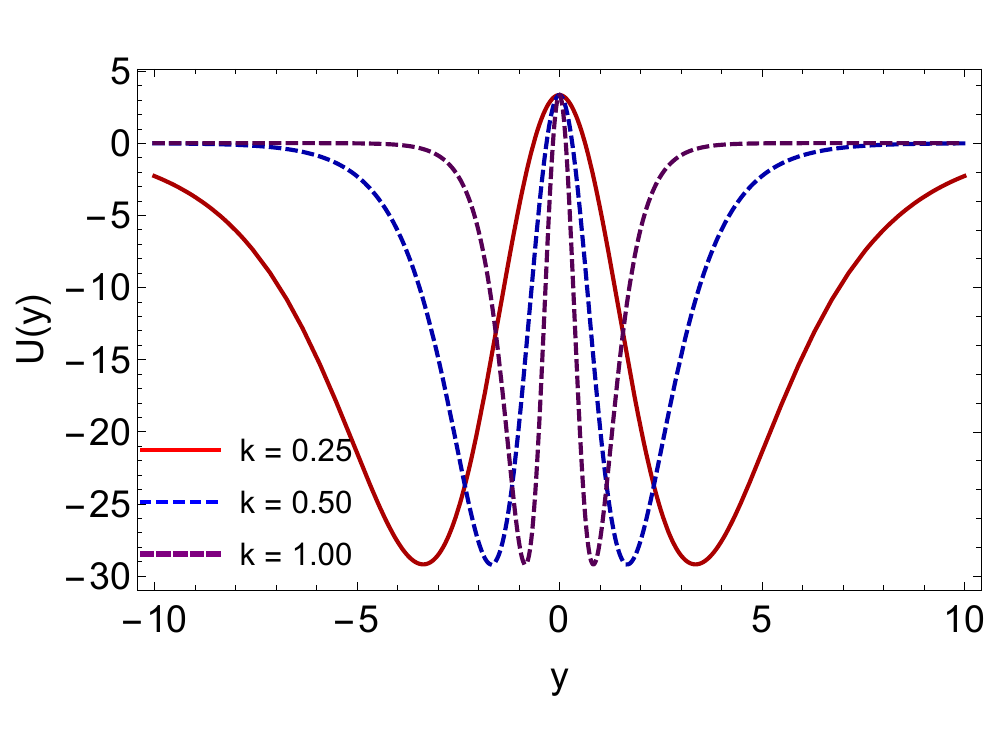}}\hfill
    \subfigure[$U(y)$ for different values of the parameter $p$.]{\includegraphics[width=2.8cm,height=3.5cm]{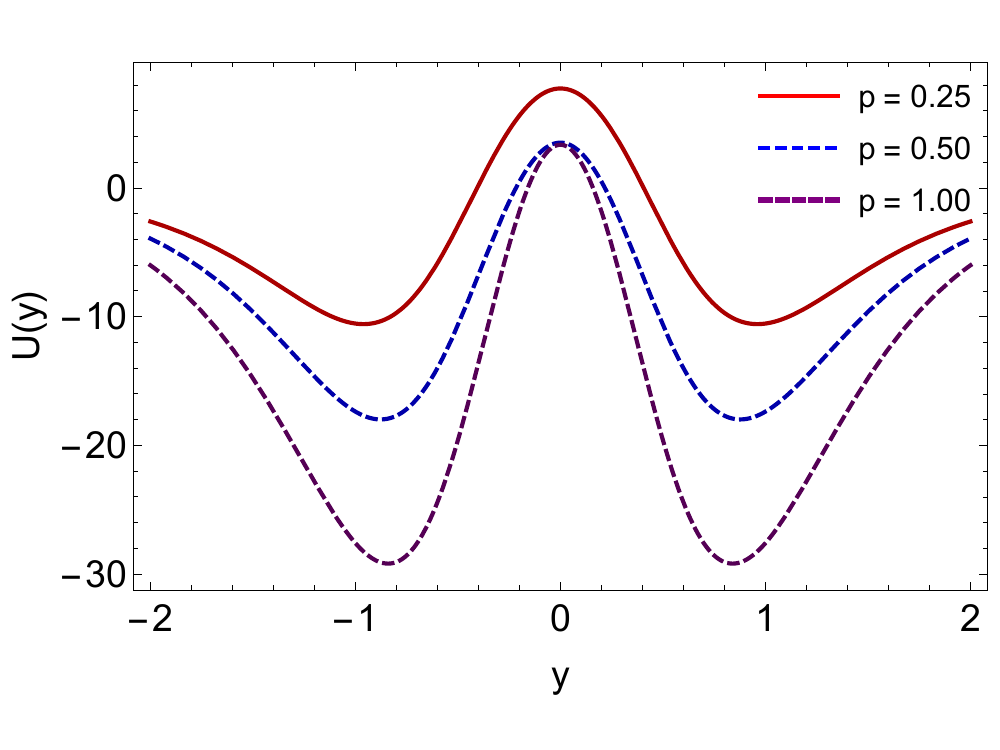}}\hfill
    \subfigure[$U(y)$ for different values of the parameter $\delta$.]{\includegraphics[width=2.8cm,height=3.5cm]{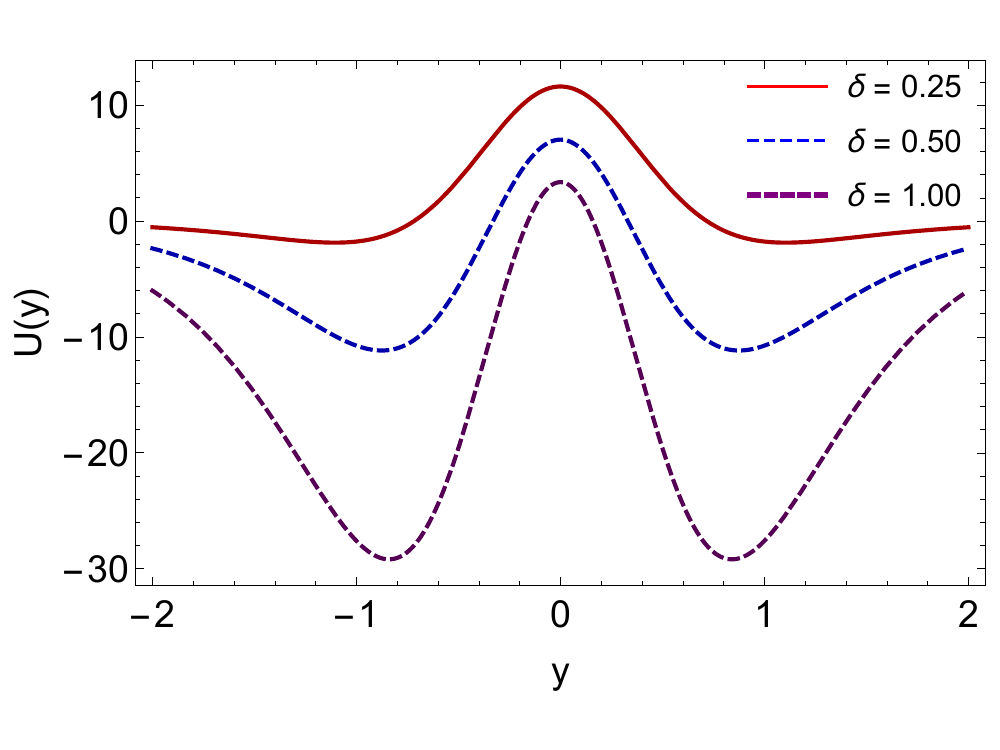}}\\
    \subfigure[$U(y)$ for different values of the parameter $\alpha$.]{\includegraphics[width=2.8cm,height=3.5cm]{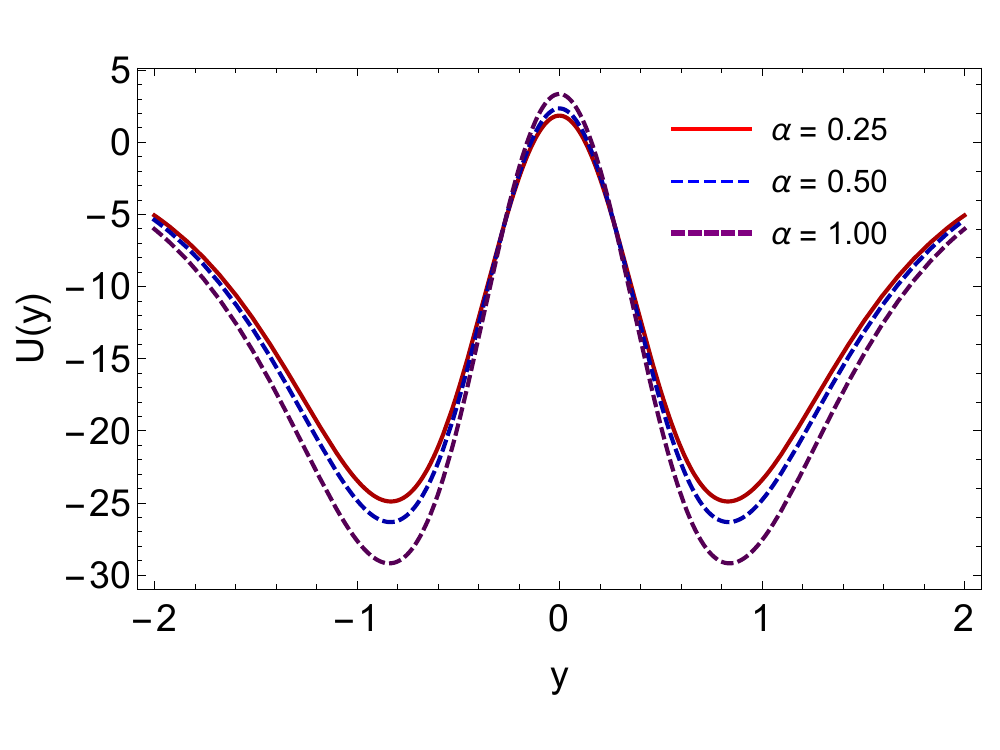}}
    \subfigure[$U(y)$ for different values of the parameter $\beta$.]{\includegraphics[width=2.8cm,height=3.5cm]{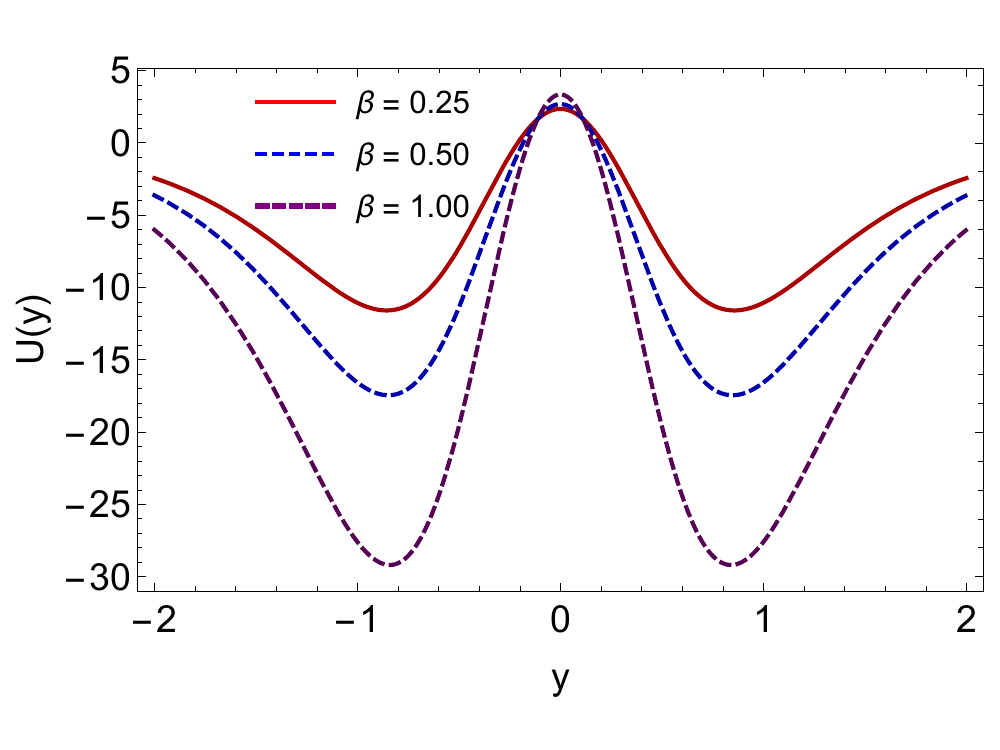}}
    \caption{Profile of the geometric scalar potential $U(y)$ vs. the extra-dimensional coordinate $y$ with $U_{\infty}=0$.}
    \label{Fig12}
\end{figure}

The integration constant $U_{\infty}$ represents the asymptotic bulk contribution and is set to zero in the numerical analysis. Unlike the potential associated with the fundamental brane, the deformed potential contains terms up to fourth order in $s(y)$. These terms produce a central barrier surrounded by symmetric lateral structures, whose separation and depth are primarily controlled by $\delta$. Increasing $k$ compresses these structures toward the brane core, whereas variations of $k$, $p$, $\delta$, $\alpha$, and $\beta$ change their relative amplitudes, as shown in Fig. \ref{Fig12}[(a)-(e)].
The profiles displayed in Figs. \ref{Fig12}[(a)-(e)] shows that the geometric potential remains smooth and symmetric under the reflection $y\to-y$, while its internal structure is strongly controlled by the deformation parameter $\delta$. In particular, increasing $\delta$ gradually transforms the single-well profile into a double-well configuration separated by a central barrier, providing the potential counterpart of the brane-splitting mechanism previously identified through the warp factor and the geometric scalars. This behavior indicates that the effective geometric interaction is no longer concentrated around a single gravitational core but instead becomes distributed between two symmetric localized regions, reflecting the emergence of an internal brane structure. By contrast, the parameters $k$ and $p$ mainly regulate the characteristic length and energy scales of the potential. As illustrated in Figs. \ref{Fig12}[(a) and (b)], increasing $k$ compresses the entire profile toward the brane core, whereas larger values of p deepen the potential wells and reinforce the overall confinement without modifying the qualitative shape of the configuration. The auxiliary coupling parameters $\alpha$ and $\beta$, shown in Figs. \ref{Fig12}[(d) and (e)], primarily alter the relative height of the central barrier and the depth of the lateral minima, demonstrating that the nonmetricity and boundary sectors control the strength of the effective geometric interaction rather than the onset of the internal structure itself. Collectively, these results show that $\delta$ acts as the genuine geometrical parameter responsible for inducing brane splitting, whereas $k$, $p$, $\alpha$, and $\beta$ provide quantitative adjustments to the localization properties while preserving the regularity of the potential and the asymptotically AdS$_5$ character of the bulk geometry.

Meanwhile, the profile of the matter scalar takes the form
\begin{align}\nonumber
 \chi'^{\,2}(y)=&k^{2}\Big\{\left[3p+12\delta-6\alpha p
 +4\beta(1-2p)\right]s\\ \nonumber
 +&[24\alpha\delta+9\alpha p+16\beta\delta+\beta(8p-6)-18\delta]s^{2}\\
 -& 2\delta(15\alpha+8\beta)s^{3}\Big\}.
 \label{eq:deformed-chi-prime}
\end{align}

At the brane core, the equation \eqref{eq:deformed-chi-prime} boils down to
\begin{align}
    \chi'^{\,2}(0)= k^{2}\left[ 3(1+\alpha)(p-2\delta)-2\beta\right],
    \label{eq:deformed-chi-core}
\end{align}
Meanwhile, in the asymptotic region, one obtains
\begin{align}
    \chi'^{\,2}(y)\simeq k^{2} \left[3p+12\delta-6\alpha p+4\beta(1-2p)\right]\sech^{2}(ky).
    \label{eq:deformed-chi-asymptotic}
\end{align}

The matter source is physically admissible only when the polynomial on the right-hand side of Eq.~\eqref{eq:deformed-chi-prime} is non-negative for every $0\leq s\leq1$. In the parameter sector $\alpha\geq0$, $\beta\geq0$, and $\delta\geq0$, the coefficient of $s^{3}$ is nonpositive, and it is sufficient to impose the endpoint conditions
\begin{align}
    3p+12\delta-6\alpha p+4\beta(1-2p)\geq0,
    \label{eq:deformed-condition-infinity}
\end{align}
and
\begin{align}
    3(1+\alpha)(p-2\delta)-2\beta\geq0.
    \label{eq:deformed-condition-core}
\end{align}

The second condition is particularly restrictive. For
$\beta\geq0$, it requires $\delta<p/2$ and therefore prevents the canonical scalar source from supporting the fully split geometrical regime. A configuration with $\delta>p/2$ can be obtained only for a sufficiently negative value of $\beta$, or by adopting different auxiliary-field profiles, in which case the complete polynomial in Eq.~\eqref{eq:deformed-chi-prime} must be checked explicitly.

For parameters satisfying these consistency conditions, the scalar field is determined numerically from
\begin{align}
    \chi(y)=\int_{0}^{y}d\bar{y}\, \sqrt{\chi'^{\,2}(\bar{y})} \qquad \mathrm{with} \qquad \chi(0)=0.
    \label{eq:deformed-chi-integral}
\end{align}

\begin{figure}[ht!]
    \centering
    \subfigure[$\chi(y)$ for different values of the parameter $k$.]{\includegraphics[width=2.8cm,height=3.5cm]{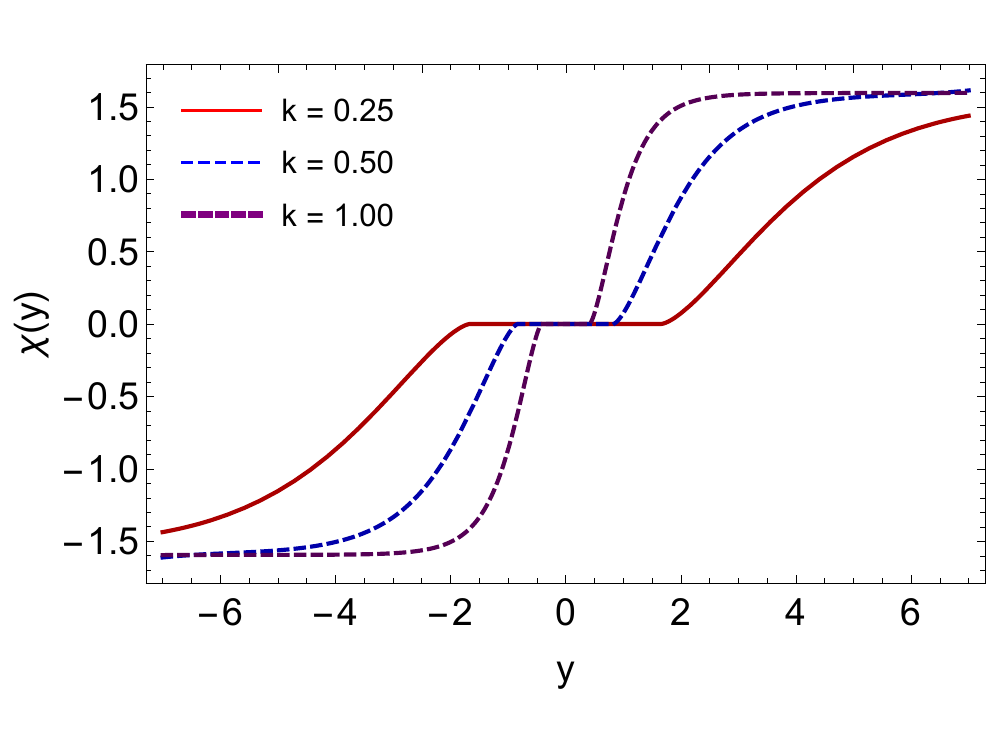}}\hfill
    \subfigure[$\chi(y)$ for different values of the parameter $p$.]{\includegraphics[width=2.8cm,height=3.5cm]{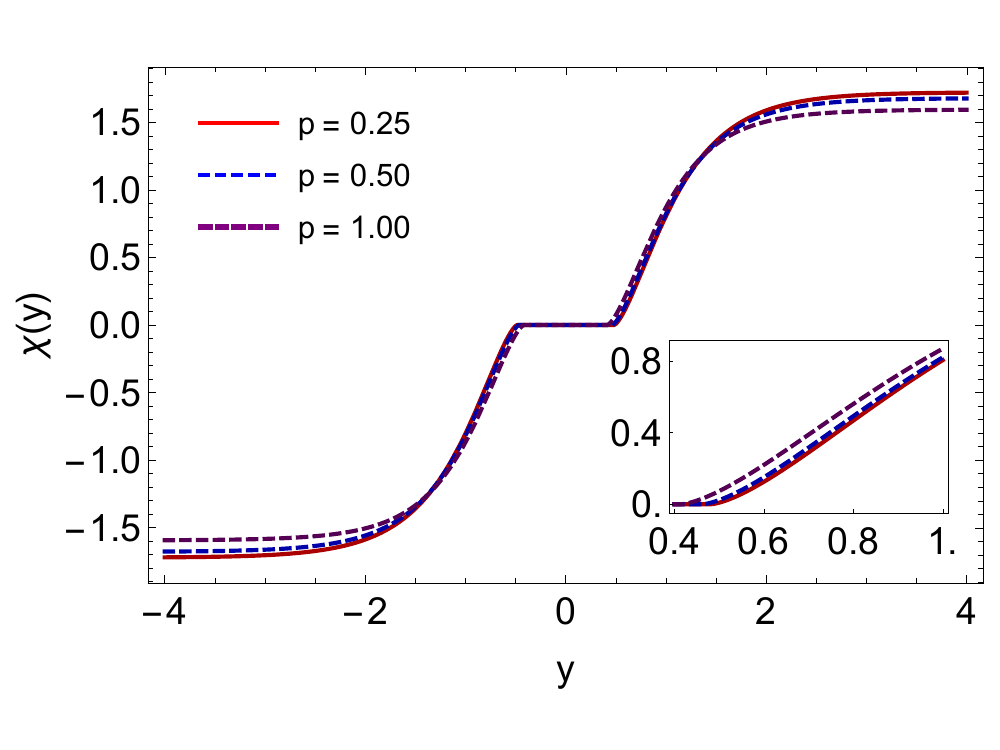}}\hfill
    \subfigure[$\chi(y)$ for different values of the parameter $\delta$.]{\includegraphics[width=2.8cm,height=3.5cm]{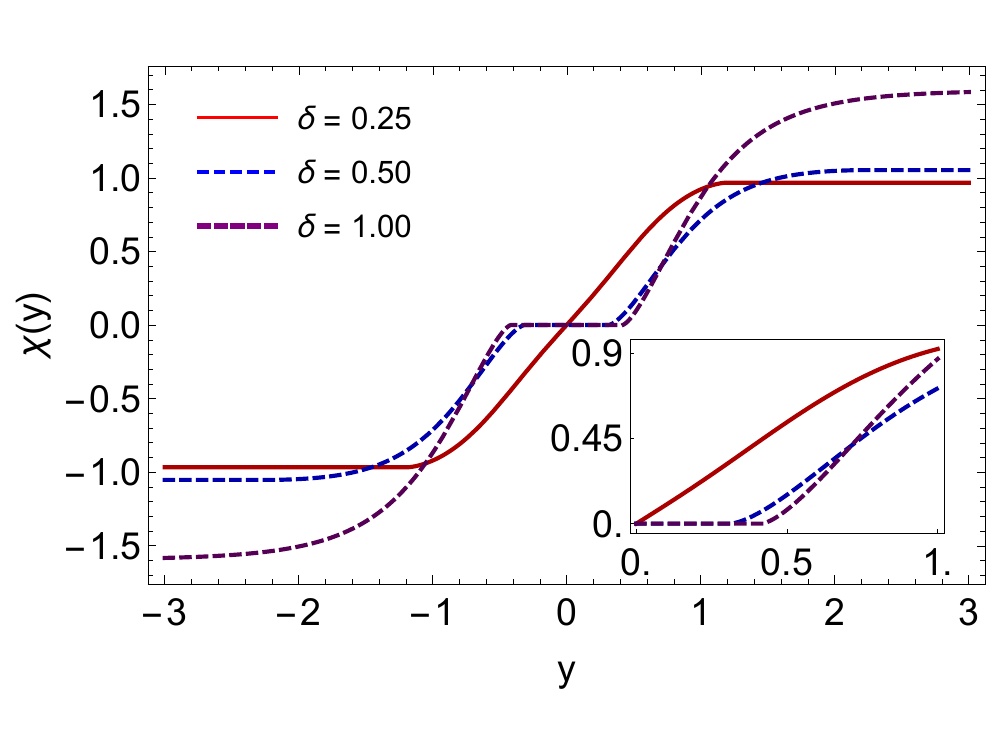}}\\
    \subfigure[$\chi(y)$ for different values of the parameter $\alpha$.]{\includegraphics[width=2.8cm,height=3.5cm]{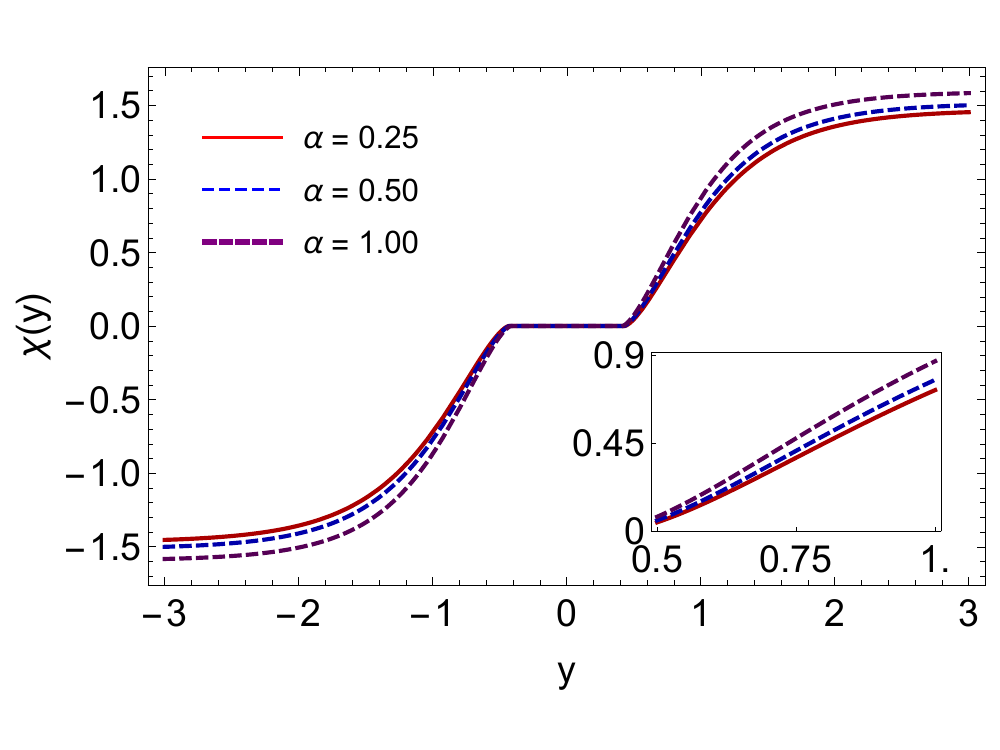}}
    \subfigure[$\chi(y)$ for different values of the parameter $\beta$.]{\includegraphics[width=2.8cm,height=3.5cm]{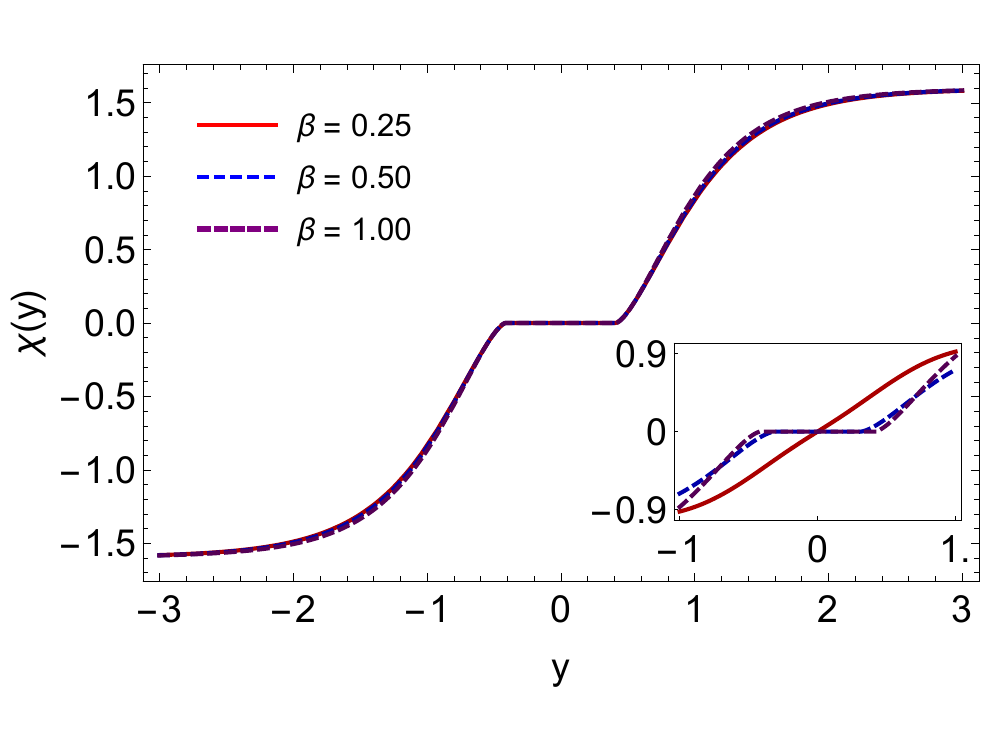}}
    \caption{Profile of the brane-generating scalar field $\chi(y)$ vs. the extra-dimensional coordinate $y$.}
    \label{Fig13}
\end{figure}

The resulting solutions are odd under $y\rightarrow-y$ and approach finite vacuum values in the asymptotic region. When the central value of $\chi'^{\,2}$ is suppressed, the scalar field develops an approximately constant region around $y=0$, followed by two localized transitions. This double-kink-like behavior is the matter-field counterpart of the internal geometrical structure. The numerical profiles shown in Fig. \ref{Fig13}[(a)-(e)] demonstrate that $k$ sets the width of the transitions, whereas $\delta$ controls the formation and extension of the central plateau. The remaining
parameters determine the field excursion and the relative strength of the two transition regions, see Figs. \ref{Fig13}[(a)-(e)].
The numerical solutions displayed in Figs. \ref{Fig13}[(a)–(e)] shows that the matter scalar field remains a smooth odd function of the extra-dimensional coordinate, interpolating continuously between two distinct asymptotic vacuum states, as expected for a topological domain-wall configuration. The parameter $k$, shown in Fig. \ref{Fig13}(a), primarily controls the characteristic thickness of the scalar transition. Thus, increasing $k$ compresses the kink profile and localizes the scalar field more strongly around the brane without modifying its asymptotic values. The parameter $p$, Fig. \ref{Fig13}(b), mainly regulates the field excursion between the vacua, producing a larger asymptotic amplitude while preserving the overall topological structure. More importantly, the deformation parameter $\delta$, illustrated in Fig. \ref{Fig13}(c), progressively suppresses the scalar gradient near the origin and generates an increasingly pronounced central plateau, signaling the onset of a double-kink configuration. This behavior constitutes the matter-sector manifestation of the geometrical brane-splitting mechanism previously identified through the warp factor, the nonmetricity scalar, and the effective geometric potential. Figs. \ref{Fig13}[(d) and (e)] show that the auxiliary coupling parameters $\alpha$ and $\beta$ do not alter the qualitative topological nature of the solution. Consequently, while $k$, $p$, $\alpha$, and $\beta$ provide quantitative adjustments to the scalar profile, the deformation parameter $\delta$ is the genuine control parameter responsible for transforming the fundamental kink-like into a double-kink-like configuration, consistently reflecting the emergence of an internal brane structure.

Finally, we examine the energy density. We exposed our results in Figs. \ref{Fig14}{(a)-(e)]. In this case, we show that the deformation redistributes the energy stored in the domain wall. For small $\delta$, the density retains a predominantly single-peak structure. As the deformation increases, the central region becomes depleted, and two symmetric maxima emerge away from the origin. This double-peak distribution constitutes the standard signature of an internal brane structure. Increasing $k$ appears to narrow the two peaks and enhances their amplitudes, while $p$ controls the overall gravitational confinement. The parameters $\alpha$ and $\beta$ modify the relative height of the peaks through the nonmetricity and boundary sectors, respectively.
\begin{figure}[ht!]
    \centering
    \subfigure[$\rho(y)$ for different values of the parameter $k$.]{\includegraphics[width=2.8cm,height=3.5cm]{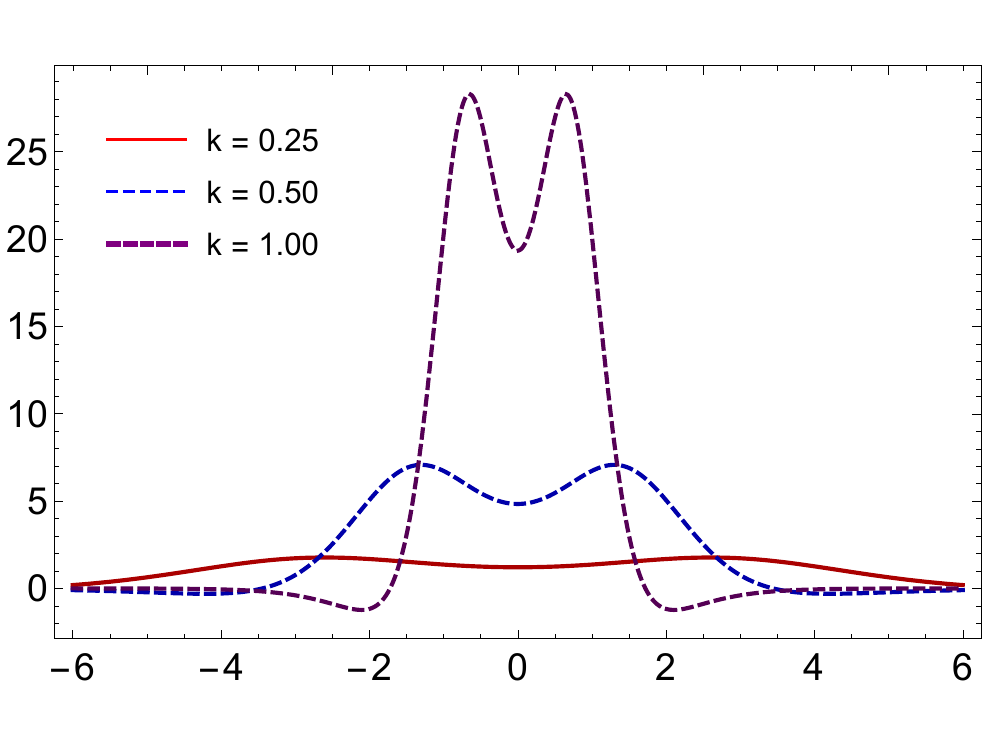}}\hfill
    \subfigure[$\rho(y)$ for different values of the parameter $p$.]{\includegraphics[width=2.8cm,height=3.5cm]{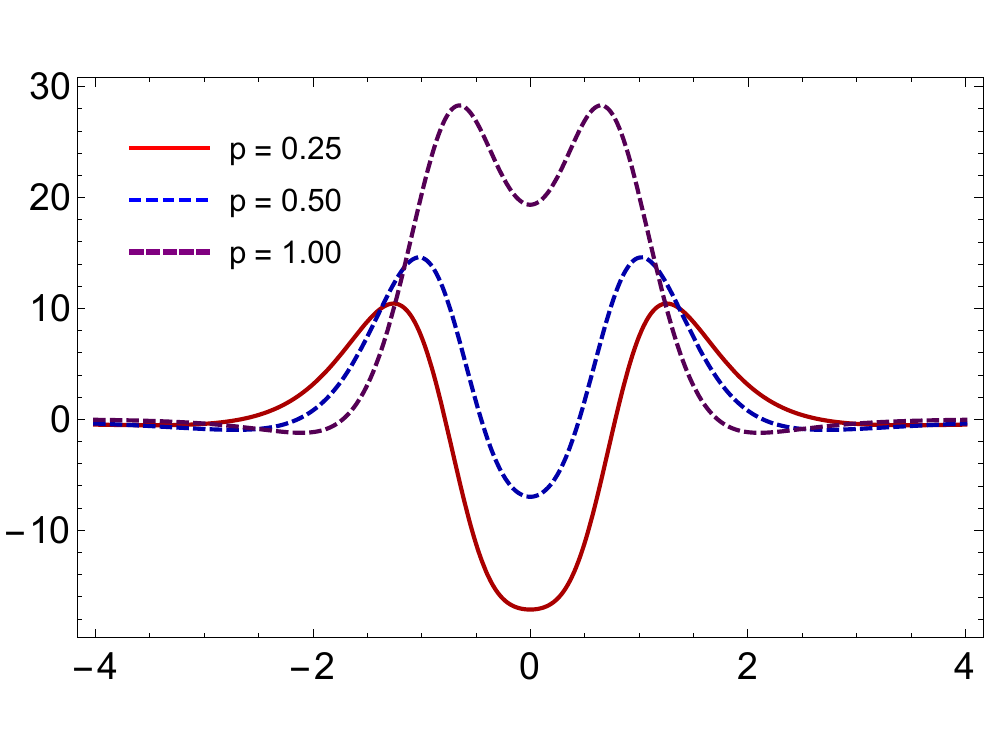}}\hfill
    \subfigure[$\rho(y)$ for different values of the parameter $\delta$.]{\includegraphics[width=2.8cm,height=3.5cm]{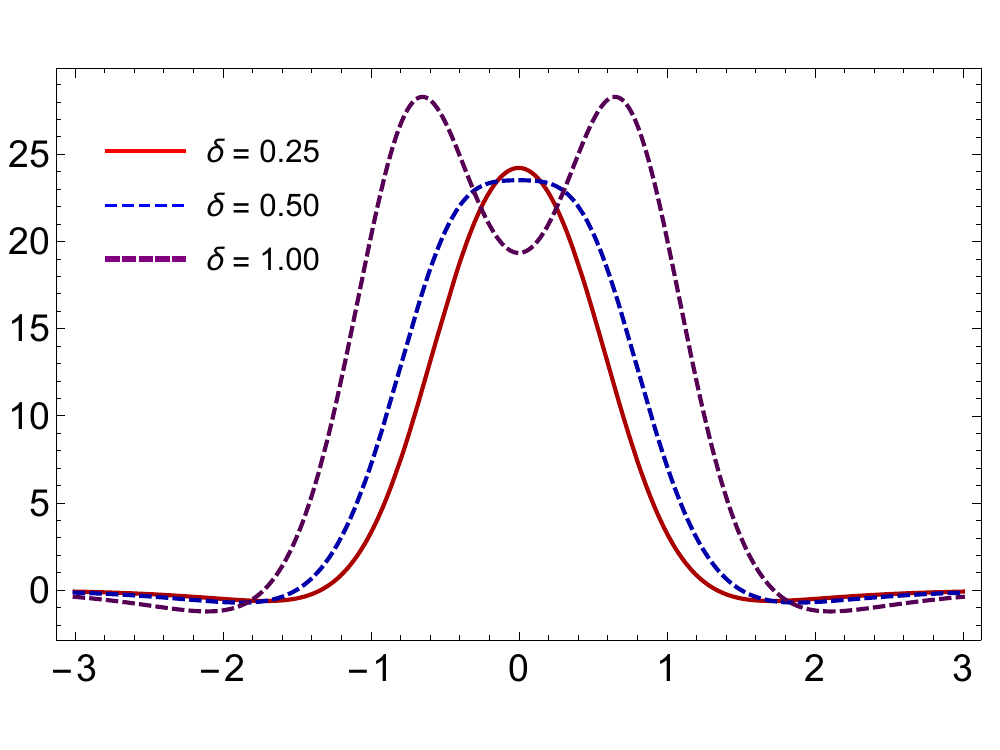}}\\
    \subfigure[$\rho(y)$ for different values of the parameter $\alpha$.]{\includegraphics[width=2.8cm,height=3.5cm]{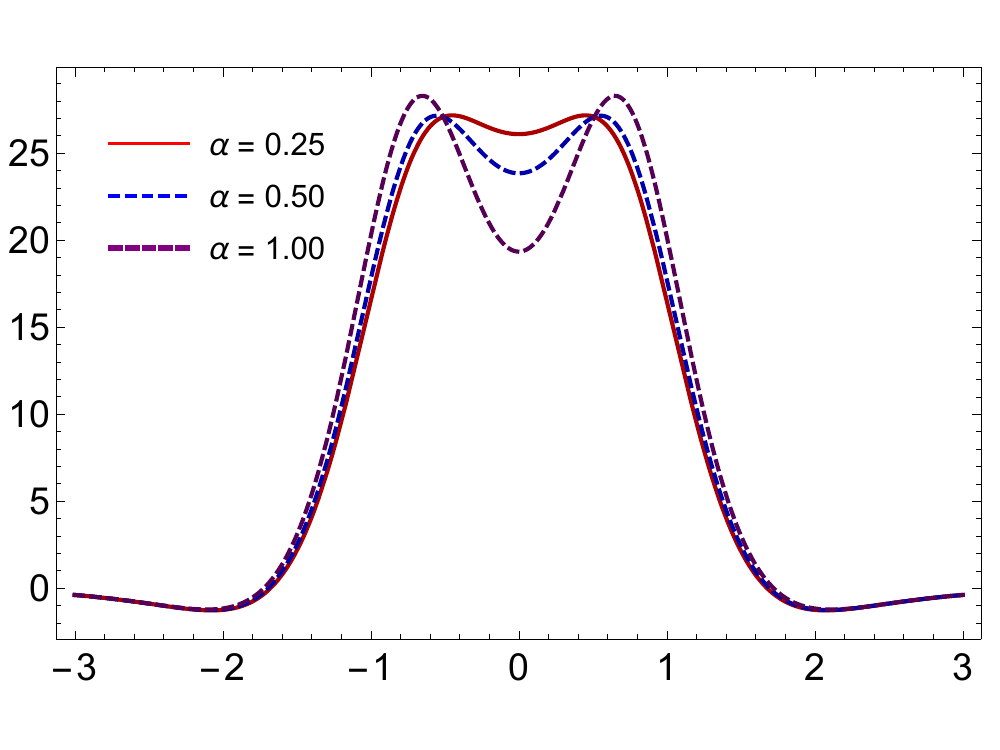}}
    \subfigure[$\rho(y)$ for different values of the parameter $\beta$.]{\includegraphics[width=2.8cm,height=3.5cm]{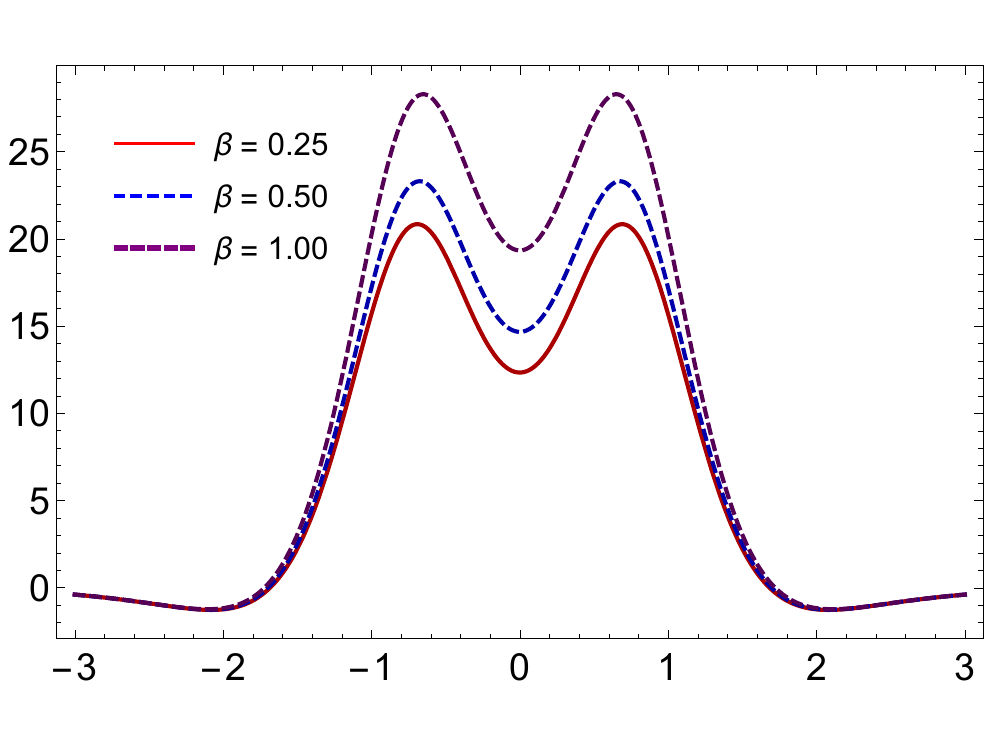}}
    \caption{Profile of the energy density $\rho(y)$ vs. the extra-dimensional coordinate $y$.}
    \label{Fig14}
\end{figure}
Furthermore, the deformed model does not guarantee the pointwise positivity of the energy density. Particularly, Fig. \ref{Fig14}(b) shows that, for sufficiently small values of $p$, such as $p=0.25$ for the parameter set considered, $\rho(y)$ develops a negative minimum around the brane core. This behavior does not indicate a ghost instability, since the condition $\chi'(y)\geq 0$ still ensures a real canonical scalar source; rather, it arises because the scalar potential becomes sufficiently negative to dominate the positive kinetic contribution in $\rho(y)=\mathrm{e}^{2A(y)}[\chi'^2(y)/2+V(y)]$. As $p$ increases, the central negative region is progressively lifted, and the energy density evolves toward a positive double-peak profile, signaling the redistribution of the matter source into two symmetric localized layers. Moreover, since $\rho(y)$ contains the asymptotic vacuum contribution, the localized density $\rho_{\mathrm{loc}}(y)$, defined in Eq. \eqref{rho_local}, provides the appropriate quantity for distinguishing a genuinely localized negative-energy region from the contribution associated with the asymptotically $\mathrm{AdS}_5$ bulk.

The deformation in Eq.~\eqref{eq:deformed-warp} thus provides a regular interpolation between a fundamental thick brane and a configuration with internal structure. Although it changes the local behavior of the geometrical scalars, the auxiliary fields, and the matter distribution, it preserves the asymptotically $\mathrm{AdS}_{5}$ character of the bulk and the localization of the effective gravitational interaction.

\section{Gravitational Tensor Modes}\label{rrm}

\subsection{The zero and massive modes}

Now, let us analyze our model linearized around Minkowski spacetime and demonstrate that the tensor sector contains the usual two transverse, massless polarization modes. At the same time, the boundary term $B$ introduces an additional scalar degree of freedom. To accomplish our purpose, we perform the tensor decomposition on the warped background within the scalar representation of $f(Q,B)$ gravity.

To study the perturbations, we introduce tensor perturbations in the four-dimensional sector, namely, \cite{Csaki,Bazeia1}
\begin{align}
    ds^{2}=\mathrm{e}^{2A(y)}\left[\eta_{\mu\nu}+h_{\mu\nu}(x,y)\right]dx^\mu dx^\nu+dy^{2},
\end{align}
adopting the transverse-traceless gauge, i.e.,
\begin{align}
\partial^\mu h_{\mu\nu}=0, \quad \eta^{\mu\nu}h_{\mu\nu}=0, \quad \mathrm{and} \quad h_{\mu 5}=h_{55}=0.
\end{align}
We noted that the scalar perturbations decouple upon imposing $\delta\chi=\delta\Phi=\delta\Psi=0$. Consequently, one obtains
\begin{align}
\Phi Q+\Psi B=(\Phi+\Psi)Q-\Psi\mathring R=\Phi\mathring R+(\Phi+\Psi)B,
\end{align}
where the coefficient of the graviton’s principal kinetic operator is $\Phi$. The terms proportional to $\Psi$ in the nonmetricity sector cancel those arising from $-\Psi \mathring{R}$. After discarding total derivative terms and using the background field equations, the quadratic tensor action takes the form
\begin{align}
S^{(2)}_{T}=\frac{1}{8}\int\,d^{4}x\,dy\,\mathrm{e}^{4A}\Phi\left[h_{\mu\nu}'h^{\prime\mu\nu}+\mathrm{e}^{-2A}\partial_\rho h_{\mu\nu}\partial^\rho h^{\mu\nu}\right].
\label{1S2}
\end{align}
Here, we adopted $\kappa_{5}=1$. Furthermore, although $\Psi$ modifies the background solutions, it does not appear explicitly in the tensor propagation operator. This explains why the appropriate no-ghost condition is $\Phi(y)>0$, rather than $\Phi+\Psi>0$.

Meanwhile, varying the action in Eq. \eqref{1S2} yields
\begin{align}
    h_{\mu\nu}''+\left(4A'+\frac{\Phi'}{\Phi}\right)h_{\mu\nu}'+\mathrm{e}^{-2A}\Box^{(4)}h_{\mu\nu}=0,
    \label{eqh}
\end{align}
where $\Box^{(4)}$ is the four-dimensional d'Alembertian operator. Notably, in the limit $\Psi=0$, one recovers the tensor perturbation equation for $f(Q)$ branes. In the limit $\Psi=-\Phi$, one recovers the tensor perturbation equation in the scalar representation of $f(\mathring{R})$ gravity.

By introducing the conformal coordinate $z$, defined by $dz=\mathrm{e}^{-A(y)}dy$, the line element can be expressed as
\begin{align}
ds^{2}=\mathrm{e}^{2A(z)}\left[(\eta_{\mu\nu}+h_{\mu\nu})dx^\mu dx^\nu+dz^{2}\right].
\end{align}
The conformal coordinate $dz=\mathrm{e}^{-A(y)}dy$, allows us to reduce the tensor equation \eqref{eqh} to
\begin{align}
    \partial_z^{2}h_{\mu\nu}+\left(3\partial_z A+\frac{\partial_z\Phi}{\Phi}\right)\partial_z h_{\mu\nu}+\Box^{(4)}h_{\mu\nu}=0.
\end{align}

By performing the Kaluza–Klein decomposition, i.e.,
\begin{align}
    h_{\mu\nu}(x,z)=\epsilon_{\mu\nu}(x)\mathrm{e}^{-\frac32A(z)}\Phi(z)^{-\frac{1}{2}}\psi(z),
\end{align}
with $(\Box^{(4)}-m^2)\epsilon_{\mu\nu}=0$, one obtains
\begin{align}\label{SE}
    \mathcal{H}\psi(z)=m^2\psi(z),
\end{align}
where
\begin{align}\label{H1}
    \mathcal{H}=\left[-\frac{d^{2}}{dz^{2}}+V_T(z)\right].
\end{align}
Here, $\mathcal{H}$ is Hamiltonian-like operator.

Within this framework, one concludes that
\begin{align}
    V_T(z)=\frac{3}{2} A_{zz}+\frac{9}{4} A_z^{2}+\frac{3}{2}A_z\frac{\Phi_z}{\Phi}+\frac{1}{2}\frac{\Phi_{zz}}{\Phi}-\frac{1}{4}\left(\frac{\Phi_z}{\Phi}\right)^{2}.
\end{align}
or, equivalently,
\begin{align}\label{nnnnw}
    V_T[y]=\mathrm{e}^{2A}\left[\frac{15}{4}A'^{\,2}+\frac{3}{2} A''+2A'\frac{\Phi'}{\Phi}+\frac{1}{2}\frac{\Phi''}{\Phi}-\frac{1}{4}\left(\frac{\Phi'}{\Phi}\right)^2\right].
\end{align}

Naturally, Eq. \eqref{H1} can be factorized, leading to
\begin{align}\label{fateq}
    S_+S_-\psi(z)=m^2\psi(z)
\end{align}
where $m^{2}\geq 0$. Therefore, the tensor sector contains no tachyonic modes.

In the massless regime, i.e., $m=0$, the factorized equation \eqref{fateq} yields
\begin{align}
    S_-\psi_0=0
\end{align}
i.e., 
\begin{align}
    \left(-\frac{d}{dz}+\mathcal{W}\right)\psi_0=0.
\end{align}
Therefore, we conclude that
\begin{align}
    \label{llllll2}\psi_0(z)=N_0\,\mathrm{e}^{\frac32A(z)}\sqrt{\Phi(z)}.
\end{align}

Imposing the normalization condition, viz.,
\begin{align}
    \int_{-\infty}^{+\infty}dz\,|\psi_0|^{2}=N_0^{2}\int_{-\infty}^{+\infty}dy\,\mathrm{e}^{2A(y)}\Phi(y)=1,
\end{align}
one obtains
\begin{align}\label{cond1}
    \int_{-\infty}^{+\infty}dy\,\mathrm{e}^{2A(y)}\Phi(y)<\infty.
\end{align}
Consequently, the condition given in Eq. \eqref{cond1} simultaneously ensures the normalizability of the zero mode and the finiteness of the effective four-dimensional Planck mass. Furthermore, one highlights that the original perturbation $h_{\mu\nu}^{(0)}$ is constant along the extra-dimensional coordinate, i.e., 
\begin{align}
    h_{\mu\nu}^{(0)}=\epsilon_{\mu\nu}^{(0)}\mathrm{e}^{-\frac32A}\Phi^{-\frac12}\psi_0=N_0\epsilon_{\mu\nu}^{(0)}.
\end{align}
This allows us to conclude that $h_{\mu\nu}^{(0)}$ is canonically normalized and, consequently, the zero-mode wave function $\psi_{0}$ is localized around the brane.

\subsubsection{The case: A smooth fundamental brane}

Let us first consider the smooth fundamental brane described by the warp function \eqref{fundamental_warp} with $k>0$ determines the characteristic energy scale and thickness of the brane, whereas $p>0$ controls the asymptotic decay of the warp factor. For the scalar field associated with the nonmetricity sector, we adopt
\begin{align}
    \Phi(y)=1+\alpha\sech^{2}(ky),
\end{align}
with $\alpha>-1$, so that $\Phi(y)>0$ throughout the bulk and the tensor sector remains free from ghost-like excitations.

The tensorial perturbations are governed by a Schr\"{o}dinger-like equation, i.e., 
\begin{align}
    \left[-\frac{d^{2}}{dz^{2}}+V_T(z)\right]\psi(z)=m^{2}\psi(z),
\end{align}
where the corresponding effective potential is
\begin{align}\nonumber
    V_T(y)=&k^{2}\sech^{2p}(ky)\Bigg\{\frac{15}{4}p^{2}\tanh^{2}(ky)-\frac{3}{2}p\sech^{2}(ky)\\ \nonumber 
    +&\frac{\alpha\sech^{2}(ky)\left[4p\tanh^{2}(ky)+2-3\sech^{2}(ky)\right]}{1+\alpha\sech^{2}(ky)}\\
    -&\frac{\alpha^{2}\sech^{4}(ky)\tanh^{2}(ky)}{\left[1+\alpha\sech^{2}(ky)\right]^{2}}\Bigg\}.
\label{VT_smooth}
\end{align}
The behavior of the potential \eqref{VT_smooth} is displayed in Figs. \ref{Fig12}[(a)-(c)]. Generally speaking, Figs. \ref{Fig12}[(a)-(c)] show the dependence of $V_T(y)$ on the parameters $k$, $p$, and $\alpha$, respectively.

The effective tensor potential is an even function of the extra-dimensional coordinate, consistently reflecting the $Z_{2}$ symmetry of the background geometry. It exhibits the characteristic volcano-like structure, with a negative well around the brane core and positive barriers separating the central region from the asymptotic bulk. Note that, at the brane, one obtains
\begin{align}
    \lim_{\vert y\vert\to 0}V_T(y)=-k^{2}\left[\frac{3p}{2}+\frac{\alpha}{1+\alpha}\right].
\end{align}
Therefore, the central well favors the localization of the massless graviton. At the same time, the asymptotically vanishing potential implies that the massive Kaluza-Klein modes form a continuous spectrum beginning at $m^{2}=0$, without a finite mass gap. The parameter $k$ sets the overall scale of the potential and controls the effective width of the localization region. Meanwhile, $p$ modifies both the asymptotic suppression and the shape of the potential barriers. The parameter $\alpha$ encodes the influence of the nonmetricity scalar $\Phi$ and directly modifies the structure of the potential near the brane core. For positive $\alpha$, the central well becomes deeper, strengthening the trapping of the tensor zero mode around the brane, see Figs. \ref{Fig12n}[(a)-(c)].
\begin{figure}[ht!]
    \centering
    \subfigure[$V_T(y)$ for different values of the parameter $k$.]{\includegraphics[width=2.8cm,height=3.5cm]{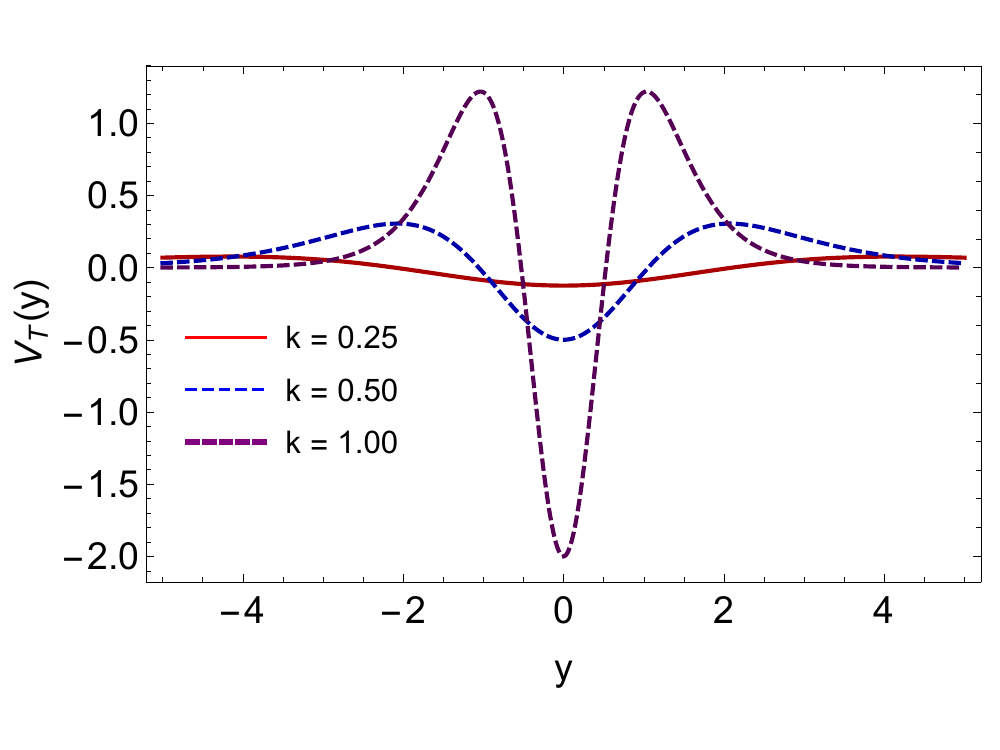}}\hfill
    \subfigure[$V_T(y)$ for different values of the parameter $p$.]{\includegraphics[width=2.8cm,height=3.5cm]{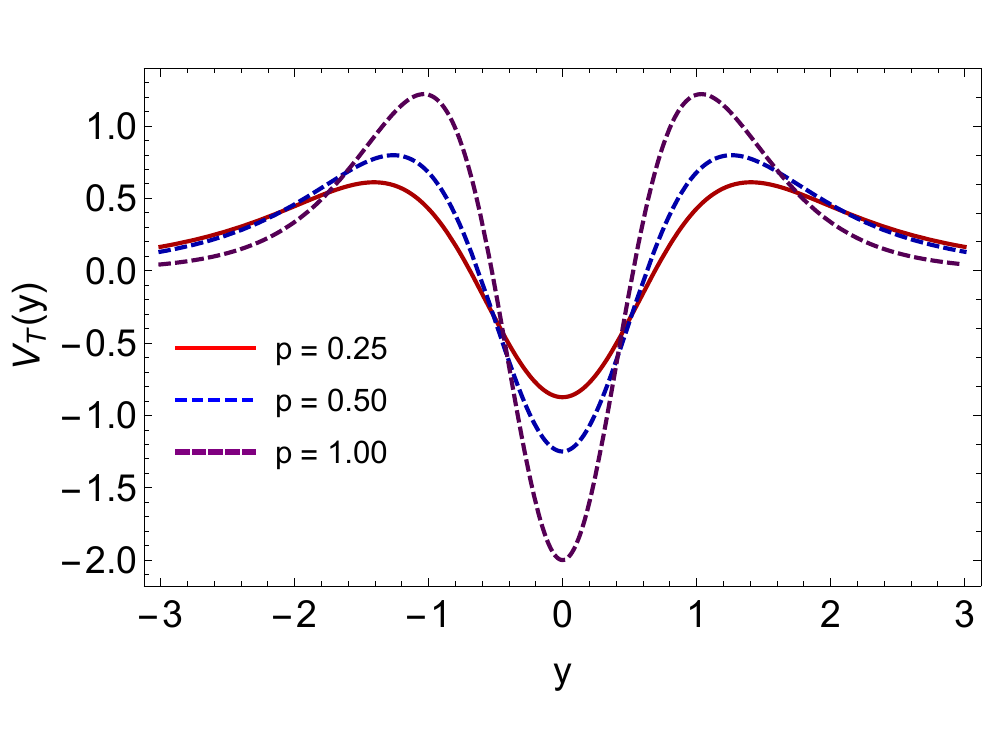}}\hfill
    \subfigure[$V_T(y)$ for different values of the parameter $\alpha$.]{\includegraphics[width=2.8cm,height=3.5cm]{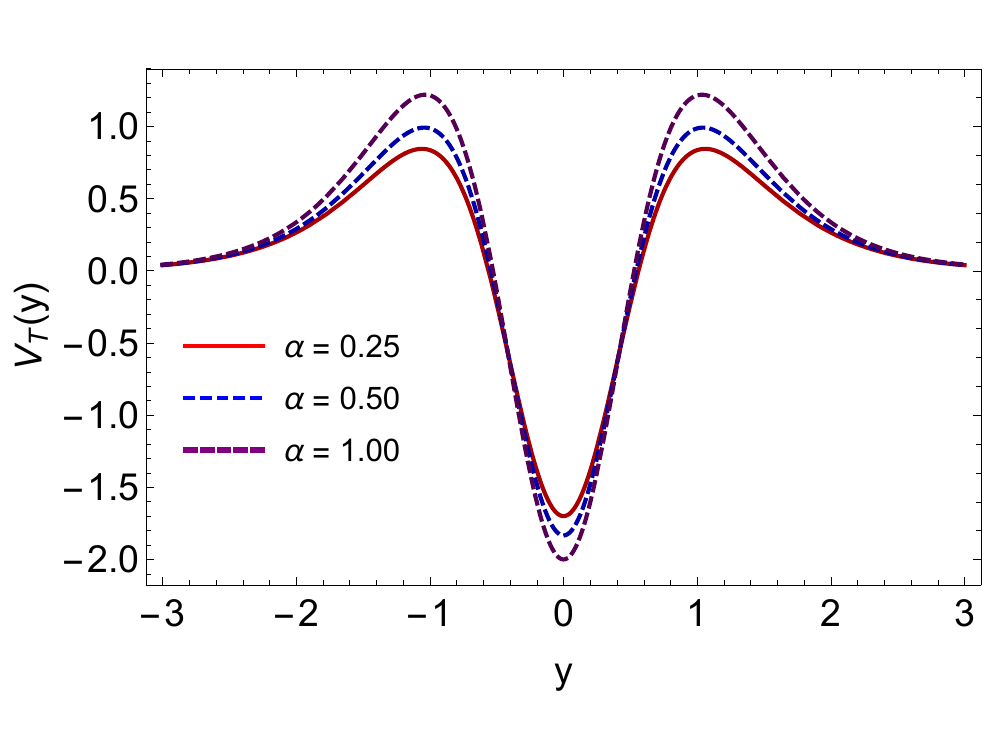}}
    \caption{Profile of the effective Schr\"{o}dinger-like potential $V_T(y)$ governing the tensor perturbations vs. the extra-dimensional coordinate.}
    \label{Fig12n}
\end{figure}

For the smooth brane, the zero-mode profile in terms of the extra-dimensional coordinate reads
\begin{align}
    \psi_{0}(y) = N_{0}\sech^{\frac{3p}{2}}(ky) \sqrt{1+\alpha\sech^{2}(ky)}, \label{tensor_zero_mode_smooth}
\end{align}
where the normalization constant is
\begin{align}
N_{0}=\left\{ \frac{\sqrt{\pi}}{k} \frac{\Gamma(p)}{\Gamma\left(p+\frac{1}{2}\right)} \left[ 1+\frac{2\alpha p}{2p+1} \right] \right\}^{-\frac{1}{2}}.
\label{normalization_constant_smooth}
\end{align}

The behavior of the zero mode \eqref{tensor_zero_mode_smooth} is displayed in Fig.~\ref{Fig13i}[(a)-(c)]. The profile is an even function of the extra-dimensional coordinate, reaches its maximum at the brane core, and vanishes asymptotically as $\vert y\vert\rightarrow\infty$. Therefore, the massless tensor excitation is concentrated around the brane and can be interpreted as the four-dimensional graviton. 
\begin{figure}[ht!]
    \centering
    \subfigure[$\psi_0(y)$ for different values of the parameter $k$.]
    {\includegraphics[width=2.8cm,height=3.5cm]{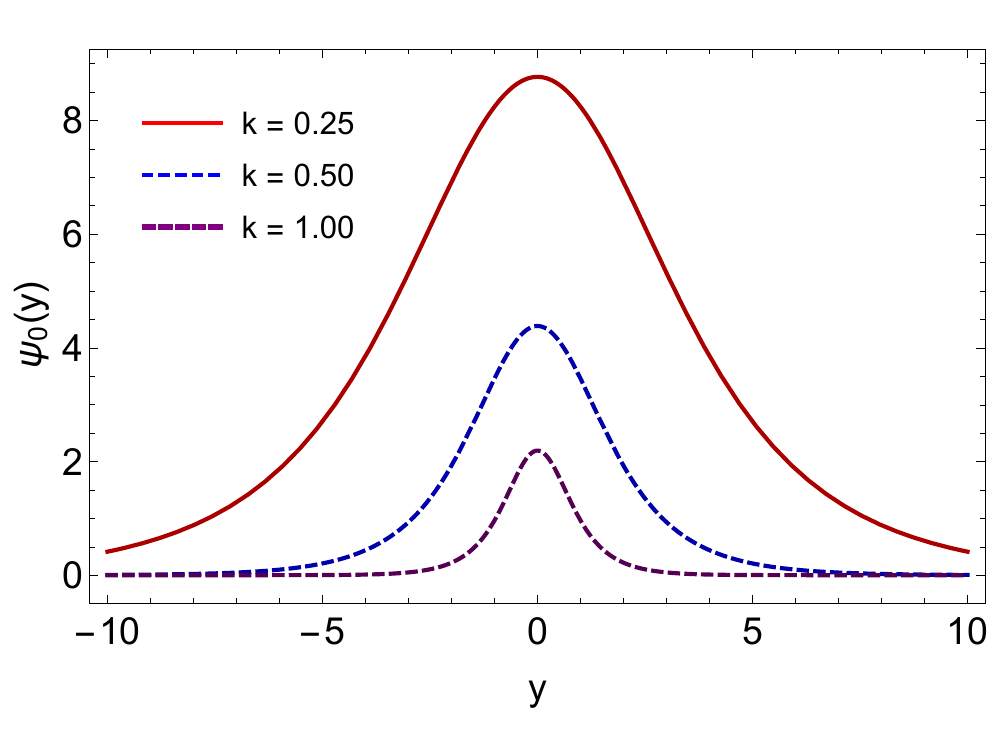}}\hfill
    \subfigure[$\psi_0(y)$ for different values of the parameter $p$.]
    {\includegraphics[width=2.8cm,height=3.5cm]{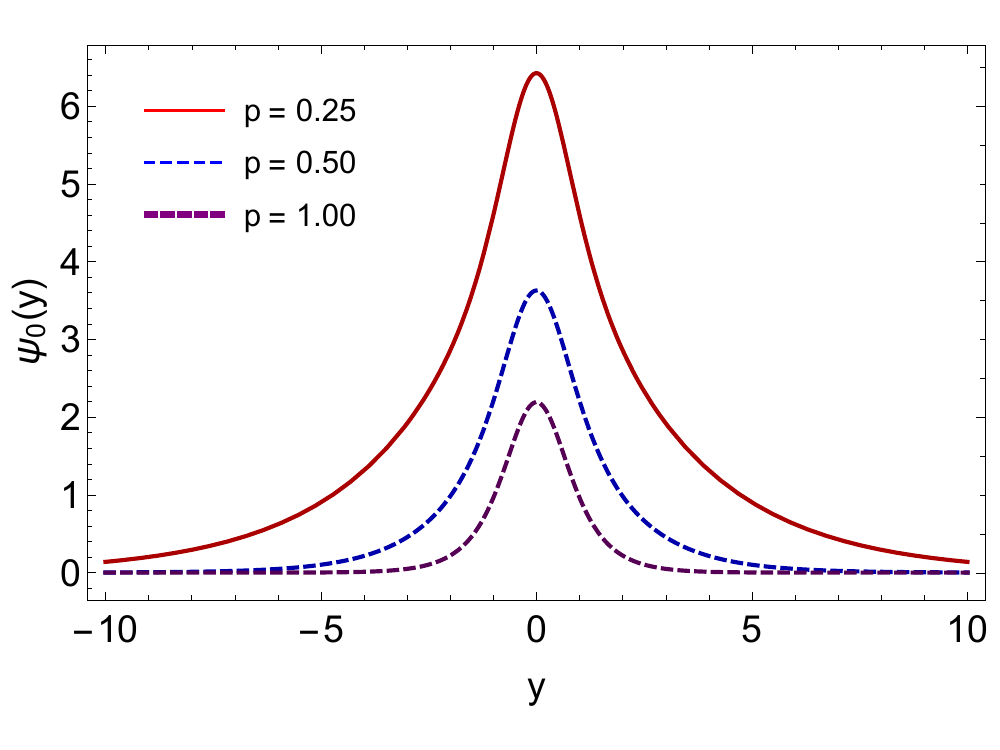}}\hfill
    \subfigure[$\psi_0(y)$ for different values of the parameter $\alpha$.]
    {\includegraphics[width=2.8cm,height=3.5cm]{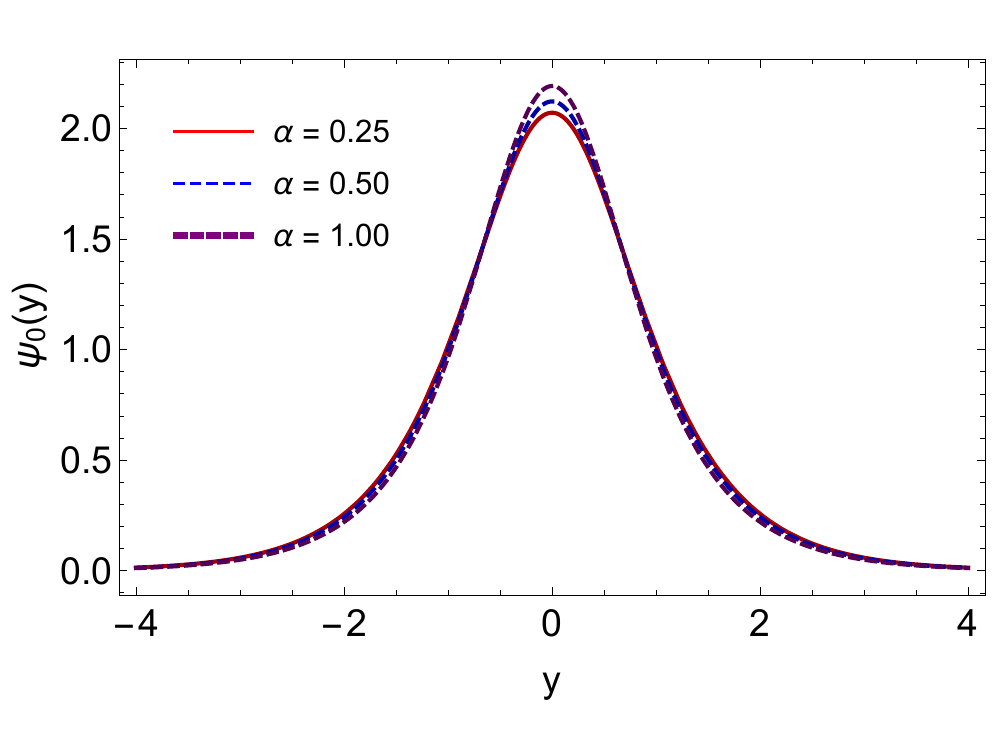}}
    \caption{Profile of the normalized zero mode $\psi_{0}(y)$ vs. the extra-dimensional coordinate.}
    \label{Fig13i}
\end{figure}
Figure \ref{Fig13i}[(a)-(b)] illustrates the behavior of the normalized tensor zero mode $\psi_{0}(y)$ under variations of the parameters $k$, $p$, and $\alpha$. In the three figures, $\psi_{0}(y)$ is an even function, concentrated around the brane core at $y=0$, and vanishes asymptotically as $\vert y\vert\to \infty$, confirming the localization of the massless four-dimensional graviton. As shown in Fig. \ref{Fig13i}(a), increasing $k$ reduces the characteristic width of the zero mode, since $k^{-1}$ determines the length scale of the brane, thereby producing stronger confinement near its core. A similar effect is noted in Fig. \ref{Fig13i}(b), where larger values of $p$ enhance the asymptotic decay of the wave function, leading to a narrower and more strongly localized gravitational state. In Fig. \ref{Fig13i}(c), the parameter $\alpha$, associated with the auxiliary geometric scalar $\Phi(y)$, primarily modifies the profile of the zero mode in the vicinity of the brane. Increasing $\alpha$ enhances the concentration of the wave function around the brane core. Meanwhile, its asymptotic decay remains essentially governed by the warp factor.

Now, let us investigate the massive tensor sector, characterized by $m^{2}>0$. Although the Schr\"{o}dinger-like equation in Eq. \eqref{SE} is naturally formulated in terms of the conformal coordinate $z$, the numerical analysis can be performed directly in terms of the original extra-dimensional coordinate $y$. This procedure is particularly useful because the transformation $dz=\mathrm{e}^{-A(y)}dy$ cannot, in general, be analytically inverted for arbitrary values of the geometrical parameters. Thus, the Schr\"{o}dinger-like equation \eqref{SE} can be rewritten in the extra-dimensional coordinate $y$ as
\small{\begin{align} 
    \left[\frac{d^2}{dy^2}+A'(y)\frac{d}{dy}-\mathrm{e}^{-2A(y)}V_T(y)\right]\psi_{m}(y)=&-\mathrm{e}^{-2A(y)}m^{2}\psi_{m}(y).
    \label{massive-y-explicit}
\end{align}}
with $V_T(y)$ announced in Eq. \eqref{VT_smooth} and $A'(y)=-p\tanh(k y)$.

Eq. \eqref{massive-y-explicit} is dynamically equivalent to the Schr\"{o}dinger-like equation formulated in the conformal coordinate. The factor $\mathrm{e}^{-2A(y)}$ multiplying $m^{2}$ arises exclusively from the coordinate transformation and does not modify the physical Kaluza-Klein spectrum.

Generally speaking, Eq. \eqref{massive-y-explicit} does not admit a closed analytical solution and must be integrated numerically. Since the background geometry is invariant under the reflection $y\rightarrow-y$, the massive eigenfunctions can be classified according to their parity. 

Once $\psi_{m}(y)$ is determined, the corresponding five-dimensional tensor perturbation is reconstructed as
\begin{align}
    h_{\mu\nu}^{(m)}(x,y)=\epsilon_{\mu\nu}^{(m)}(x)
    \mathrm{e}^{-\frac{3}{2}A(y)}\Phi^{-\frac{1}{2}}(y)\psi_{m}(y),
\end{align}
where the four-dimensional tensor satisfies
\begin{align}
    \left(\Box^{(4)}-m^{2}\right)\epsilon_{\mu\nu}^{(m)}(x)=0.
\end{align}

In the asymptotic region, one obtains that the tensor potential vanishes, i.e., 
\begin{align}
    \lim_{\vert y\vert\to\infty}V_{T}(y)=0,
\end{align}
Consequently, the massive wave functions expressed directly in terms of $y$ satisfy
\begin{align}\nonumber
    \psi_{m}(y)\sim&\,C_{1}\exp\left[\frac{im\,\operatorname{sgn}(y)}{2^{p}pk}\mathrm{e}^{pk|y|}\right]+C_{2}\exp\Bigg[-\frac{im}{2^{p}pk}\times\\
    &\operatorname{sgn}(y)\mathrm{e}^{pk|y|}\Bigg],
    \label{e94}
\end{align}
with $\vert y\vert\to\infty$.

Thus, the massive wave functions become increasingly oscillatory toward the asymptotic bulk. Since the effective potential approaches zero at infinity, the Kaluza-Klein massive modes form a continuous spectrum
starting at $m^{2}=0$, and no finite mass gap separates the massless graviton from the massive sector. Furthermore, discrete normalizable massive bound states are not expected. Nevertheless, the positive barriers of the volcano-like potential may temporarily confine some massive modes around the brane, giving rise to quasi-localized resonant
states. Such resonances can be interpreted as metastable massive gravitons with an enhanced probability density near the brane and a finite lifetime before escaping into the five-dimensional bulk.

To examine the massive modes, we numerically investigate the solution equation \eqref{massive-y-explicit} for representative values of the Kaluza-Klein mass and of the geometrical parameters. The resulting eigenfunctions are displayed in Figs. \ref{Figx1} and \ref{Figx2}[(a)-(b)]. Figure \ref{Figx1} shows the profiles of $\psi_m$ for the first massive modes, i.e, when $m=0.447$ and $m=0.548$, keeping $k=p=1$. For these cases, the massive modes exhibit a bounded behavior in the neighborhood of the brane core, followed by increasingly rapid oscillations as they propagate toward the asymptotic bulk. The larger mass produces a stronger phase accumulation and, consequently, a shorter local oscillation scale away from the brane. This behavior is consistent with the asymptotic form given in Eq. \eqref{e94}, according to which the phase of the massive wave function grows rapidly with $\mathrm{e}^{pk|y|}$. Therefore, the increase in the oscillation frequency at large $|y|$ is a consequence of the warped geometry and of expressing the Schr\"{o}dinger problem in terms of the original extra-dimensional coordinate $y$.

An important feature of Fig. \ref{Figx1} is that the massive eigenfunctions do not exhibit exponential suppression in the asymptotic region. Instead, they remain oscillatory and extend into the five-dimensional bulk. This result is consistent with $V_T(y)\rightarrow0$ and confirms that the states with $m^2>0$ belong to the continuous Kaluza-Klein spectrum rather than forming discrete massive bound states. Moreover, the difference between the two profiles is manifested through their oscillation phase and wavelength. Increasing $m$ enhances the effective asymptotic wave number, as expected from the mass-dependent term $m^2\mathrm{e}^{-2A(y)}$ in Eq. \eqref{massive-y-explicit}. However, the volcano-like potential can distort the wave functions near the brane and generate nontrivial scattering through its barriers, sufficiently far from the core the dynamics is dominated by the mass term amplified by the warp factor.

The influence of the geometrical parameters on the massive sector is illustrated in Figs. \ref{Figx2}[(a) and (b)], where the mass is fixed at $m=1.84$. In Fig. \ref{Figx2}(a), the parameter $k$ is varied while $p=1$. Since $k^{-1}$ determines the characteristic thickness of the brane, increasing $k$ corresponds to a thinner gravitational configuration and modifies both the width of the effective interaction region and the rate at which the asymptotic regime is reached. The numerical solutions show that the two modes remain qualitatively similar around the brane core but develop an increasing phase separation as $\vert y\vert$ grows. Furthermore, one notes that the larger value of $k$ produces more rapid oscillations in the outer bulk. This result follows directly from the asymptotic behavior $A(y)\simeq-pk|y|$, for which the effective mass contribution scales as $m^2\mathrm{e}^{-2A(y)}\propto m^2\mathrm{e}^{2pk|y|}$.

In Fig. \ref{Figx2}(b), the parameter $p$ controls the asymptotic strength of the warp factor affecting the phase accumulated by a massive Kaluza-Klein mode along the extra-dimensional coordinate. Larger values of $p$ increase the exponential suppression of $\mathrm{e}^{2A(y)}$ and, equivalently, enhance the factor $\mathrm{e}^{-2A(y)}$ entering the massive-mode equation. Consequently, the oscillations become progressively more rapid away from the brane. Meanwhile, the differences between the solutions are comparatively small close to $y=0$. The numerical behavior in Figs. \ref{Figx2}(a) and \ref{Figx2}(b) show that $k$ and $p$ affect the massive spectrum in complementary ways, e.g.,  $k$ controls the characteristic geometric length scale of the brane, whereas $p$ controls the asymptotic strength of the warping.
\begin{figure}[ht!]
    \centering
    \includegraphics[width=6cm,height=5cm]{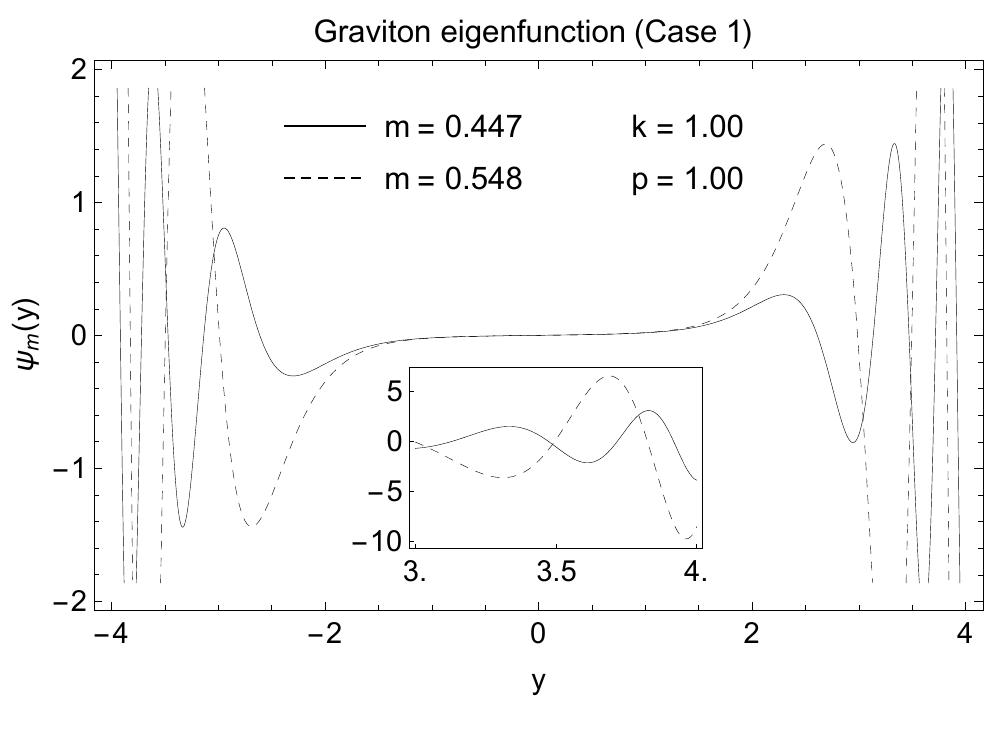}
    \caption{Numerical solutions of the graviton eigenfunctions $\psi_m$ for the smooth fundamental brane with $k=p=1$.}
    \label{Figx1}
\end{figure}
\begin{figure}[ht!]
    \centering
    \subfigure[Massive graviton eigenfunction $\psi_m$ for different values of the parameter $k$ with $p=1$ and $m=1.84$.]{\includegraphics[width=4.3cm,height=3.7cm]{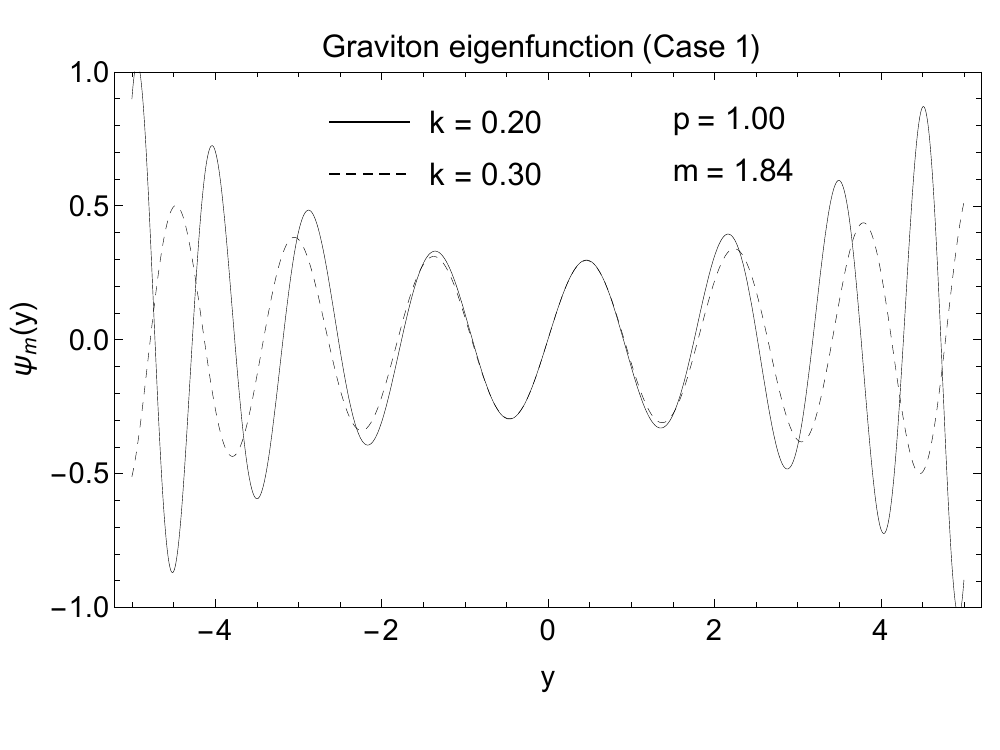}}\hfill
    \subfigure[Massive graviton eigenfunction $\psi_m$ for different values of the parameter $p$ with $k=1$ and $m=1.84$.]{\includegraphics[width=4.3cm,height=3.7cm]{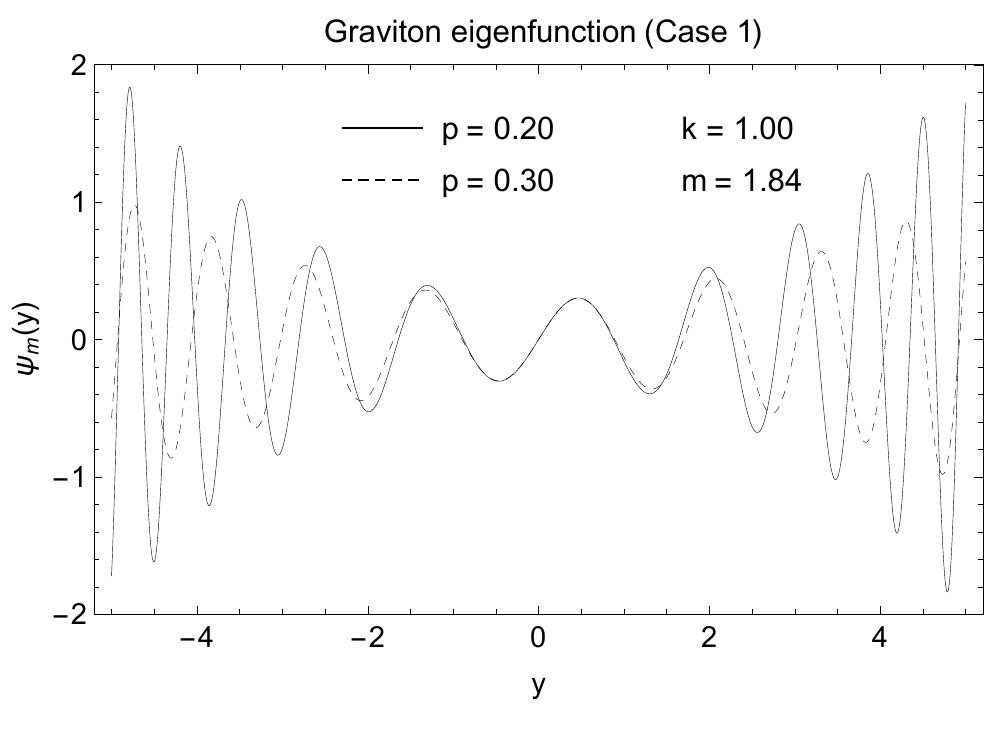}}
    \caption{Numerical profiles of the massive tensor eigenfunctions $\psi_m(y)$ vs. the extra-dimensional coordinate: the case of the smooth fundamental brane.}
    \label{Figx2}
\end{figure}

\subsubsection{The case: Deformed brane}

Now, let us examine how the internal structure generated by the deformed brane affects the localization of the tensor modes. Within this framework, we consider the warped background introduced in Eq. \eqref{eq:deformed-warp}, i.e., $A_{\delta}(y)=-p\mathrm{ln}\,[\cosh(ky)]+\delta\tanh^{2}(ky)$, while keeping the auxiliary scalar associated with the nonmetricity sector in the form $\Phi(y)=1+\alpha,\sech^{2}(ky)$, with $\alpha>-1$. Therefore, the condition $\Phi(y)>0$ is satisfied throughout the bulk, and the tensor sector is free of ghost-like excitations. Note that the boundary scalar $\Psi$ does not enter explicitly into the tensor propagation equation. Its effects are encoded indirectly through the background geometry.

For convenience, let us define the compact notation $s(y)\equiv \sech^{2}(ky)$ and $t(y)\equiv \tanh(ky)$, so that 
\begin{align}
	\mathrm{e}^{2A_{\delta}(y)}= \mathrm{sech}^{2p}(ky) \exp\left[2\delta\tanh^{2}(ky)\right].
\end{align}
Thus, by using
\begin{align}
    A'_\delta=k \tanh(ky)[-p+2\delta\,\mathrm{sech}^2\,(k y)]
\end{align}
and
\begin{align}
    A''_\delta(y)=k^2\mathrm{sech}\,(ky)[-p-4\delta+6\delta \mathrm{sech}^{2}(ky)],
\end{align}
together with
\begin{align}\label{Ph1}
    &\Phi(y)=1+\alpha\,\mathrm{sech}^2\,(k y),\\ \label{Phi2}
    &\Phi'(y)=-2\alpha k\,\mathrm{sech}^2(ky)\mathrm{tanh(k y)},\\ 
    &\Phi''(y)=-2\alpha k^2\,\mathrm{sech}^2(ky)[2-3\sech^2(k y)].
    \label{Phi3}
\end{align}

Therefore, the effective potential in Eq. \eqref{nnnnw} becomes
\begin{align}\nonumber
    V_{T}^{(\delta)}(y)=&k^2\,\mathrm{sech}^{2p}(ky)\mathrm{e}^{2\delta\,\tanh^2(k y)}\Bigg\{\frac{15}{4}\tanh^{2}(k y)[p-\\ \nonumber
    2&\delta\,\mathrm{sech}^{2}(ky)]^2+\frac{3}{2}\mathrm{sech}^{2}(k y)[-p-4\delta+6\delta\mathrm{sech}^2(ky)]\\ \nonumber
    +&\frac{\alpha\mathrm{sech}(ky)[4\tanh^2(k y)(p-2\delta \sech^{2}(k y)]+2]}{1+\alpha\,\mathrm{sech}^2(k y)}-\\
    &\frac{3\alpha\mathrm{sech}^4(ky)}{1+\alpha\,\mathrm{sech}^2(ky)}-\frac{\alpha^2\,\mathrm{sech}^4(k y)\tanh^{2}(k y)}{[1+\alpha\,\mathrm{sech}^2(k y)]^2}.
\label{EqVTdeformed}
\end{align}
Note that the limit $\delta\to 0$  continuously reproduces the effective potential obtained for the fundamental smooth brane. Therefore, the deformation does not introduce a distinct tensor sector, but rather modifies the localization potential through the internal restructuring of the warped geometry. Meanwhile, the profiles of $V_{T}^{(\delta)}(y)$ are displayed in Figs. \ref{Fig13x}[(a)-(d)]. The potential remains an even function of the extra-dimensional coordinate. The $\mathbb{Z}_{2}$-symmetry of the deformed background is consistently present. Nevertheless, its internal structure is considerably richer than that obtained for the fundamental brane. Briefly, the deformation produces additional wells and barriers around the brane core, reflecting the redistribution of the gravitational structure into internal layers.
\begin{figure}[ht!]
    \centering
    \subfigure[$V_T(y)$ for different values of the parameter $k$.]{\includegraphics[width=4.3cm,height=3.7cm]{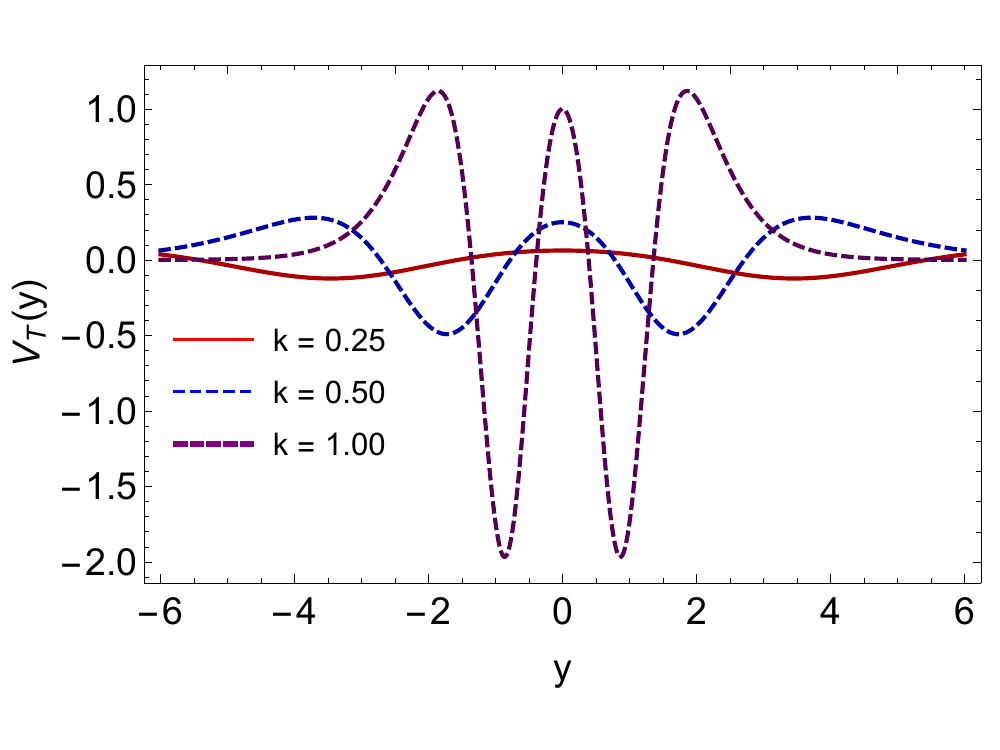}}\hfill
    \subfigure[$V_T(y)$ for different values of the parameter $p$.]{\includegraphics[width=4.3cm,height=3.7cm]{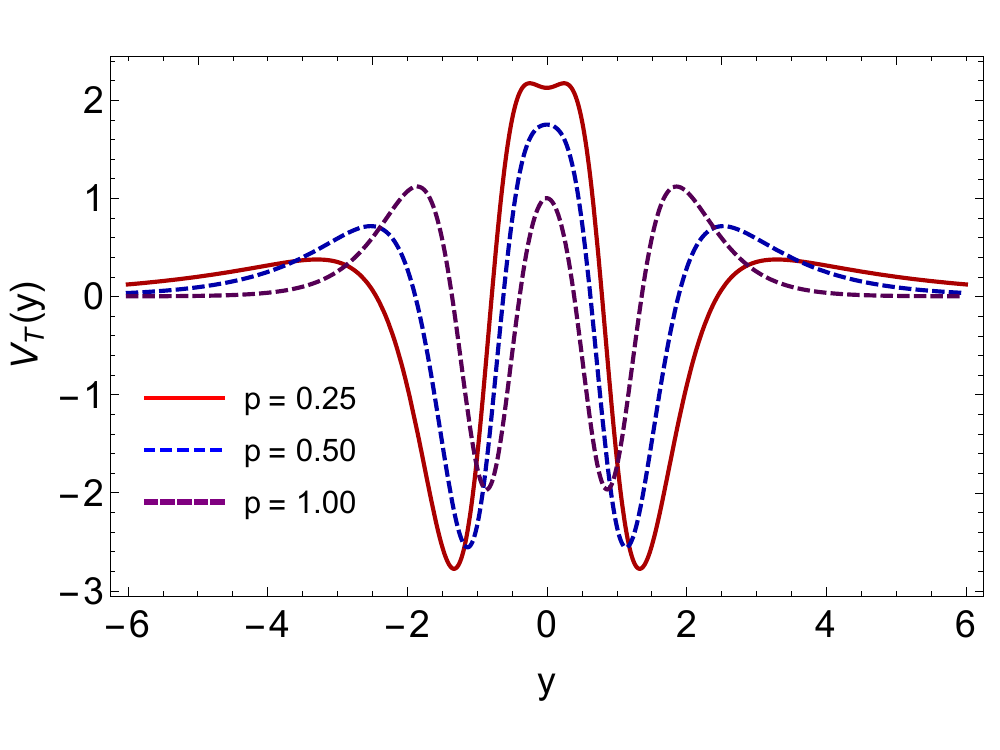}}
    \subfigure[$V_T(y)$ for different values of the parameter $\alpha$.]{\includegraphics[width=4.3cm,height=3.7cm]{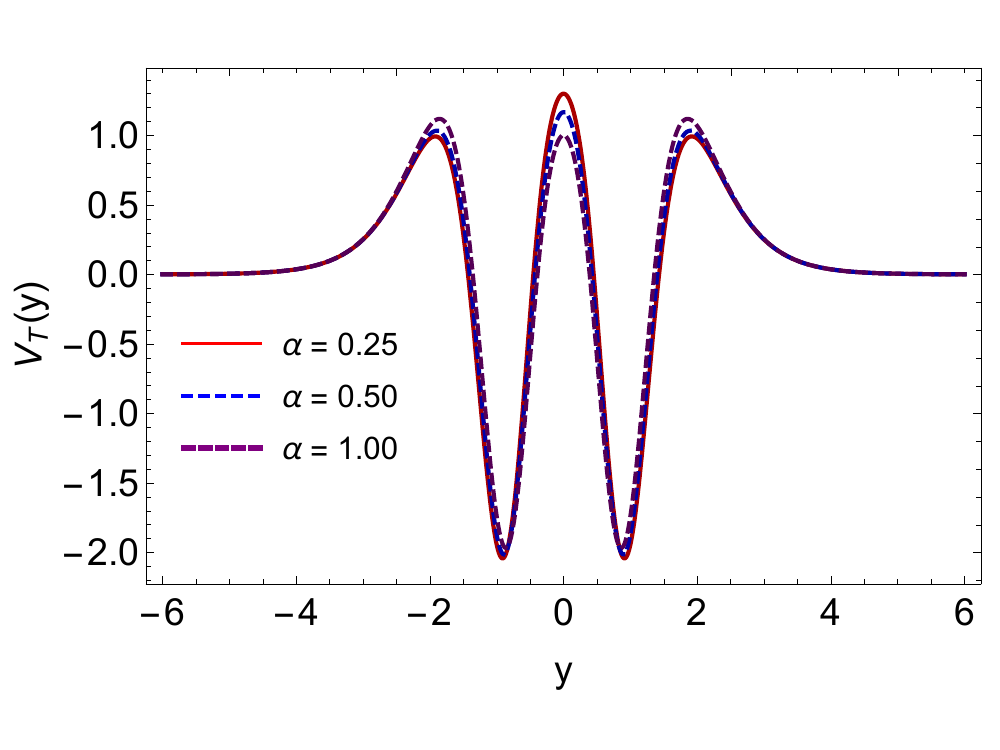}}\hfill
    \subfigure[$V_T(y)$ for different values of the deformation parameter $\delta$.]{\includegraphics[width=4.3cm,height=3.7cm]{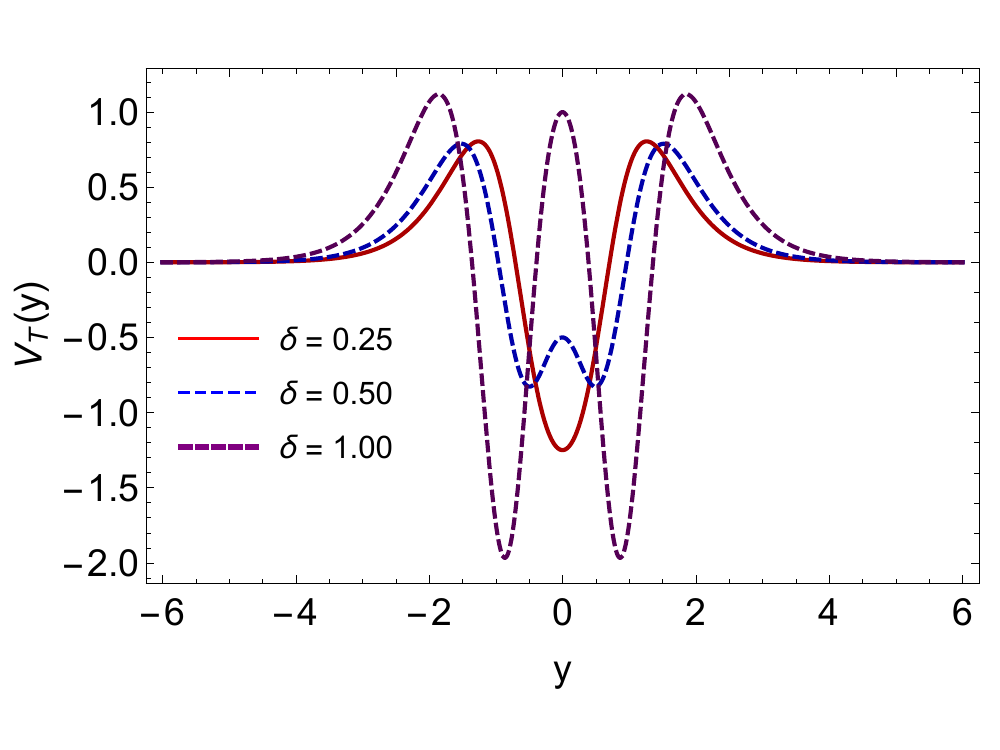}}
    \caption{Profile of the effective Schrödinger-like potential $V_T(y)$ governing tensor perturbations for the deformed brane vs. the extra-dimensional coordinate $y$.}
    \label{Fig13x}
\end{figure}

At the brane center, the potential assumes the values
\begin{align}
V_{T}^{(\delta)}(0)\simeq k^{2}\left[\frac{3}{2}(2\delta-p)-\frac{\alpha}{1+\alpha}\right].
\label{EqVTdeformedOrigin}
\end{align}
This expression exhibits the competition between the geometrical deformation, controlled by $\delta$, and the nonmetricity contribution encoded in $\alpha$. For the fundamental configuration, the first contribution favors the usual central potential well. However, increasing $\delta$ progressively raises the central region and redistributes the trapping structure toward finite values of $\vert y \vert$. The contribution proportional to $\alpha$ acts in the opposite direction for $\alpha>0$, tending to preserve a deeper trapping region around the origin. This behavior can be identified in Figs. \ref{Fig13x}[(a)-(d)]. Fig. \ref{Fig13x}(a) shows that increasing $k$ compresses the internal structures toward the brane while simultaneously increasing the characteristic amplitude of the potential. This follows from the fact that $k^{-1}$ determines the characteristic length scale along the extra dimension. The parameter $p$, displayed in Fig. \ref{Fig13x}(b), modifies the strength of the warped geometry. Thus, one can adjust the depths of the wells and the heights of the surrounding barriers. Meanwhile, Figure \ref{Fig13x}(c) exhibits the effect of the nonmetricity-sector parameter $\alpha$. Since $\Phi(y)-1=\alpha\sech^{2}(ky)$ is strongly localized around the brane, variations of $\alpha$ predominantly affect the central region of the tensor potential, while leaving its asymptotic behavior unchanged. Particularly, for positive $\alpha$, increasing this parameter enhances the negative contribution to $V_T^{(\delta)}(y)$ at the origin, thereby deepening the central trapping region.

The most significant modification is produced by the deformation parameter $\delta$, see Fig. \ref{Fig13x}(d). Increasing $\delta$ changes the potential from the conventional volcano-like configuration into a multi-structured profile characterized by separated wells and barriers. This behavior is the tensor-sector counterpart of the brane-splitting mechanism identified previously through the warp factor, the nonmetricity scalar, the boundary term, the matter field, and the energy density. Thus, the internal structure of the background is directly inherited by the potential governing gravitational fluctuations.

One highlights that the massive Kaluza-Klein sector remains continuous and starts at $m^{2}=0$, exactly as in the fundamental-brane configuration. Furthermore, the factorization of the Hamiltonian-like operator remains valid and guarantees $m^{2}\geq0$. Therefore, the deformation does not introduce tachyonic tensor instabilities.

Let us examine the massless mode. From the general solution in Eq. \eqref{llllll2}, the zero modes are 
\begin{align}\nonumber
    \psi_0^{(\delta)}(y)=&N_{\delta}\,\sech^{\frac{3p}{2}}(ky)\,\exp\,\left[\frac{3\delta}{2}\tanh^{2}(ky)\right]\times \\
    &\sqrt{1+\alpha\sech^{2}(ky)},
    \label{pppp3}
\end{align}
where the normalization constant is\footnote{Here, $_1F_1(a,b;x)$
is the well-known confluent hypergeometric function.}
\begin{align}\nonumber
    N_\delta=&\Bigg[\frac{\sqrt{\pi}}{k}\frac{\Gamma(p)}{\Gamma\left(p+\frac{1}{2}\right)}\Bigg\{\,_1F_1\left(\frac{1}{2},p+\frac{1}{2};2\delta\right)+\frac{2\alpha p}{2p+1}\times\\
    &_{1}F_{1}\left(\frac{1}{2}, p+\frac{3}{2};2\delta\right)\Bigg\}\Bigg]^{-2}
\end{align}
We display the massless modes in Figs. \ref{Figq}[(a)-(d)].
\begin{figure}[ht!]
    \centering
    \subfigure[$\psi_0(y)$ for different values of the parameter $k$.]
    {\includegraphics[width=4.3cm,height=3.7cm]{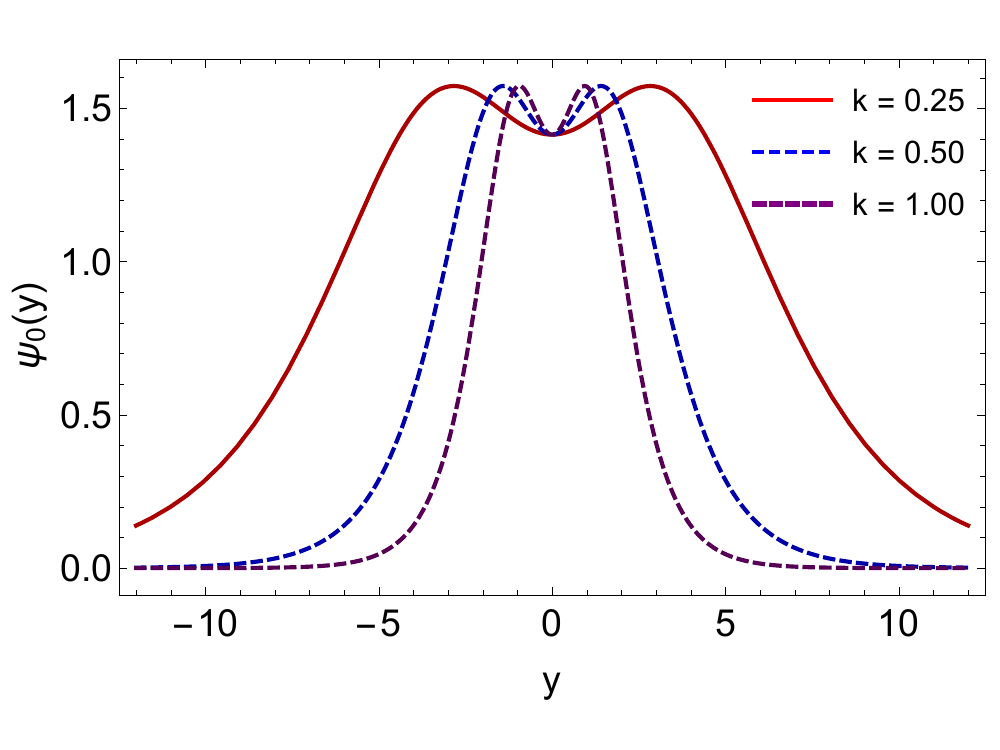}}\hfill
    \subfigure[$\psi_0(y)$ for different values of the parameter $p$.]
    {\includegraphics[width=4.3cm,height=3.7cm]{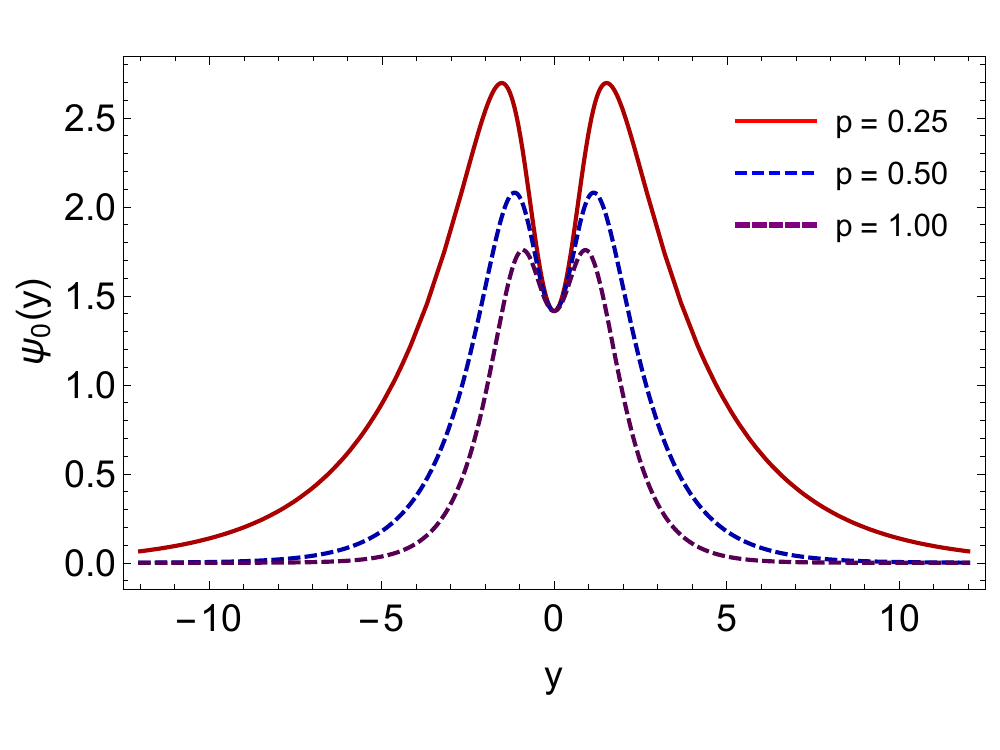}}\\
    \subfigure[$\psi_0(y)$ for different values of the parameter $\alpha$.]
    {\includegraphics[width=4.3cm,height=3.7cm]{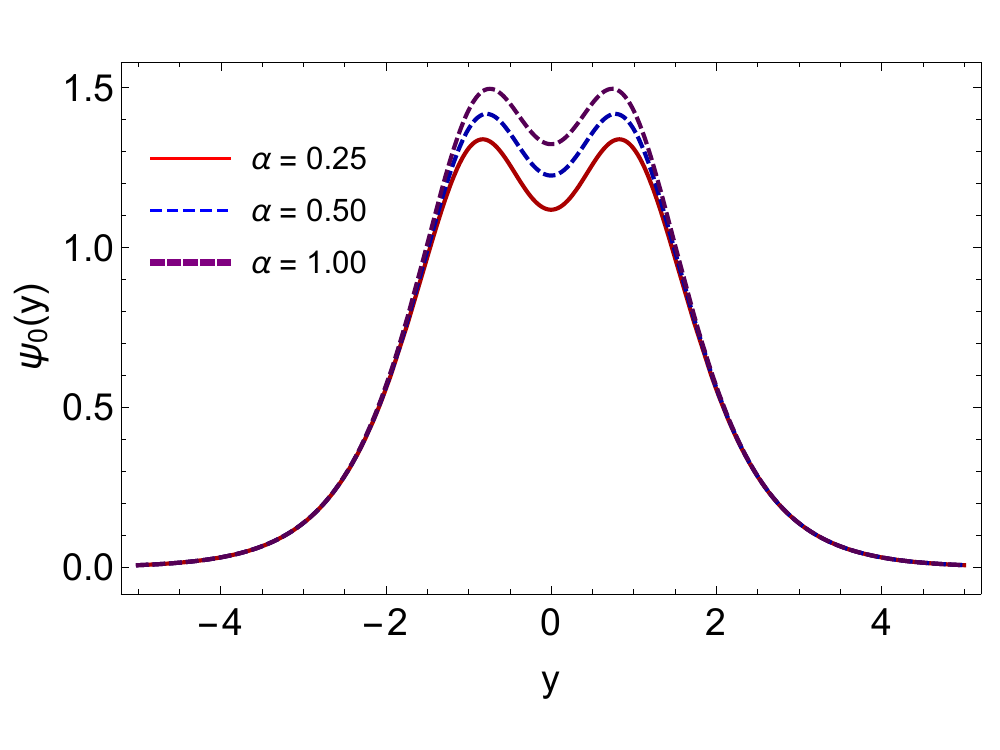}}\hfill
    \subfigure[$\psi_0(y)$ for different values of the parameter $\delta$.]
    {\includegraphics[width=4.3cm,height=3.7cm]{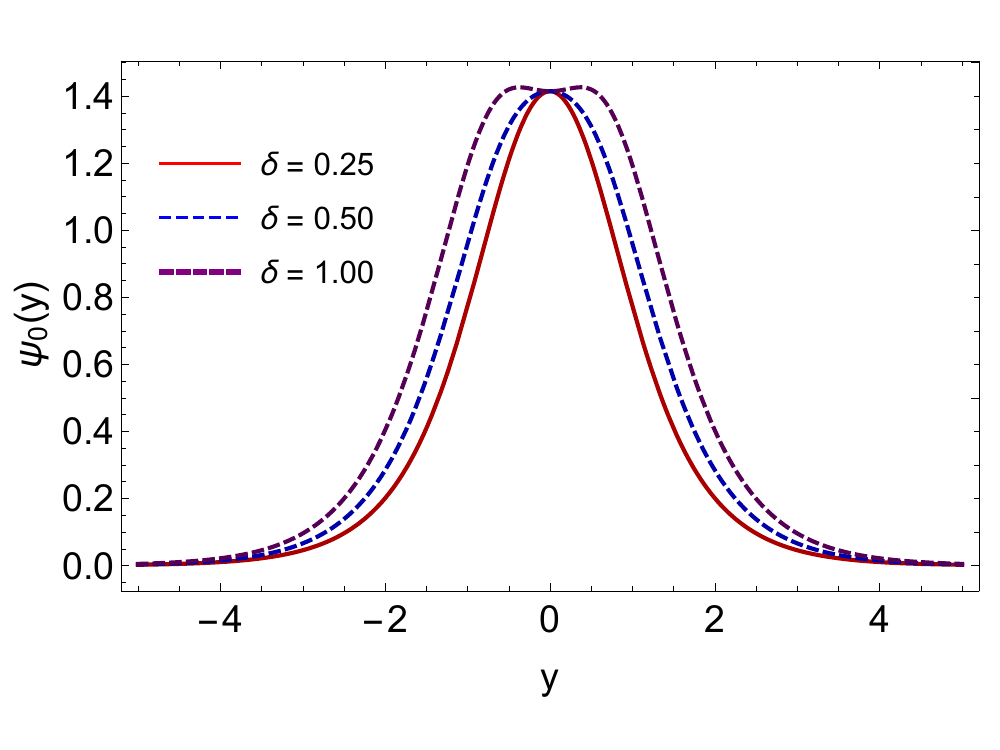}}
    \caption{Profile of the normalized zero mode $\psi_{0}(y)$ vs. the extra-dimensional coordinate: the deformed brane.}
    \label{Figq}
\end{figure}

The profiles displayed in Figs. \ref{Figq}[(a)–(d)] show that the massless tensor mode remains normalizable and localized around the deformed brane, despite the nontrivial internal structure induced by the geometrical deformation. Indeed, from Eq. \eqref{pppp3}, the asymptotic behavior is dominated by the factor $\sech^{3p/2}(ky)$, so that $\psi_{0}^{(\delta)}(y)\sim \mathrm{e}^{-3pk|y|/2}$ for $|y|\rightarrow\infty$. Therefore, neither the deformation parameter $\delta$ nor the localized nonmetricity correction $\alpha$ modifies the exponential suppression of the zero mode in the asymptotic bulk. Fig. \ref{Figq}(a) shows that increasing $k$ compresses the wave function toward the brane, reducing both its characteristic width and the separation between the lateral maxima, consistently with $k^{-1}$ setting the geometrical thickness of the configuration. Similarly, Fig. \ref{Figq}(b) indicates that increasing $p$ strengthens the asymptotic warping. Consequently, it enhances the confinement of the massless graviton, producing a narrower profile and reducing the relative prominence of the split structure. The effect of the nonmetricity-sector parameter $\alpha$, displayed in Fig. \ref{Figq}(c), is essentially localized near the brane core because it enters through $\Phi(y)=1+\alpha\sech^{2}(ky)$. Increasing $\alpha$ enhances the zero-mode amplitude in the central region while leaving its asymptotic decay practically unchanged. By contrast, $\delta$ directly controls the redistribution of the graviton wave function within the brane, see Fig. \ref{Figq}(d). Furthermore, increasing the deformation progressively broadens and flattens the central profile, and favors the displacement of the gravitational amplitude toward finite values of $|y|$. In fact, an expansion of Eq. \eqref{pppp3} around $y=0$ shows that the origin changes from a local maximum to a local minimum when $\delta>p/2+\alpha/[3(1+\alpha)]$, providing a direct criterion for the appearance of a double-peaked graviton zero mode. Thus, the internal structure generated by the deformed geometry can be directly inherited by the localized four-dimensional graviton. Meanwhile, its exponential localization and, consequently, the recovery of effective four-dimensional gravity remain preserved.

Finally, by examining the massive modes through the solution of Eq. \eqref{massive-y-explicit} for the deformed brane, we obtain the solutions displayed in Figs. \ref{Fig13xi}[(a)–(d)].
\begin{figure}[ht!]
    \centering
    \subfigure[Graviton eigenfunctions $\psi_m$ for the smooth fundamental brane with $k=p=\delta=1$.]{\includegraphics[width=4.3cm,height=3.7cm]{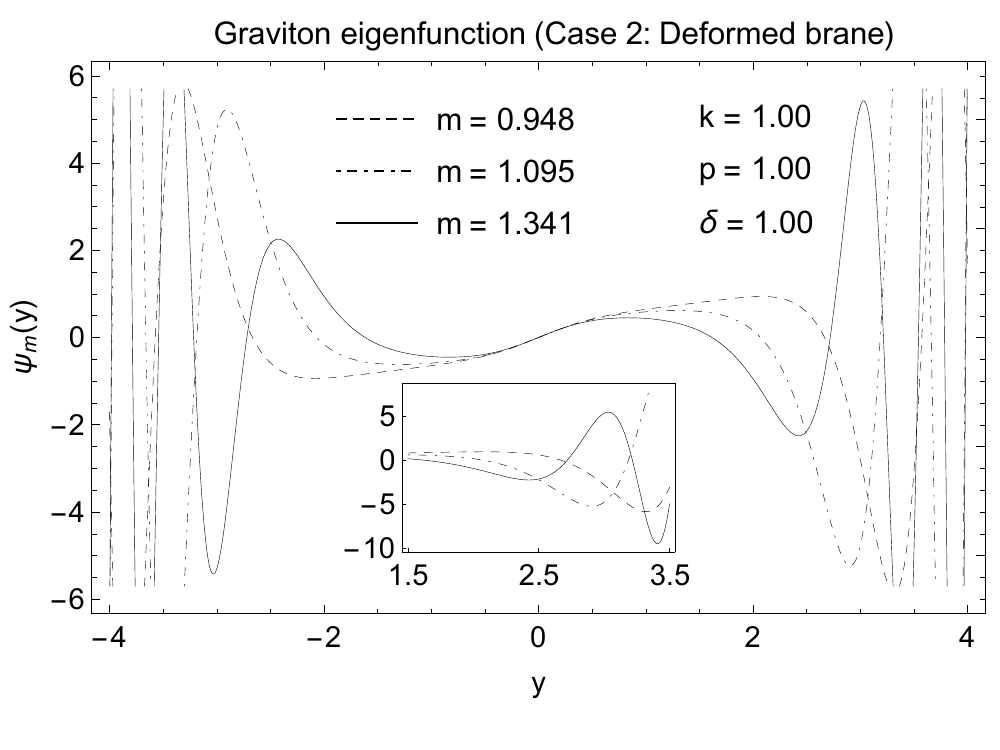}}\hfill
    \subfigure[Solution for the graviton eigenfunctions $\psi_m$ for the smooth fundamental brane with $k=p=\delta=1$.]{\includegraphics[width=4.3cm,height=3.7cm]{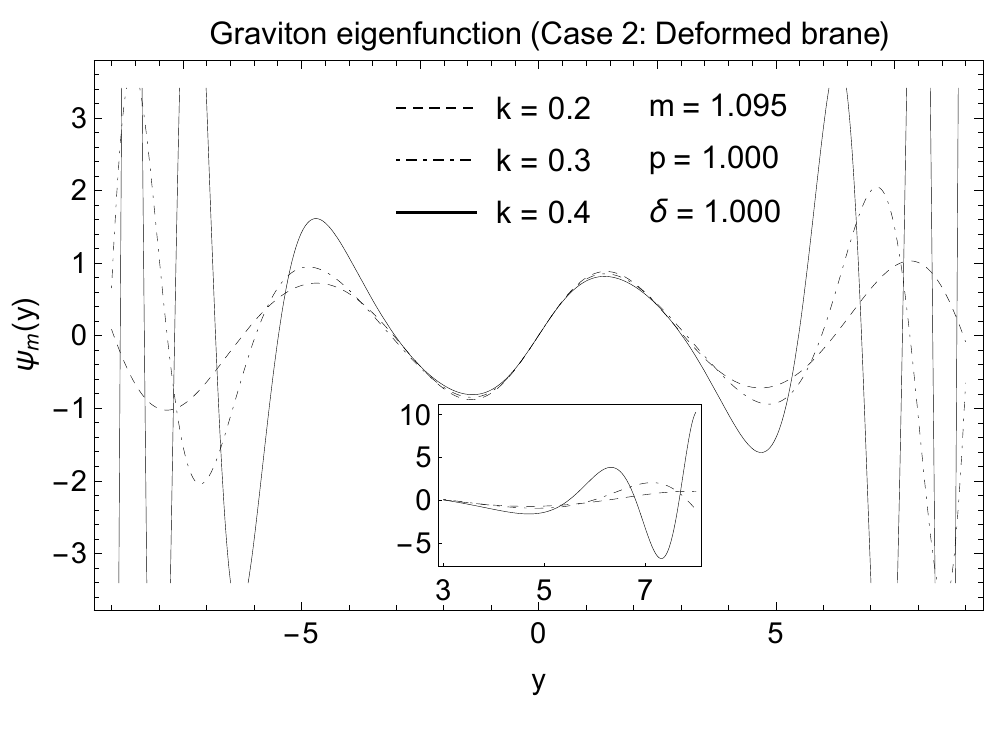}}
    \subfigure[[Massive graviton eigenfunction $\psi_m$ for different values of the parameter $k$ with $p=\delta=1.00$ and $m=1.095$.]{\includegraphics[width=4.3cm,height=3.7cm]{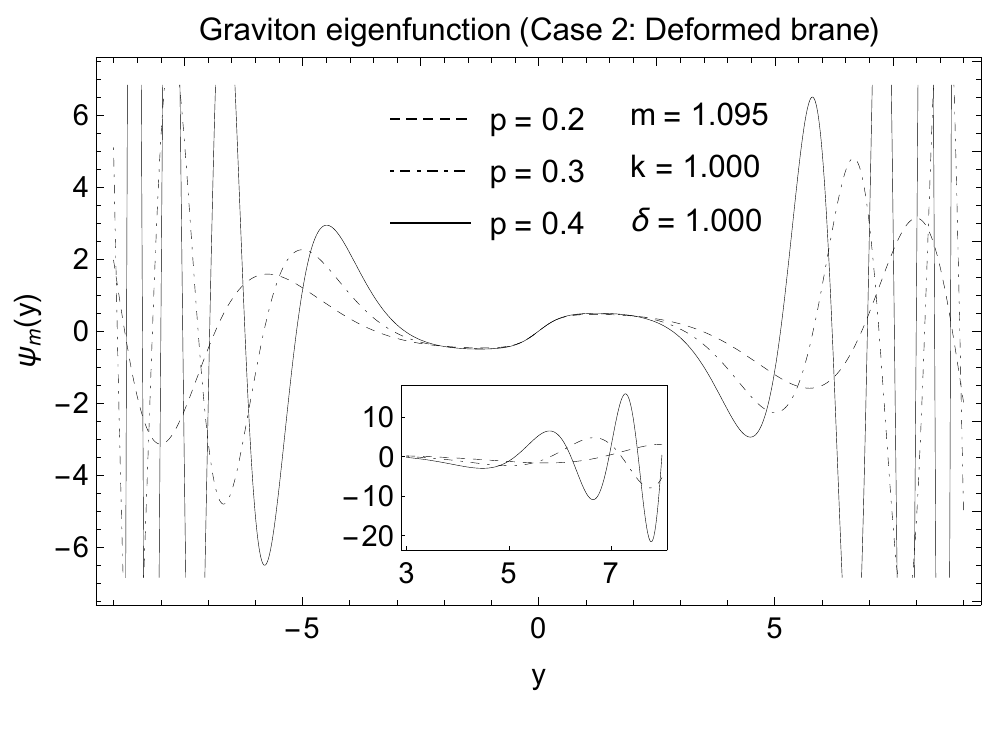}}\hfill
    \subfigure[Massive graviton eigenfunction $\psi_m$ for different values of the parameter $p$ with $k=\delta=1.00$ and $m=1.095$.]{\includegraphics[width=4.3cm,height=3.7cm]{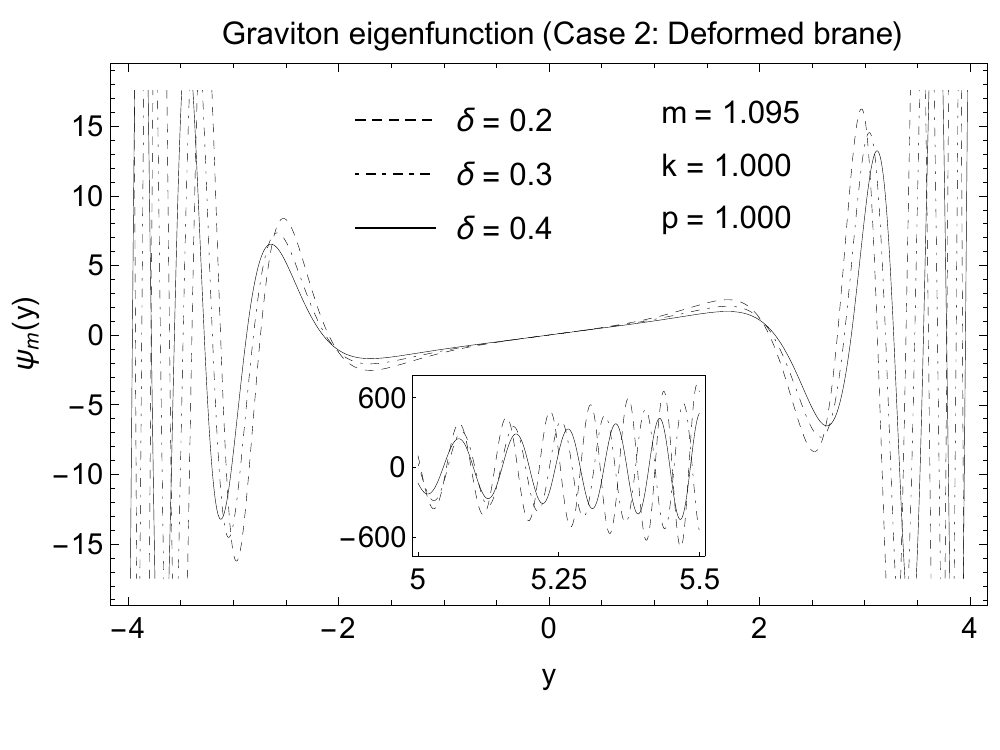}}
    \caption{Numerical profiles of the massive tensor eigenfunctions $\psi_m(y)$ vs. the extra-dimensional coordinate: the case of the deformed brane.}
    \label{Fig13xi}
\end{figure}

Figures \ref{Fig13xi}[(a)-(d)] display the behavior of the massive tensor Kaluza-Klein modes in the deformed-brane background and show how their propagation is affected by both the Kaluza-Klein mass and the geometrical parameters. In Fig. \ref{Fig13xi}(a), varying $m$ while keeping $k=p=\delta=1$ reveals that larger masses lead to a stronger phase accumulation and shorter local oscillation wavelengths, particularly away from the brane, as expected from the effective contribution $m^{2}\mathrm{e}^{-2A_{\delta}(y)}$ entering the massive-mode equation. The nontrivial modulation of the eigenfunctions near the origin reflects scattering by the multiple wells and barriers of the internal structure. Figures \ref{Fig13xi}(b) and \ref{Fig13xi}(c), obtained for fixed $m=1.095$, show that increasing $k$ and $p$ enhances the oscillatory behavior of the massive modes in the outer bulk. $k$ controls the characteristic thickness of the brane compressing the interaction region, whereas $p$ strengthens the asymptotic warping and increases the effective wave number of the Kaluza-Klein excitations. Finally, Fig. \ref{Fig13xi}(d) exhibits the specific influence of the deformation parameter $\delta$. Since $\delta$ primarily reorganizes the geometry in the neighborhood of the brane without changing its asymptotic $\mathrm{AdS}_5$ decay rate, its effect is manifested through pronounced changes in the amplitude, phase, and local oscillatory structure of the massive eigenfunctions within and around the brane core. Nevertheless, in all four figures the modes remain oscillatory and extend into the asymptotic bulk rather than being exponentially suppressed, consistently identifying them as propagating states of the continuous massive Kaluza-Klein spectrum.

\section{Conclusions}
\label{conclusion}

In this work, we have investigated five-dimensional thick braneworlds within the scalar representation of $f(Q, B)$ gravity, where the nonmetricity scalar $Q$ and the boundary term $B$ are promoted to two independent auxiliary scalar degrees of freedom through a Legendre transformation. This formulation converts the original higher-order gravitational theory into an equivalent scalar-tensor representation described by the geometric fields $\Phi=f_Q$ and $\Psi=f_B$, together with the interaction potential $U(\Phi,\Psi)$. Working in the coincident gauge, we derived the complete set of background equations for warped geometries sourced by a canonical scalar field and explicitly verified that the formalism reproduces the scalar representations of both $f(Q)$ and $f(R)$ gravity in the appropriate limits. These consistency checks establish the robustness of the scalar representation and provide a unified framework for studying braneworld solutions in generalized symmetric teleparallel gravity.

Rather than assuming a specific functional form for the gravitational Lagrangian, we adopted a reconstruction procedure in which the warp factor and the auxiliary geometric scalars are prescribed, while the matter scalar field, its self-interaction potential, the geometric potential, and the on-shell gravitational function are reconstructed directly from the field equations. This approach allows one to disentangle the individual contributions of the nonmetricity and boundary sectors to the brane dynamics while preserving the physical consistency.

For the fundamental thick-brane configuration, we found regular asymptotically AdS$_5$ solutions supported by smooth kink-like scalar fields. The nonmetricity scalar and the boundary term remain finite over the entire bulk and approach constant values asymptotically, confirming that the modified geometric effects are localized around the domain wall. Meanwhile, one notes that the bulk preserves its anti-de Sitter character. The auxiliary scalar $\Phi$ governs the effective nonmetricity coupling and smoothly approaches its general-relativistic limit away from the brane, whereas $\Psi$ controls the localized boundary contribution. The reconstructed geometric potential is everywhere regular, and the corresponding energy density exhibits a single localized maximum centered at the brane core, confirming the existence of a regular fundamental thick brane. Furthermore, our analysis shows that the parameter $k$ primarily determines the localization scale, and $p$ controls the strength of the warped geometry. Moreover, the auxiliary parameters $\alpha$ and $\beta$ regulate independently the localized contributions associated with the nonmetricity and boundary sectors.

Subsequently, we generalized the model by introducing a localized deformation of the warp factor through the parameter $\delta$. Unlike the remaining parameters, $\delta$ modifies only the internal geometry of the brane without changing the asymptotic AdS$_5$ behavior of the bulk spacetime. A particularly important result is the identification of the critical value $\delta=p/2$, at which the warp factor changes its local structure and the geometry undergoes a continuous transition from a single gravitational core to a split configuration. This geometrical transition is consistently reflected by every physical quantity analyzed in this work. The warp factor develops two localized maxima, the nonmetricity scalar acquires a multi-peak distribution, the boundary term changes sign at the brane center, the geometric potential evolves from a single-well into a double-well configuration, the matter scalar develops a double-kink profile, and the energy density splits into two symmetric localized peaks. These results demonstrate that the internal structure of the brane is entirely induced by geometrical effects encoded in the interplay between the nonmetricity and boundary sectors of the theory.

Regarding the gravitational tensor sector, we found that the effective Schrödinger-like operator admits a factorized form, ensuring $m^{2}\geq 0$ and the absence of tachyonic tensor instabilities. The massless graviton mode is normalizable and localized around the brane, with its normalization condition directly related to the finiteness of the effective four-dimensional Planck mass. Meanwhile, for the deformed configuration, the internal structure of the brane is inherited by the tensor potential and by the corresponding zero mode, whose profile can develop a double-peaked structure. Furthermore, the massive modes remain oscillatory in the asymptotic bulk and present a continuous Kaluza-Klein spectrum. Meanwhile, their phase and amplitude are significantly modified by the geometrical parameters and by the deformation parameter $\delta$.

Overall, our results demonstrate that the scalar representation of $f(Q, B)$ gravity provides a powerful and flexible framework for constructing exact thick-brane solutions while simultaneously separating the physical effects associated with nonmetricity and boundary contributions. The reconstruction formalism developed considerably simplifies the analysis of higher-order modified gravity theories and reveals how independent geometric sectors govern the localization properties and internal structure of braneworld configurations.

\section*{Acknowledgment}

This work was supported in part by the São Paulo Research Foundation (FAPESP) through the grants  2025/05176-7 (FCEL), and by the National Council for Scientific and Technological Development (CNPq), grant 151845/2025-5 (FMB) and 420854/2025-8 (CASA).

\section*{Conflicts of interest/Competing interest}

The authors declared that there is no conflict of interest in this manuscript. 

\section*{Data availability}

No data was used for the research described in this article.

\bibliographystyle{unsrt}
\bibliography{refs}

\end{document}